\documentclass{article}

\usepackage{arxiv}

\usepackage[utf8]{inputenc} 
\usepackage[T1]{fontenc}    
\usepackage{hyperref}       
\usepackage{url}            
\usepackage{booktabs}       
\usepackage{amsfonts}       
\usepackage{nicefrac}       
\usepackage{microtype}      
\usepackage{lipsum,enumerate,bm,mathrsfs,yfonts}
\usepackage{graphicx,amsmath,multicol,multirow}
\usepackage{url}
\usepackage{amsthm}
\usepackage[style=apa, backend=biber, natbib=true]{biblatex}
\usepackage{adjustbox,xr,comment} 

\usepackage{amsmath}
\usepackage{amssymb}
\usepackage{amsthm,tikz}
\usepackage{colortbl,caption,xr}
\usepackage{adjustbox}  
\usepackage{graphicx}
\usepackage{lscape}
\usepackage{rotating}
\usepackage{graphicx}
\usepackage{float}
\usepackage{url}
\usepackage{subfigure}
\usepackage{dcolumn}
\usepackage{multirow}
\usepackage{verbatim,booktabs, array,arydshln}
\usepackage{color}
\usepackage{setspace}
\usepackage{bbm}
\usepackage{booktabs}
\usepackage{hhline}
\usepackage{bm}
\usepackage{dsfont,import}
\usepackage{enumerate}
\usepackage{threeparttable}
\usepackage{array}
\usepackage{caption}

\theoremstyle{definition}

\theoremstyle{remark}
\newtheorem*{remark}{Remark}

\makeatletter

\title{Model-robust standardization in stepped wedge cluster randomized trials}

\author{
 Xi Fang \\
    Department of Biostatistics \\
    Yale School of Public Health \\
    New Haven, CT, USA\\
   \And
Xueqi Wang\\
    Department of Biostatistics \\
    Yale School of Public Health \\
    New Haven, CT, USA\\
\And
Patrick J. Heagerty\\
Department of Biostatistics\\
University of Washington \\
Seattle, WA, 98195 \\
\And
Bingkai Wang \\
Department of Biostatistics \\
School of Public Health \\
University of Michigan \\
Ann Arbor, MI, USA\\
  \And
 Fan Li \\
Department of Biostatistics \\
Yale School of Public Health \\
New Haven, CT, USA\\
  \texttt{fan.f.li@yale.edu} \\
}

\begin{document}
\maketitle
\begin{abstract}
Stepped-wedge cluster-randomized trials (SW-CRTs) are widely used in healthcare and implementation science, enabling all clusters to receive the intervention through a staggered rollout. Traditional model-based methods, including generalized estimating equations and mixed models, yield estimates that depend on implicit weighting schemes and parametric assumptions, and therefore may target ambiguous estimands under model misspecification. In this article, we propose a model-robust standardization framework for SW-CRTs that generalizes existing methods from parallel-arm CRTs to address informative sizes. We define causal estimands including horizontal-individual, horizontal-cluster, vertical-individual, and vertical-cluster average treatment effects under a super population framework and introduce a simple procedure that standardizes parametric and semiparametric working models for estimand-aligned analysis. For any specified working model, the resulting estimators remain consistent for their target estimands even if the working regression model is misspecified; moreover, their efficiency improves as the working model more closely approximates the true data-generating process. We evaluate the finite-sample properties of our proposed estimators through extensive simulations. Finally, we illustrate the application of our methods through reanalyses of two real-world SW-CRTs.
\end{abstract}


\section{Introduction}\label{intro}

Stepped-wedge cluster randomized trials (SW-CRTs) are a type of unidirectional crossover design that has gained widespread use in healthcare delivery research \citep{nevins2024adherence}. In SW-CRTs, clusters such as hospitals, schools, or communities sequentially transition from a control condition to an intervention in a staggered and randomized manner. All clusters begin in the control condition, and at predetermined time points, known as steps, a randomly assigned subset of clusters switches from control to intervention. This process continues until all clusters have eventually received the intervention, making the SW-CRT a fully rolled-out design \citep{hussey2007design}. Throughout the trial, data are collected at each time point for all clusters, irrespective of their intervention status, enabling both within-cluster and between-cluster comparisons for possible efficiency improvement over traditional parallel-arm designs. SW-CRTs are particularly useful for evaluating cluster-level interventions that are expected to be beneficial and for which it is practically desirable that all clusters receive the intervention within the study. The staggered crossover scheme permits randomized evaluation without leaving any cluster in control indefinitely, and has become attractive in pragmatic and implementation clinical trials. 

In the general context of cluster randomized trials (CRTs), both \citet{imai2009essential} and \citet{kahan2023estimands} pointed out that a single parallel-arm CRT can be associated with multiple estimands, each answering a different research question. As two typical examples, the individual-average treatment effect (i-ATE) represents the expected difference in outcome due to treatment for the population of individuals across clusters, whereas the cluster-average treatment effect (c-ATE) estimates the expected difference in outcome for the population of clusters. These two estimands not only differ conceptually by the unit of inference but also numerically when the cluster sizes are informative \citep{kahan2023informative,kahan2024demystifying}; that is, cluster size is associated with the outcome and/or the treatment effect. Such differences can arise due to variations in cluster-level characteristics, differential access to resources, or systematic differences in participant baseline risk factors. In the presence of multiple periods, estimand considerations are more complex and require other features beyond the distinction between i-ATE and c-ATE. \citet{lee2025should} recently clarified the conditions when different working models can consistently estimate i-ATE and c-ATE under informative cluster sizes in parallel-arm CRTs with a baseline period. Under SW-CRTs, \citet{chen2024model} defined a class of weighted average treatment effect estimands to account for informative cluster-period sizes, and identified i-ATE, cluster-period ATE (cp-ATE), and period ATE (p-ATE) as three members of typical interest. Furthermore, under a two-period cluster randomized crossover design, \citet{lee2025estimated} inherited the estimands definition in \citet{chen2024model}, but has additionally noted that the c-ATE is the fourth estimand of interest. These prior works have suggested the importance in addressing informative sizes (including informative cluster sizes, period sizes, and cluster-period sizes) for defining estimands in more complex CRTs.


There has been increasing emphasis that the analysis of CRTs should align with the target estimand, particularly in the presence of informative sizes. While traditional model-based approaches, such as generalized estimating equations (GEE) and linear mixed models (LMM), are commonly used to analyze CRTs, without modification, their implicit precision weighting schemes can lead to ambiguity in the estimands interpretation. For example, under parallel-arm randomization, \citet{wang2022two} demonstrated that using an exchangeable correlation structure with inverse cluster size weighting in GEE can result in biased estimates for c-ATE under informative cluster sizes. Similarly, \citet{kahan2024demystifying} explained that LMMs can correspond to implicit estimands that depend on the intracranial correlation coefficient (ICC), leading to biased estimation of both the i-ATE and the c-ATE. In SW-CRTs, these challenges are further complicated by the staggered rollout of interventions. \citet{nevins2024adherence} found that most analyses in SW-CRTs rely on generalized linear mixed models (GLMM) \citep{hussey2007design},
for which assumptions of random effects may directly affect the interpretation of treatment effect coefficient. Although GEE separates the mean model from the working correlation model \citep{li2018}, the correct cluster-period size weighting scheme for any target estimand under a complex correlation structure may be cumbersome to derive and a general estimand-aligned formulation remains less clear in SW-CRTs \citep{wang2024achieve}. 

To address these issues, recent methodological advancements have focused on developing model-robust (also referred to as model-assisted) methods. In parallel-arm CRTs, \citet{su2021model} introduced model-assisted linear regression estimators that provide valid inference without requiring modeling assumptions. \citet{li2025model} introduced a model-robust standardization procedure to consistently estimate i-ATE and c-ATE based on more general parametric regression models, and \citet{wang2024achieve} developed triply-robust and debiased machine learning estimators that additionally address informative within-cluster subsampling. Additionally, \citet{balzer2023two,balzer2019new} developed targeted maximum likelihood estimation (TMLE) to improve the efficiency and robustness of c-ATE estimation while accounting for within-cluster dependencies and missing data. In comparison, there have been much fewer developments on model-robust analysis of SW-CRTs. 
Under a finite-population framework, \citet{chen2024model} proposed model-assisted analysis of covariance (ANCOVA) estimators; they have introduced individual-level data weights that allow targeted estimation of i-ATE, cp-ATE and p-ATE and provided design-based variance estimators. However, their ANCOVA estimators are weighted ordinary least squares estimators derived under the working independence assumption and do not easily generalize beyond other types of working models. Recently, \citet{wang2024achieve} provided a first attempt to clarify that linear mixed model and GEE, when combined with a g-computation procedure, can consistently estimate the c-ATE estimand regardless of working model misspecification in SW-CRTs; they have also extended their model-robustness results to accommodate exposure and/or calendar time-dependent treatment effect structures. Beyond these efforts, there exist few discussions on similar model-robust standardization procedures that can target i-ATE, c-ATE, p-ATE and cp-ATE based on any parametric or semiparametric working models that are routinely used in practice for analyzing SW-CRTs.

In this article, we develop a model-robust standardization framework for analyzing SW-CRTs, addressing key challenges in estimand-aligned inference in the presence of informative sizes. We first define key causal estimands including the horizontal-individual, horizontal-cluster, vertical-individual, and vertical-cluster treatment effects,  extending the definitions of \citet{chen2024model} and \citet{lee2025estimated}. To estimate these quantities, we then characterize a novel class of augmented estimators that standardize the output from standard parametric or semiparametric regression models. These estimators remain consistent for the causal estimands and robust to informative sizes even when the working regression model differs from the true data generating model. However, the magnitude of efficiency gains depends on how closely the working model approximates the true data generating process. Section 5 presents a series of simulations to evaluate the properties of the proposed standardization estimators; in them, we evaluate their robustness by intentionally misspecifying the working model relative to the true data generating mechanism. 
Furthermore, we detail the steps for implementing model-robust estimation in SW-CRTs, enabling researchers to continue applying familiar multilevel regression-based analysis without compromising estimand clarity. To summarize, a key contribution of this work is the development of a generalizable strategy for model-robust inference in SW-CRTs, bridging the gap between theoretical estimand definitions and conventional statistical modeling practice. Our approach is implemented in the \texttt{MRStdLCRT} R package at \url{https://github.com/fancy575/MRStdLCRT}. 


\section{Formalizing the treatment effect estimands in SW-CRTs under informative sizes}\label{estimand}
\subsection{Notation and general definition}
We first consider a cross-sectional SW-CRT consisting of \(I\) clusters (indexed by \(i\)) and \(J\) periods (indexed by \(j\)), and will provide a generalization to closed-cohort designs in Web Appendix A. The trial includes a baseline period (\(j=1\)), during which no clusters received treatment, and a complete rollout period (\(j=J\)), by which all clusters have adopted treatment. The treatment status of cluster \(i\) in period \(j\) is denoted by \(Z_{ij} \in \{0,1\} \), where \(Z_{ij} = 1\) indicates the cluster is treated, and \(Z_{ij}=0\) otherwise. The number of clusters receiving treatment in each period, denoted by \(I_j\), is fixed and known for all \(j\). We define \(N_{ij}\) as the number of individuals included in cluster \(i\) during period \(j\), with the total number of observation in cluster \(i\) given by \(N_i = \sum_{j=1}^{J} N_{ij}\), the total number of observations in period \(j\) by \(N_j = \sum_{i=1}^{I} N_{ij}\), and the overall sample size by \(N = \sum_{i=1}^{I} \sum_{j=1}^{J} N_{ij}\). The treatment adoption follows a randomized staggered rollout pattern, whereby each cluster \(i\) is randomly assigned to initiate treatment at one of the prespecified adoption times \(A_i = a \in \mathcal{A} = \{2,\dots, J\}\). In other words, we consider a simple staggered rollout randomization without additional stratification or covariate-constrained component. As a result, the treatment status of cluster \(i\) in period \(j\) is given by \(Z_{ij} = \mathbb{I}(A_i\leq j) \), indicating that once a cluster initiates treatment, it remains exposed in all subsequent periods. We follow the potential outcomes notation in \citet{chen2024model} and define \(Y_{ijk}(z)\) as the potential outcome under the treatment for individual \(k \in \{1,\dots, N_{ij} \}\) in period \(j\) of cluster \(i\), had the cluster-period been assigned to treatment condition \(Z_{ij}=z\), where \(z=1\) denotes treatment and \(z=0\) denotes control. With this notation, we make the assumption of no anticipation, nor exposure-time varying treatment effect; see Assumptions 1 and 2 in \citet{chen2024model} which formalize this notation with rigorous mathematical statements. 

We denote \(e_j = I_j/I\) as the known propensity score in period $j$ that is fixed by design. In SW-CRT, since all clusters start with control in period \(j=1\), and end with treatment in period \(j=J\), the propensity score \(e_1 = 0\) and \(e_J=1\), violating the treatment positive condition in causal inference \citep{chen2024model}. Under the potential outcomes framework, the extreme propensity score implies that there is no possibility to observe the potential outcome under the opposite treatment at the first and last period, thus the identifiable causal contrasts are only well-defined such that \(e_j \wedge (1-e_j) >0 \), where \(a \wedge b = \min (a,b)\). Let \(f(a,b)\) denote a pre-specified contrast function, for example, \(f(a,b) = a-b\) leads to a mean or risk difference estimand, and \(f(a,b) = a/b\) leads to a risk ratio estimand. We define a general class of weighted average treatment effects in stepped-wedge cluster-randomized trials (SW-CRTs) as  
\begin{equation}
\tau_{\omega} = f\{\mu_{\omega}(1), \mu_{\omega}(0) \}, \label{weight_causal_effect}    
\end{equation}
where \(\mu_{\omega}(z)\) represents the weighted average potential outcome under treatment condition \(Z_{ij}=z\), given by  
\begin{equation}
    \mu_{\omega}(z) = \frac{\mathbb{E}\left[ \sum_{j=2}^{J-1} \omega_{ij} \overline{Y}_{ij}(z) \right] }{\mathbb{E}\left[\sum_{j=2}^{J-1} \omega_{ij} \right]}, \label{weight_potential}
\end{equation}
and \(\overline{Y}_{ij}(z)\) denotes the weighted cluster-period mean potential outcome:
\begin{equation} \label{cp_mean}
  \overline{Y}_{ij}(z) = \frac{\sum_{k=1}^{N_{ij}} \omega_{ijk} Y_{ijk}(z) }{\sum_{k=1}^{N_{ij}}\omega_{ijk}},  
\end{equation}
for \(z\in \{0,1\}\). 

In the above definition, \(\omega_{ijk} > 0\) is a pre-specified individual-level weight that determines the contribution of each individual to the overall estimand and \( \omega_{ij} = \sum_{k=1}^{N_i} \omega_{ijk} \) is the grand weight at cluster-period level. The weights can be further aggregated such that the grand weights at cluster level and period level are defined as \(\omega_i = \sum_{j=2}^{J-1} \omega_{ij}\) and \(\omega_{j} = \sum_{i=1}^{I}\omega_{ij}\), respectively. In most cases, we assume a uniform weighting scheme within each cell such that \(\omega_{ijk}\) is constant for all \(k\in \{1,\dots, N_{ij}\}\) within each cluster-period. The expectation in \eqref{weight_potential} is taken over the super-population of clusters, considering the observed clusters as a random sample, but without a further subsampling step within each cluster. Finally, the summation in \eqref{weight_potential} is from \(j=2\) to \(J-1\), as the baseline period \(j=1\) consists entirely of untreated clusters and therefore does not contribute to the treatment effect estimands definition; this also applies to the final rollout period \(j=J\), where all clusters have received treatment.

\subsection{Choices of weight $\omega_{ijk}$ and specific estimands to address informative sizes}\label{sec:estimands_def}
\begin{figure}[htbp!]
  \centering
  \begin{adjustbox}{angle=90}
    \begin{minipage}{0.95\textheight}
      \centering
      \includegraphics[width=0.85\textwidth]{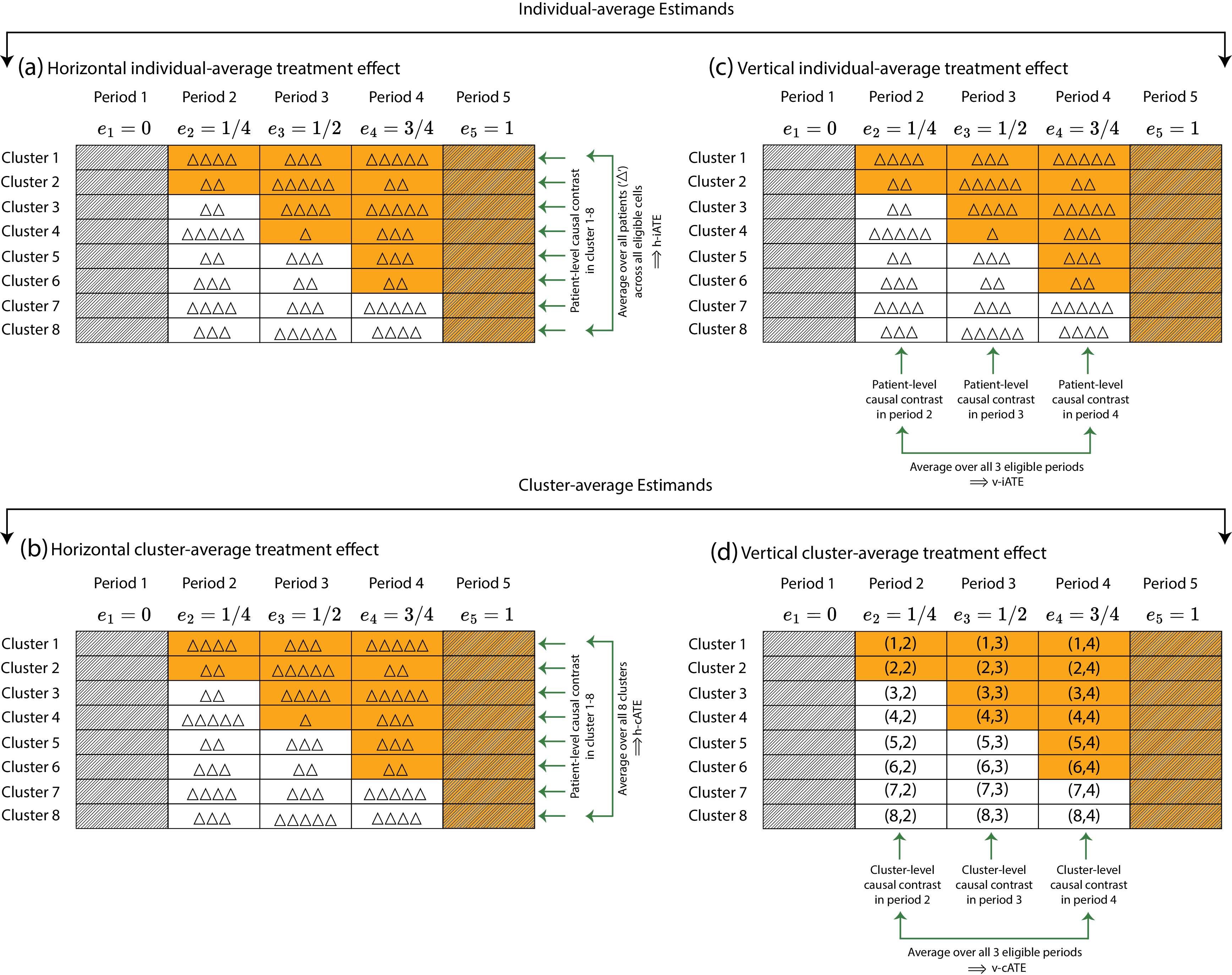}

     \begingroup
    \refstepcounter{figure}
    \tiny         
    \parbox{0.87\textwidth}{%
      \textbf{Figure \thefigure.      \label{fig:ATE}} A schematic illustration of different treatment effect definitions in a hypothetical cross‐sectional SW‐CRT with 8 clusters and 5 periods. The hatch‐mark shaded cells in periods 1 and 5 correspond to periods where the propensity score is 0 or 1, hence the unobserved potential outcomes are not identifiable under our nonparametric causal model framework. White (non‐shaded) cells in periods 2–4 indicate control cluster‐periods and colored cells in periods 2–4 indicate intervention cluster‐periods that are eligible in a sense that the propensity score is neither 0 nor 1 (hence satisfying the overlap assumption). (a) Horizontal individual‐average treatment effect (h‐iATE): the average is computed across all individuals in eligible cluster‐periods, treating each individual equally. (b) Horizontal cluster‐average treatment effect (h‐cATE): outcomes are averaged within each cluster across periods, and then across clusters, giving each cluster equal weight. (c) Vertical individual‐average treatment effect (v‐iATE): individual‐average treatment effects are first computed within each period and then averaged across periods, giving equal weight to each period. (d) Vertical cluster‐average treatment effect (v‐cATE): cluster‐period means are averaged across all cluster‐periods, giving equal weight to each cluster‐period. This specification} corresponds to the cluster‐period average treatment effect described in \citet{chen2024model,lee2025estimated} and generalizes the unit average treatment effect (UATE) from \citet{wang2022two} to longitudinal SW‐CRTs.
    
  \endgroup

    \end{minipage}
  \end{adjustbox}
\end{figure}

In SW-CRTs, different choices of \(\omega_{ij}\) give different estimands of interest, showing different levels of aggregation \citep{chen2024model,lee2025estimated}. As in standard parallel-arm CRTs \citep{kahan2023estimands}, estimands in SW-CRTs can be categorized into two broad types based on their underlying unit of inference: individual-average estimands, where inference is drawn for the population of individuals, and cluster-average estimands, where inference is drawn for the population clusters. Due to the multiple time-period structure of SW-CRTs, where each period can be considered as a mini-CRT, the individual-level and cluster-level aggregation for these mini-CRTs will lead to subtypes of the estimands. These categories can be further subclassified according to how outcomes are aggregated across periods and clusters, resulting in four distinct estimands. We consider time periods as the horizontal dimension and clusters as the vertical dimension shown in Figure \ref{fig:ATE}. Accordingly, individual-average estimands include the horizontal individual-average treatment effect (h-iATE) and the vertical individual average treatment effect (v-iATE). Similarly, cluster-average estimands also have two subtypes: the horizontal cluster-average treatment effect (h-cATE) and the vertical cluster-average treatment effect (v-cATE). Each of these estimands can be generated with different weights \(\omega_{ijk}\) as we explain below.

\emph{Horizontal individual-average treatment effect}. The h-iATE estimand is obtained by pooling individual-level potential outcomes first horizontally across periods and then across clusters. This estimand aligns with its counterpart defined in standard parallel-arm CRTs \citep{kahan2024demystifying}, where each individual is assigned an equal weight in estimand's definition, irrespective of their cluster-period membership (excluding $j=1$ and $j=J$). In the general definition \eqref{weight_potential}, this estimand is generated by setting $\omega_{ijk} = 1$, and $\omega_{ij} = N_{ij}$, which ensures that all individual-level potential outcomes are directly pooled across clusters and periods before defining the final contrast. Formally, the h-iATE is defined as: $\tau_{\text{I}}^{\text{h}} = f\{\mu_{\text{I}}^{\text{h}}(1),\mu_{\text{I}}^{\text{h}}(0)\}$, where
\[
\mu_{\text{I}}^{\text{h}}(z) = \frac{\mathbb{E}\left[\sum_{j=2}^{J-1} N_{ij} \overline{Y}_{ij}(z)   \right]}{ \mathbb{E}\left[ \sum_{j=2}^{J-1} N_{ij} \right] } = \frac{\sum_{j=2}^{J-1} \mathbb{E}\left[\sum_{k=1}^{N_{ij}} Y_{ijk}(z) \right] }{ \sum_{j=2}^{J-1} \mathbb{E}[N_{ij}] }.
\]
As illustrated in Figure \ref{fig:ATE}(a), h-iATE measures the expected change in outcome due to treatment across all individuals pooled from all cluster-periods during the roll-out, and targets an estimand similar to a hypothetical individually-randomized trial with the same set of individuals.

\emph{Horizontal cluster-average treatment effect}. A different horizontally aggregated estimand is the h-cATE, which extends the definition of cluster-average treatment effect estimand in a standard parallel-arm CRT \citep{kahan2023estimands}. In this estimand, individual outcomes are first aggregated across time periods within each cluster, and then the clusters themselves serve as the unit of inference. A key characteristic of h-cATE is that each cluster contributes equally to the final estimand, regardless of its total number of observations across time periods. In the general definition \eqref{weight_potential}, the weights are $\omega_{ijk}=1/\sum_{j=2}^{J-1}N_{ij}$ and $\omega_{ij} = N_{ij}/\sum_{j=2}^{J-1} N_{ij}$; this ensures that each cluster-period is weighted according to its proportion of the total cluster size across periods $j=2$ to $J-1$. As a result, after aggregating across the roll-out periods, each cluster contributes equally to the overall estimand. Formally, h-cATE is defined as: $\tau_{\text{C}}^{\text{h}} = f\{\mu_{\text{C}}^{\text{h}}(1),\mu_\text{C}^{\text{h}}(0)\}$, where
\[
\mu_\text{C}^{\text{h}}(z) = \mathbb{E}\left[ \frac{ \sum_{j=2}^{J-1} N_{ij} \overline{Y}_{ij}(z)  }{\sum_{j=2}^{J-1} N_{ij} } \right] = \mathbb{E}\left[ \frac{\sum_{j=2}^{J-1} \sum_{k=1}^{N_{ij}} Y_{ijk}(z) }{\sum_{j=2}^{J-1}N_{ij} } \right].
\]
As illustrated in Figure \ref{fig:ATE}(b), the h-cATE differs from the h-iATE by balancing the information contribution at the cluster level. This estimand mimics the cluster-average treatment effect in a standard parallel-arm CRT which recruits the same set of individuals observed from period $2$ to $J-1$.

\emph{Vertical individual-average treatment effect}. Different from the h-iATE estimand, the v-iATE estimand is defined by separately constructing the iATE estimand within each roll-out period, before averaging across periods. Thus, the v-iATE assigns equal weight to each time period, ensuring that the contributions to the treatment effect estimand are uniformly averaged over calendar time; within each period, the v-iATE also assigns equal weight to each individual and hence considers the individuals as the unit of inference. In the general definition \eqref{weight_potential}, the specific weighting scheme is defined as: \( \omega_{ijk} = \frac{1}{I \mathbb{E}(N_{ij}) } \), \( \omega_{ij} = \frac{N_{ij}}{I \mathbb{E}[N_{ij}]}, \) leading to the formal definition:
\(\tau_{\text{I}}^{\text{v}} = f\{\mu_{\text{I}}^{\text{v}}(1),\mu_{\text{I}}^{\text{v}}(0) \},
\) where
\[
\mu_{\text{I}}^{\text{v}}(z) = \frac{\mathbb{E}\left[\sum_{j=2}^{J-1} \displaystyle\frac{N_{ij}\overline{Y}_{ij}(z)}{I\mathbb{E}[N_{ij}] }   \right]   }{\mathbb{E}\left[ \sum_{j=2}^{J-1} \displaystyle\frac{N_{ij}}{I\mathbb{E}[N_{ij}]} \right] } = \frac{1}{J-2} \sum_{j=2}^{J-1} \frac{\mathbb{E}\left[ \sum_{k=1}^{N_{ij}}Y_{ijk}(z) \right]}{\mathbb{E}[N_{ij}]}.
\]
As shown in Figure \ref{fig:ATE}(c), different from the h-iATE estimand, the v-iATE estimand targets a quantity that uniformly averages the iATE estimands across several standard parallel-arm CRTs, each defined over the individuals recruited in a distinct roll-out period. By summarizing a period-specific iATE, the v-iATE estimand acknowledges potential treatment effect variation across periods \citep{lee2025analysis}.

\emph{Vertical cluster-average treatment effect}. As a final subtype, v-cATE is obtained by first aggregating the cluster-level mean potential outcomes within each period and then uniformly averaging across periods. Mathematically, this is also equivalent to averaging the mean potential outcomes across all cluster-periods during the roll-out. Therefore, this estimand can also be referred to as the cluster-period average treatment effect estimand \citep{chen2024model,lee2025estimated}. 
In the general definition \eqref{weight_potential}, the weights are defined as $\omega_{ijk} = {1}/{N_{ij}}$, $\omega_{ij} = 1$, such that each cluster-period is given equal weight, and each cluster is given equal weight within each period. The v-cATE is thus formally defined as:
\(
\tau_{\text{C}}^{\text{v}} = f\{\mu_{\text{C}}^{\text{v}}(1), \mu_{\text{C}}^{\text{v}}(0)\},
\)
where
\[
\mu_{\text{C}}^{\text{v}}(z) = \frac{\mathbb{E}\left[ \sum_{j=2}^{J-1} \overline{Y}_{ij}(z) \right]}{J-2} = \frac{1}{J-2} \sum_{j=2}^{J-1} \mathbb{E}\left[ \frac{\sum_{k=1}^{N_{ij}} Y_{ijk}(z) }{N_{ij}}\right].
\]
As shown in Figure \ref{fig:ATE}(d), the v-cATE estimand also implicitly acknowledges potential treatment effect variation across periods, but considers a cluster-level average rather an individual-level average. This estimand bears the spirit of the (cluster-average) period-specific treatment effect estimand in \citet{wang2024achieve}, and targets a quantity that uniformly averages the cATE estimands across several standard parallel-arm CRTs, each defined over a distinct roll-out period.

The four estimands represent distinct perspectives on expressing the treatment effects, with varying degrees of aggregation across individuals, clusters, and calendar periods. 
The horizontal estimands, including the h-iATE and h-cATE, appear to implicitly assume constant treatment effects across rollout periods, aggregating outcomes over time as if the treatment effect were homogeneous. This concept aligns well with that of parallel-arm CRTs with stable and immediate treatment implementation \citep{kahan2024demystifying}. As summarized in Table \ref{tab:estimands}, the h-iATE assigns equal weight to each individual, whereas the h-cATE gives equal weight to each cluster, recovering the familiar weighting schemes previously discussed in simpler parallel-arm CRTs. By contrast, the vertical estimands, including the v-iATE and v-cATE, appear to recognize that the treatment effects may vary across calendar periods (this is to be distinguished from exposure periods as discussed in \citet{kenny2022analysis} and \citet{maleyeff2022assessing}), and thus uniformly aggregate information from period-specific individual-average or cluster-average treatment effect estimands. That is, these estimands first compute either the individual-average or the cluster-average effects within each period and then average across periods, thereby aggregating over several ``mini-parallel CRTs''. Hence, the v-iATE gives equal weight to each period whereas the v-cATE gives equal weight to each cluster-period. 
It is helpful to note that, when the expected cluster-period size remains constant across periods such that $\mathbb{E}[N_{ij}]=\overline{N}$, the v-iATE coincides with the h-iATE since
\[
\mu_{\text{I}}^{\text{h}}(z) = \frac{\sum_{j=2}^{J-1} \mathbb{E}\left[\sum_{k=1}^{N_{ij}} Y_{ijk}(z) \right] }{ (J-2)\overline{N}}=\mu_{\text{I}}^{\text{v}}(z).
\]
Furthermore, under conditions where the potential outcomes have identical marginal means across cluster-periods and cluster-period sizes are noninformative---i.e., independent of the potential outcomes---the h-iATE, h-cATE, v-iATE, and v-cATE all converge to the same estimand, which can be broadly referred to as the average treatment effect; also see \citet{lee2025estimated} for a discussion on connections between these estimands in simpler two-period cluster randomized crossover designs.

\begin{table}[htbp]
    \centering
    \caption{Summary of four different types of treatment effect estimands \(\tau_\omega\) in SW-CRTs.}
    \label{tab:estimands}
\resizebox{\textwidth}{!}{%
    \begin{tabular}{ccccp{3.5cm}}  
    \toprule
        \textbf{Aggregation} & \textbf{Estimand} & \textbf{Weights} & \textbf{Definition} & \textbf{Remarks} \\
    \midrule
       \multirow{8}{*}{Horizontal} 
        & \multirow{4}{*}{h-iATE} 
        & $\omega_{ijk} = 1$ 
        & \multirow{2}{*}{$\mu_{\text{I}}^{\text{h}}(z)  = \frac{\sum_{j=2}^{J-1} \mathbb{E}\left[\sum_{k=1}^{N_{ij}} Y_{ijk}(z) \right] }{ \sum_{j=2}^{J-1} \mathbb{E}[N_{ij}] }$} 
        & \multirow{4}{3.5cm}{Equal weight for each individual} \\  
        & & $\omega_{ij} = N_{ij}$ & & \\
        & & $\omega_{j} = \sum_{i=1}^{I} N_{ij}$ 
        & \multirow{2}{*}{$\tau_{\text{I}}^{\text{h}} = f\{\mu_{\text{I}}^{\text{h}}(1),\mu_{\text{I}}^{\text{h}}(0)\}$} & \\
        & & $\omega_i = \sum_{j=2}^{J-1}N_{ij}$ & & \\
        \cdashline{2-5} 
        & \multirow{4}{*}{h-cATE} 
        & $\omega_{ijk} = 1/\sum_{j=2}^{J-1} N_{ij}$ 
        & \multirow{2}{*}{$\mu_{\text{C}}^{\text{h}}(z) = \mathbb{E}\left[ \frac{\sum_{j=2}^{J-1} \sum_{k=1}^{N_{ij}} Y_{ijk}(z) }{\sum_{j=2}^{J-1}N_{ij} } \right]$} 
        & \multirow{4}{3.5cm}{Equal weight for each cluster} \\  
        & & $\omega_{ij} = N_{ij} / \sum_{j=2}^{J-1} N_{ij}$ & & \\
        & & $\omega_{j} = \sum_{i=1}^{I} \left(N_{ij} / \sum_{s=2}^{J-1} N_{is} \right)$ 
        & \multirow{2}{*}{$\tau_{\text{C}}^{\text{h}} = f\{\mu_{\text{C}}^{\text{h}}(1),\mu_{\text{C}}^{\text{h}}(0)\}$} & \\
        & & $\omega_i = 1$ & & \\
    \midrule
        \multirow{8}{*}{Vertical} 
        & \multirow{4}{*}{v-iATE} 
        & $\omega_{ijk} = 1/\{I\mathbb{E}[N_{ij}]\}$ 
        & \multirow{2}{*}{$\mu_{\text{I}}^{\text{v}}(z) = \frac{1}{J-2} \sum_{j=2}^{J-1} \frac{\mathbb{E}\left[ \sum_{k=1}^{N_{ij}}Y_{ijk}(z) \right]}{\mathbb{E}[N_{ij}]} $} 
        & \multirow{4}{3.5cm}{Equal weight for each period} \\  
        & & $\omega_{ij} = N_{ij} / \{I\mathbb{E}[N_{ij}]\}$ & & \\
        & & $\omega_{j} = 1$ 
        & \multirow{2}{*}{$\tau_{\text{I}}^{\text{v}} = f\{\mu_{\text{I}}^{\text{v}}(1),\mu_{\text{I}}^{\text{v}}(0)\}$} & \\
        & & $\omega_i = \sum_{j=2}^{J-1} \left(  \frac{N_{ij}}{I\mathbb{E}[N_{ij}]} \right)$ & & \\
        \cdashline{2-5} 
        & \multirow{4}{*}{v-cATE} 
        & $\omega_{ijk} = 1/N_{ij}$ 
        & \multirow{2}{*}{$\mu_{\text{C}}^{\text{v}}(z) = \frac{1}{J-2} \sum_{j=2}^{J-1} \mathbb{E}\left[ \frac{\sum_{k=1}^{N_{ij}} Y_{ijk}(z) }{N_{ij}}\right]$} 
        & \multirow{4}{3.5cm}{Equal weight for each cluster-period} \\  
        & & $\omega_{ij} = 1$ & & \\
        & & $\omega_{j} = I$ 
        & \multirow{2}{*}{$\tau_{\text{C}}^{\text{v}} = f\{\mu_{\text{C}}^{\text{v}}(1),\mu_{\text{C}}^{\text{v}}(0)\}$} & \\
        & & $\omega_i = J-2$ & & \\
    \bottomrule
    \end{tabular}
}
\end{table}

\section{Model-robust standardization}\label{method}
\subsection{Constructing the estimator}

In SW-CRTs, only one of the potential outcomes \(\{Y_{ijk}(1), Y_{ijk}(0)\}\) is observed for each individual. To target the estimands introduced in Section \ref{estimand}, we make several key assumptions that are plausible in typical stepped wedge designs. First, we assume the cluster-level Stable Unit Treatment Value Assumption (SUTVA), which implies that the treatment is well-defined and no interference between clusters. We also assume no anticipation prior to the treatment adoption and no exposure-time treatment effect heterogeneity such that the individual-level potential outcomes from cluster \(i\) in period \(j\) depend only on the treatment received at this period, \(Z_{ij}\); also see Assumption 2 of \citet{chen2024model} for a rigorous presentation of this assumption. Under these assumptions, the observed outcome follows \(Y_{ijk} = Z_{ij} Y_{ijk}(1) + (1-Z_{ij})Y_{ijk}(0)\). Second, we assume staggered randomization, which implies that marginally, for each period $j\in\{2,\ldots,J-1\}$, $Z_{ij}$ is independent of $\{\bm{Y}_{ij}(0), \bm{Y}_{ij}(1) \}$, possibly given the entire baseline covariates for a cluster $\{\underline{\bm{X}}_{ij}, N_{ij} \}$, where for cluster \(i\) in period \(j\), \(\bm Y_{ij}(z) = \{Y_{ij1}(z),\dots, Y_{ijN_{ij}}(z) \}^\top \) denotes the collection of potential outcomes under condition \(Z_{ij}=z\), and \(\underline{\bm{X}}_{ij} = \{\bm{X}_{ij1},\dots, \bm{X}_{ijN_{ij}}  \}^\top \) denotes a set of baseline covariates across all cluster members during that period. The baseline covariates include \(\bm{X}_{ijk} = \bm{C}_{ij} \cup \bm{W}_{ijk}\), where \(\bm{W}_{ijk}\) is a vector of individual-level covariates. The exogenous cluster-period covariates \(\bm{C}_{ij}\) may include the cluster-period mean of individual covariates, \(\overline{\bm{W}}_{ij} = \sum_{k=1}^{N_{ij}}\bm{W}_{ijk}/N_{ij}\), and the baseline mean outcome \(\overline{Y}_{i0} = \sum_{k=1}^{N_{ij}}Y_{i0k} \), to account for any potential contextual effects. 

Under this general setup, the simple and intuitive approach to estimate \(\mu_\omega(z) \) is through the unadjusted nonparametric moment estimator:
\begin{equation}
    \widehat{\mu}_\omega^{\text{unadj}}(z) = \frac{1}{\sum_{j=2}^{J-1} \omega_{j}}\sum_{j=2}^{J-1} \omega_{j}  \frac{\sum_{i=1}^{I} \omega_{ij} \mathbb{I}(Z_{ij}=z) \overline{Y}_{ij}  }{\sum_{i=1}^{I} \omega_{ij} \mathbb{I}(Z_{ij}=z) }, \label{non_adj}
\end{equation}
where \(\overline{Y}_{ij}  = Z_{ij} \overline{Y}_{ij}(1) + (1-Z_{ij})\overline{Y}_{ij}(0) \) is the observed average outcome in cluster-period cell \((i,j)\). In Web Appendix B, we prove that \(\widehat{\mu}_\omega^{\text{unadj}}(z)\) is a consistent estimator for \(\mu_\omega(z)\). The estimator for \(\tau_\omega\) takes the form of \(\widehat{\tau}_\omega^{\text{unadj}} = f\{\widehat{\mu}_\omega^{\text{unadj}}(1),\widehat{\mu}_\omega^{\text{unadj}}(0)  \} \). When $f$ takes the identity function, this estimator is a generalization of the nonparametric difference-in-means moment estimator in individually randomized trials to SW-CRTs, and we will label this estimator as the unadjusted estimator as it does not leverage information from the baseline covariates for possible efficiency improvement. 

To address this potential inefficiency, we propose an augmented estimator that incorporates a working outcome regression model: \(\widehat{\tau}_\omega^{\text{aug}} = f\{\widehat{\mu}_\omega^{\text{aug}}(1),\widehat{\mu}_\omega^{\text{aug}}(0) \}  \), where the covariate-adjusted estimator of \(\mu_\omega(z)\) for \(z=0,1\) take the form of a general class of augmented average estimators:  
\begin{align}
    \widehat{\mu}_{\omega,j}^{\text{aug}}(z) = \frac{\sum_{i=1}^{I }\omega_{ij} m_{zj}(\underline{\bm{X}}_{ij},N_{ij})  }{\omega_j} +  \frac{\sum_{i=1}^{I} \omega_{ij} \mathbb{I}(Z_{ij} =z) \{\overline{Y}_{ij} - m_{zj} (\underline{\bm{X}}_{ij},N_{ij})  \}  }{\sum_{i=1}^{I}\mathbb{I}(Z_{ij}=z) \omega_{ij} }, \label{aug_mu}
\end{align}
where \(m_{zj} (\underline{\bm{X}}_{ij},N_{ij}) \) is a working regression function based on baseline information \(\underline{\bm{X}}_{ij} \) and \(N_{ij}\). This estimator combines two sources of information: the observed outcome via data (via residuals), and the fitted values from the model. The expression \eqref{aug_mu} can be equivalently rewritten as:
\begin{align}
    \widehat{\mu}_{\omega, j}^{\text{aug}} =  \underbrace{\frac{\sum_{i=1}^{I} \omega_{ij} \mathbb{I}(Z_{ij}=z) \overline{Y}_{ij}  }{\sum_{i=1}^{I} \omega_{ij} \mathbb{I}(Z_{ij}=z) }}_{\text{Unadjusted estimator in period $j$}} + \underbrace{\left[ \frac{\sum_{i=1}^{I} \omega_{ij} m_{zj} (\underline{\bm{X}}_{ij},N_{ij})   }{\omega_{j}} - \frac{\sum_{i=1}^{I} \omega_{ij} \mathbb{I}(Z_{ij} = z) m_{zj} (\underline{\bm{X}}_{ij},N_{ij}) }{\sum_{i=1}^{I} \omega_{ij} \mathbb{I}(Z_{ij}=z) }\right]}_{\text{Augmentation term}},\label{aug_mu_2}
\end{align}
which expresses the augmented estimator as an unadjusted estimator in \eqref{non_adj} plus a regression function correction. This correction captures the difference between the model-based estimator under the full population and the model-based estimator within the treatment group of interest.
As a result, the estimator leverages baseline adjustments without compromising consistency and could further improve efficiency from the unadjusted estimator. In particular, such an estimator extends Equation (4) from \citet{li2025model} to SW-CRTs, which can be expressed as:
\begin{align*}
\widehat{\mu}_\omega(z)=\sum_{i=1}^m \frac{\omega_{ij}}{\omega_{j}}\left\{m_{zj}(\underline{\bm{X}}_{ij},N_{ij})+\frac{I(Z_i=z)\left(\overline{Y}_i-m_{zj}(\underline{\bm{X}}_{ij},N_{ij})\right)}{\pi_{ij}(z)}\right\}.
\end{align*}
Under fixed design settings, where \(\sum_{i=1}^{I} \omega_{ij} \mathbb{I}(Z_{ij}=z) = \sum_{i=1}^{m} \omega_j\pi_{ij}(z) \), and \(\pi_{ij}(z)\) is a weighted proportion of treatment assignment for cluster \(i\) at period \(j\), this expression reduces to the form of \eqref{aug_mu}. Furthermore, if we set \(\omega_{ij}=1\) and treat each cluster as an independent unit, the estimator mimics the classical augmented inverse probability weighting (AIPW) estimator in independent data applied to randomized clinical trials \citep{robins1994estimation}.

The final covariate-adjusted estimator of \(\mu_{\omega}(z)\) is obtained by averaging over periods:
\[
\widehat{\mu}_\omega^{\text{aug}}(z) = \frac{\sum_{j=2}^{J-1} \omega_j \widehat{\mu}_{\omega,j}^{\text{aug}}(z)  }{ \sum_{j=2}^{J-1} \omega_j }.
\]
This estimator is consistent and asymptotically normal for any choice of working function \(m_{zj}(\underline{\bm{X}}_{ij},N_{ij})\), regardless of whether it is correctly specified. The proof is provided in Web Appendix C. The choice of \(m_{zj}(\underline{\bm{X}}_{ij}, N_{ij})\) affects efficiency, and we prove in Web Appendix D that the optimal function that minimizes the asymptotic variance of \(\widehat{\tau}_\omega^{\text{aug}}\) for each period \(j\) is given by the cluster-period potential outcome prediction, 
$$m_{zj}(\underline{\bm{X}}_{ij}, N_{ij})  = \mathbb{E}\left[\overline{Y}_{ij}(z) \mid \underline{\bm{X}}_{ij},N_{ij} \right]= \mathbb{E}[\overline{Y}_{ij} \mid Z_{ij}=z,\underline{\bm{X}}_{ij}, N_{ij}].$$ 
In what follows, we give specific examples on how to specify working outcome regression models to predict the potential outcomes $m_{zj}(\underline{\bm{X}}_{ij}, N_{ij})$, for the purpose of constructing the model-robust standardization estimator.

\subsection{Specifying the working outcome regression model in SW-CRTs}

The working model \(m_{zj}(\underline{\bm{X}}_{ij}, N_{ij})=\mathbb{E}[\overline{Y}_{ij} \mid Z_{ij}=z,\underline{\bm{X}}_{ij}, N_{ij}]\) can be estimated using a broad class of parametric or semiparametric regression methods applied to the full dataset, with predictions extrapolated to the target period \(j\). For example, one can use mixed models or GEE to appropriately account for within-cluster correlation and heterogeneity in cluster-period sizes, as is commonly in the conventional literature on model-based inference for SW-CRTs \citep{li2021mixed}. There are two common strategies for specifying \(m_{zj}(\cdot)\). One approach is to model the cluster-period mean outcome \(\overline{Y}_{ij}\) directly, for example by regressing it on cluster-period level covariates in a linear or nonlinear way \citep{li2022marginal}. Alternatively, an individual-level outcome model can be specified for \(Y_{ijk}\), with predicted values subsequently aggregated to the cluster-period level. These choices of approaches could involve different modeling complexity and computational cost, and in some cases, can depend on software availability. In the examples that follow, we elaborate on different working models allowing for baseline adjustment. While such models offer simplicity and interpretability, more flexible specifications---incorporating higher-order terms, interactions, or nonparametric components---could be accommodated with some modifications (even though they are less commonly seen in practice). 
Importantly, the validity of the augmented estimators does not hinge on correct specification of the working outcome regression. Because consistency is guaranteed by the randomization and the weighting scheme, the working model is used solely to improve efficiency and to avoid ad hoc model selection. As a result, efficiency gains will be greatest when the working model coincides with the true outcome-generating process, whereas substantial misspecification may yield little improvement relative to unadjusted estimators. We provide some examples of model specifications to connect this work with the vast possibilities on model specification in the existing SW-CRT literature. Before we give specific examples, we first provide two remarks below.


\begin{remark}[General consideration for fitting an outcome model]\label{rem1}
{Since the definition of potential outcome estimands (h-iATE, h-cATE, v-iATE and v-cATE) are confined to the roll-out periods $2$ to $J$ due to the treatment positivity assumption, our model-robust standardization estimator $\widehat{\mu}_\omega^{\text{aug}}(z) = {\sum_{j=2}^{J-1} \omega_j \widehat{\mu}_{\omega,j}^{\text{aug}}(z)  }/{ \sum_{j=2}^{J-1} \omega_j }$ is defined accordingly as a sum across the roll-out periods only. However, when fitting the subsequent working models to the observed data to generate \(m_{zj}(\underline{\bm{X}}_{ij}, N_{ij})=\mathbb{E}[\overline{Y}_{ij} \mid Z_{ij}=z,\underline{\bm{X}}_{ij}, N_{ij}]\), one can still fit each model to the entire data collected including the control period $1$ and the full roll-out period $J$ as the standardization procedure remains consistency without restrictions of the model specification or fitting procedure. But fitting each model for data collected from periods $1$ and $J$ can potentially improve efficiency of the prediction. Finally, regardless of how the outcome models are specified or fitted, it is only necessary to generate predictions for \(m_{zj}(\underline{\bm{X}}_{ij}, N_{ij})=\mathbb{E}[\overline{Y}_{ij} \mid Z_{ij}=z,\underline{\bm{X}}_{ij}, N_{ij}]\) during $j=2,\ldots, J-1$, as the estimands and the estimator $\widehat{\mu}_{\omega,j}^{\text{aug}}(z) $ are only defined through these roll-out periods.}

\end{remark}
\begin{remark}[Condition under which the augmented estimator and nonparametric estimator coincide] \label{rem:NP_wo_cov}
For a fixed period \(j\) and treatment level \(z\), the augmentation term in estimator \(\widehat{\mu}_{\omega,j}^{\text{aug}}(z)\) 
\[ \frac{\sum_{i=1}^{I} \omega_{ij} m_{zj}(\underline{\bm{X}}_{ij},N_{ij})}{\omega_j} - \frac{\sum_{i=1}^{I} \omega_{ij} \mathbb{I}(Z_{ij}=z)\, m_{zj}(\underline{\bm{X}}_{ij},N_{ij})} {\sum_{i=1}^{I} \omega_{ij} \mathbb{I}(Z_{ij}=z)} \]
is identically zero if and only if there exists a constant \(\mu_{zj} \in \mathbb{R}\) such that
\[ m_{zj}(\underline{\bm{X}}_{ij},N_{ij}) =  \mathbb{E}\left[\overline{Y}_{ij} \mid Z_{ij}=z,\underline{\bm{X}}_{ij},N_{ij}\right] = \mu_{zj} \quad \text{for all } i=1,\dots,I.\]
This condition is satisfied when the regression function does not vary with baseline covariates or cluster-period size within each period $j$ and treatment level $z$.  In this case, \(\widehat{\mu}_{\omega,j}^{\text{aug}}(z) = \widehat{\mu}_{\omega,j}^{\text{unadj}}(z)\) for all \(j,z\), and hence \(\widehat{\tau}_\omega^{\text{aug}} = \widehat{\tau}_\omega^{\text{unadj}}\); that is, the MRS estimator coincides with the unadjusted estimator.
\end{remark}

The proof of Remark \ref{rem:NP_wo_cov} is given in Web Appendix E. A direct consequence is that if the working outcome model only depends on treatment and period indicators but not further on individual covariates, there is essentially no augmentation even after model-robust standardization. For instance, if one assumes the following mean model
\[
\mathbb{E}\left( Y_{ijk} \mid Z_{ij}, \underline{\bm{X}}_{ij}, N_{ij} \right) = \beta_j + \tau Z_{ij},
\]
we have $m_{zj}(\underline{\bm{X}}_{ij},N_{ij}) = \mathbb{E}\left( \overline{Y}_{ij} \mid Z_{ij}=z, \underline{\bm{X}}_{ij}, N_{ij} \right) = \mathbb{E}\left( \overline{Y}_{ij} \mid Z_{ij}=z \right) = \beta_j + \tau z =\mu_{zj}$, which is constant in \((\underline{\bm{X}}_{ij},N_{ij})\) for fixed \((z,j)\). This would lead to \(\widehat{\tau}_\omega^{\text{aug}} = \widehat{\tau}_\omega^{\text{unadj}}\). From this standpoint, the augmented estimator will only likely improve efficiency when individual covariates and cluster-period sizes are adjusted for. For this reason, we will focus our examples with working regression models with baseline covariate adjustment.

\subsubsection{Linear mixed model with individual-level observations}\label{example1}

A popular approach for modeling outcomes in SW-CRTs is to specify the individual-level outcome \(Y_{ijk}\) using a linear mixed model. A general specification can be written as
\begin{align} \label{lmm_ind}
     Y_{ijk} = \beta_j + Z_{ij}\tau_j + g(\underline{\bm{X}}_{ij}, N_{ij}, Z_{ij}) + RE_{ij} + \epsilon_{ijk}, 
\end{align}
where \(\beta_j\) represents the fixed effect for period \(j\), and \(\tau_j\) denotes the period-specific treatment effect, which in practice often gets simplified to a constant treatment effect parameter \(\tau_j = \tau\) \citep{li2021mixed}. The term \(g(\underline{\bm{X}}_{ij}, N_{ij}, Z_{ij})\) is a general function that captures the contribution of covariates, cluster-period size, and potential treatment-by-covariate interactions \citep{li2024planning}. To specify the term \(g(\cdot)\), it is also possible to allow for the decomposition of individual-level covariates into between-cluster, contextual (cluster-period), and individual-level components, as well as higher-order or nonlinear terms; hence this formulation allows for flexible choices of baseline adjustment. The term \(RE_{ij}\) denotes a general random effect term, capturing deviations from the fixed effects due to the hierarchical design, and \(\epsilon_{ijk} \sim \mathcal{N}(0, \sigma_\epsilon^2)\) is the independent individual-level residual error. A commonly used specification for \(RE_{ij}\) is the nested exchangeable correlation structure, under which  \(RE_{ij} = \alpha_i + \gamma_{ij}\), where \(\alpha_i \sim \mathcal{N}(0, \tau_\alpha^2)\) is the random effect for cluster \(i\), and \(\gamma_{ij} \sim \mathcal{N}(0, \tau_\gamma^2)\) is the cluster-by-period interaction \citep{li2021mixed}. This structure induces a within-period ICC \(\rho_w = (\tau_\alpha^2+\tau_\gamma^2)/(\tau_\alpha^2 + \tau_\gamma^2 + \sigma_\epsilon^2)\) and a between period ICC \(\tau_\alpha^2/(\tau_\alpha^2 + \tau_\gamma^2 + \sigma_\epsilon^2)\). To allow the correlation between cluster-period to decay over time,  \citet{kasza2019impact} proposed an alternative specification in which the cluster-period random effect for cluster \(i\), \(\bm \gamma_i = (\gamma_{i1},\dots ,\gamma_{iJ})^\top\), follow \(\bm{\gamma}_i \sim  \mathcal{N}(\bm{0}, \tau_\gamma^2 \widetilde{\bm{M}})\)  with covariance matrix \(\widetilde{\bm{M}}\) having off-diagonal elements decay exponentially with the temporal distance between periods. Specifically, let \(r_0 \in (0,1)\) denote the correlation between adjacent periods, and \(r \in (0,1)\) denote the exponential decay rate between period, then the correlation between \(\gamma_{ij}\) and  \(\gamma_{il}\) is  \(r_0r^{|j-l|}\). This induces a lag-\(|j - l|\) between-period ICC of \(\rho_{b,|j - l|} = \frac{\tau_\gamma^2 r_0 r^{|j - l|}}{\tau_\gamma^2 + \sigma_\epsilon^2}\),
and a within-period ICC of \(\rho_b = \tau_\gamma^2 / (\tau_\gamma^2 + \sigma_\epsilon^2)\) \citep{li2021mixed}. Given model \eqref{lmm_ind}, the cluster-period mean outcome prediction is
\[
\widehat{m}_{zj}(\underline{\bm{X}}_{ij}, N_{ij})=\widehat{\mathbb{E}}[\overline{Y}_{ij} \mid Z_{ij} = z, \underline{\bm{X}}_{ij}, N_{ij}]
= \widehat{\beta}_j + z \widehat{\tau}_j + \widehat{g}(\underline{\bm{X}}_{ij}, N_{ij}, z),
\]
where \(\widehat{g}(\cdot)\) represents the estimated marginal contribution of covariates to the cluster-period mean.

Parameters in the linear mixed model can be estimated via maximum likelihood or restricted maximum likelihood using standard software. For example, in \texttt{R}, the \texttt{lme4}, \texttt{nlme}, and \texttt{lmerTest} packages support modeling nested exchangeable correlation structures but do not accommodate structured residual correlation. In contrast, the \texttt{glmmTMB} package allows for modeling residual correlations with exponential decay. In \texttt{SAS}, nested exchangeable and exponential decay correlation structures can both be specified using \texttt{PROC MIXED}, \texttt{PROC GLIMMIX}, or \texttt{PROC HPMIXED}. These procedures allow for flexible specification of the correlation through \texttt{RANDOM} statement. Among them, \texttt{PROC HPMIXED} is optimized for high-performance computing and can offer faster computation; see \citet{ouyang2023estimating} for example code to fit such models. A different modeling strategy involves applying a linear mixed-effects model to the cluster-period mean outcomes. Further details are provided in Web Appendix F.

\subsubsection{Generalized linear mixed model with individual-level observations}\label{example3}

A natural extension of the linear mixed-effects model for binary and count outcomes in SW-CRT is the GLMM, which incorporates a link function to model the conditional mean of the outcome. Let \( \mu_{ijk} = \mathbb{E}[Y_{ijk} \mid \bm{X}_{ijk}, Z_{ij}, N_{ij}] \), and define regression model as:
\begin{align} \label{glmm_ind}
    \eta(\mu_{ijk}) = \beta^{\#}_j + Z_{ij} \tau^{\#}_j + g^{\#}(\underline{\bm{X}}_{ij}, N_{ij}, Z_{ij}) + RE_{ij},
\end{align}
where \( \eta(\cdot) \) is a pre-specified link function, such as the logit for binary outcomes or the log link for count outcomes. The function \( g^{\#}(\cdot) \) could capture the contribution of covariates across multiple levels, including individual, cluster-period, and between-cluster summaries, along with cluster size and potential interactions with treatment. The random effect \(RE_{ij}\) captures between-cluster and cluster-period variability and can be specified just like those in Section \ref{example1}. 
The parameter \(\tau^{\#}_j\) denotes the conditional treatment effect in period \(j\) on the link function scale; however, due to the nonlinearity of the link function, it does not always correspond to a marginal effect of interest such as a risk difference or relative risk. To obtain the marginal cluster-period mean, the conditional expectation must be averaged over both the individual-level outcomes and the random effects:
\begin{align} \label{glmm_marginal}
    \widehat{m}_{zj}(\underline{\bm{X}}_{ij}, N_{ij}) = \mathbb{E}[\overline{Y}_{ij} \mid Z_{ij} = z, \bm{X}_{ijk},N_{ij}] = \frac{1}{N_{ij}} \sum_{k=1}^{N_{ij}} \int \eta^{-1}\left\{ \widehat{\theta}_{ijk} + RE_{ij} \right\} f(RE_{ij}) \, dRE_{ij},
\end{align}
where \( \widehat{\theta}_{ijk} = \widehat{\beta}^{\#}_j + z \widehat{\tau}^{\#}_j + \widehat{g}^{\#}(\underline{\bm{X}}_{ij}, N_{ij}, z) \), and \( f(RE_{ij}) \) denotes the distribution of the random effect. Due to the nonlinearity of \( \eta(\cdot) \), this integral generally lacks a closed-form solution and is commonly approximated using adaptive Gaussian-Hermite quadrature or Monte Carlo integration \citep{heagerty2000marginalized}. Closed-form approximations are available for certain link functions. For example, for the logit link, \citet{hedeker2018note} proposed the following approximation for nested exchangeable structure:
\[
\widehat{\mathbb{E}}[\overline{Y}_{ij} \mid Z_{ij} = z, \bm{X}_{ijk},N_{ij}] \approx \frac{1}{N_{ij}} \sum_{k=1}^{N_{ij}} \text{expit}\left\{ \widehat{\theta}_{ijk} / \sqrt{ (\widehat{\tau}_\alpha^2 + \widehat{\tau}_\gamma^2 + \pi^2/3)/(\pi^2/3) } \right\},
\]
where \(\text{expit}(x) = \exp(x)/\{1+\exp(x)\}\).
For log-link models (e.g., Poisson or log-binomial regression), the marginal mean is approximated by:
\[
\widehat{\mathbb{E}}[\overline{Y}_{ij} \mid Z_{ij} = z, \bm{X}_{ijk},N_{ij}] \approx \frac{1}{N_{ij}} \sum_{k=1}^{N_{ij}} \exp\left\{ \widehat{\theta}_{ijk} + \frac{1}{2}(\widehat{\tau}_\alpha^2 + \widehat{\tau}_\gamma^2) \right\}.
\]
When $RE_{ij}$ specifies an exponential decay structure, the matrix \( \widetilde{\bm{M}} \) induces dependence across \( j \), and the random effects \( \gamma_{ij} \) are no longer independent across periods. However, for a single cluster-period mean \( \mathbb{E}[\overline{Y}_{ij} \mid Z_{ij}=z, \bm{X}_{ijk}] \), the marginal variance of \( RE_{ij} \) remains \( \tau_\gamma^2 \), since \( \widetilde{\bm{M}}_{jj} = 1 \) on the diagonal. Thus, a Hedeker-style approximation may still be applied by substituting \( \tau_\alpha^2 + \tau_\gamma^2 \) with this marginal variance \(\tau_{\gamma}^2\) without the need for additional modifications. Parameter estimation in GLMMs is typically performed via maximum likelihood, integrating over the random effects. Common numerical techniques include the Laplace approximation and adaptive Gaussian-Hermite quadrature, which give accurate inference when an adequate number of quadrature points is used \citep{jin2020note}.

For implementation of these models, the \texttt{R} package \texttt{lme4} (via \texttt{glmer}) supports GLMMs for binary and count outcomes using canonical link functions and flexible random effects but does not support structured residual correlation. The \texttt{glmmTMB} package extends this functionality by allowing specification of correlation structures such as exponential decay \citep{brooks2017glmmtmb}. In \texttt{SAS}, correlation structure across cluster-periods can be modeled using the \texttt{RANDOM} statement in \texttt{PROC GLIMMIX}.

\subsubsection{Marginal model with individual-level observations}\label{example4}
In contrast to conditional models that require explicit specification of random effects, marginal models estimated via GEE directly model the marginal expectation of the outcome, providing population-average estimates without conditioning on latent variables. There is an increasing body of literature devoted to marginal model-based design and analysis of SW-CRTs, with particular attention to structured multilevel working correlation models \citep{li2018,li2020design,tian2022impact,zhang2023general,zhang2023geemaee}. With an exponential family outcome, the individual-level marginal mean model is
\begin{equation}\label{eq:GEE}
\eta(\mu_{ijk}) =  \alpha_j + Z_{ij}\nu_j + g(\underline{\bm{X}}_{ij}, N_{ij},Z_{ij}),
\end{equation}
where \(\alpha_j\) represents period fixed effects, \(\nu_j\) denotes period-specific treatment effects which can again be simplified under the constant treatment effect assumption $\nu_j=\nu$. In GEE, within-cluster correlations are accounted for through a working correlation matrix for outcomes. In cross-sectional SW-CRTs, commonly used correlation structures include the nested exchangeable model, where within-period and between-period correlations are parameterized separately, and the exponential decay structure, which allows correlation to decrease as the temporal distance between periods increases; also see Table 1 of \citet{tian2022impact} for explicit examples of these two working correlation models. Once the marginal means are estimated, the standardized cluster-period mean is computed by averaging the inverse link-transformed individual-level means:
\[
\widehat{m}_{zj}(\underline{\bm{X}}_{ij}, N_{ij}) = \widehat{\mathbb{E}}[\overline{Y}_{ij} \mid Z_{ij} = z, \bm{X}_{ijk},N_{ij}] = \frac{1}{N_{ij}} \sum_{k=1}^{N_{ij}} \eta^{-1} \left\{ \widehat{\alpha}_j + z\widehat{\nu}_j +  \widehat{g}(\underline{\bm{X}}_{ij}, N_{ij},z)\right\}.
\]
Typical GEE methods can be implemented in standard software, including the \texttt{geepack} package in \textsf{R}, which supports individual-level GEE models with nested exchangeable correlation structures. In the context of SW-CRTs, GEE methods have now been further developed and the \texttt{geeCRT} package could also implement the nested exchangeable correlation structure, along with a set of finite-sample corrections to the correlation estimating equations (via the so-called matrix-adjusted estimating equations, MAEE \citep{preisser2008}). \citet{zhang2023geemaee} further developed the GEEMAEE SAS macro for fitting GEE and MAEE under nested exchangeable and exponential decay correlation structures for multiple-period CRTs.

Under the working independence assumption and assuming an identity link function, the marginal mean model \eqref{eq:GEE} is equivalent to the class of analysis of covariance (ANCOVA) models introduced by \citet{chen2024model}. Under a finite-population framework, they proved that the class of ANCOVA estimators, fitted by weighted least squares, are consistent and asymptotically normal, under arbitrary model misspecification and hence are model-assisted rather than model-based. In Web Appendix G, we prove an interesting result that the form of the augmented estimator in \eqref{aug_mu} in fact generalizes the ANCOVA model-assisted estimators proposed by \citet{chen2024model}.
Therefore, each one of the ANCOVA model-assisted estimators for \(\tau_{\omega}\) by \citet{chen2024model} is a special case of the proposed augmented estimator. An advantage of our proposed estimator, however, is that we do not confine to independence GEE and can accommodate a wide class of working models.  We will also compare with their methods in the ensuing simulation studies. Similarly to the discussion in Section \ref{example1}, a GEE model can also be specified for the cluster-period mean outcome. Details are provided in Web Appendix F.


\subsection{Variance estimation via jackknifing} \label{sec:var_jack}
For variance estimation of the model-robust standardization estimators, we use a set of jackknife variance estimators, initially proposed by \citet{efron1982jackknife} and examined by \citet{li2025model} for parallel-arm CRTs. Jackknife variance can provide improved control over test size, especially in studies with limited number of clusters, compared to more simplistic variance approximation or bias-corrected robust variance estimators. Additionally, jackknifing offers substantial computational efficiency relative to other resampling methods such as bootstrapping or permutation techniques, which are usually more computationally intensive.

Specifically, we define a leave-one-cluster-out (LOCO) estimator of \(\mu_\omega(z)\) for each cluster \(g \in \{1,\dots,I\}\), denoted as \(\widehat{\mu}_\omega^{-g}(z)\) to estimate the average potential outcome when leaving cluster \(g\) out of the analysis. This LOCO estimator is defined as follows:
\[\widehat{\mu}_\omega^{-g}(z) = \left( \sum_{j=2}^{J-1}\omega_j^{-g}  \right)^{-1} \sum_{j=2}^{J-1} \omega_j^{-g} \widehat{\mu}_{\omega,j}^{-g}(z),  \]
with $\omega_j^{-g} = \sum_{i\neq g}\omega_{ij}$ and

\[ \widehat{\mu}_{\omega,j}^{-g} = \frac{\widehat{m}_{zj}^{-g} (\underline{\bm{X}}_{ij},N_{ij})  }{w_j^{-g}} + \frac{\sum_{j\neq g} \omega_{ij} \mathbb{I}(Z_{ij}=z) \{\overline{Y}_{ij} - \widehat{m}_{zj}^{-g} (\underline{\bm{X}}_{ij},N_{ij})  \}  }{\sum_{i\neq g} \omega_{ij} \mathbb{I} (Z_{ij}=z)  } .   \]
Here \(\widehat{m}_{zj}^{-g} (\underline{\bm{X}}_{ij},N_{ij})  \) denotes prediction from the outcome regression model fitted without cluster \(g\). The standard jackknife variance estimator, centered around the mean of LOCO estimators for \(\{\mu_\omega(1),\mu_\omega(0) \}\) is 

\begin{align*}
    \widehat{\Sigma}= \frac{I-1}{I}\sum_{g=1}^I \left(\begin{matrix} \left\{\widehat\mu_\omega^{-g}(1)-\overline\mu_\omega(1)\right\}^2 & \left\{\widehat\mu_\omega^{-g}(1)-\overline\mu_\omega(1)\right\}\left\{\widehat\mu_\omega^{-g}(0)-\overline\mu_\omega(0)\right\} \\ \left\{\widehat\mu_\omega^{-g}(1)-\overline\mu_\omega(1)\right\}\left\{\widehat\mu_\omega^{-g}(0)-\overline\mu_\omega(0)\right\} & \left\{\widehat\mu_\omega^{-g}(0)-\overline\mu_\omega(0)\right\}^2 \end{matrix}\right).
\end{align*}
where \(\overline{\mu}_\omega(z)  = I^{-1} \sum_{g=1}^{I} \widehat{\mu}_\omega^{-g}(z) \). and $\omega_j^{z(-g)} = \sum_{i\neq g}\omega_{ij}\mathbb{I}(Z_{ij}=z)$. Hypothesis tests and confidence intervals can be constructed using the \(t\)-distribution with \(I-1\) degrees of freedom. The use of \(t\)-quantiles potentially addresses bias arising from the finite number of clusters, providing improved control of type I error rate compared to normal approximations.

\section{Implied hypothesis testing procedures for informative sizes}\label{info_test}

The development of the proposed estimators naturally motivates hypothesis tests for the presence of informative sizes, including informative cluster sizes, informative period sizes and informative cluster-period sizes. We view such estimation and testing as providing evidence regarding the strength of association between cluster sizes and outcomes, but do not encourage use of these tests as ``pre-tests'' for model selection that may exclude cluster size from the working model. As mentioned in Section \ref{sec:estimands_def}, when the cluster-period sizes are only randomly varying, aggregating outcomes using different weighting schemes should give equivalent ATE estimates. Thus, potential discrepancy in the estimands construction serves as a natural basis for hypothesis testing. 

We first construct a global hypothesis test to assess the presence of any informative sizes. Let \(\bm{\tau} = (\tau_{\text{I}}^{\text{h}}, \tau_{\text{C}}^{\text{h}}, \tau_{\text{I}}^{\text{v}}, \tau_{\text{C}}^{\text{v}})^\top\) denote the vector of ATEs across horizontal and vertical directions at both individual and cluster levels. The global null hypothesis tests equality of all four estimands:
\[
\mathcal{H}_0: \ \tau_{\text{I}}^{\text{h}} = \tau_{\text{C}}^{\text{h}} = \tau_{\text{I}}^{\text{v}} = \tau_{\text{C}}^{\text{v}},
\]
which can be expressed compactly as a linear contrast \(\bm{C} \bm{\tau} = \bm{0}\), where
\[
\bm{C} = 
\begin{pmatrix}
    1 & -1 & 0 & 0 \\
    0 & 0 & 1 & -1 \\
    1 & 0 & -1 & 0
\end{pmatrix}
\]
has rank \(p=3\), representing the three linearly independent contrasts of interest. The global test statistic is defined as
\begin{align*}
 F =  \frac{(\bm{C} \widehat{\bm{\tau}})^\top (\bm{C} \widehat{\bm{V}} \bm{C}^\top)^{-1} (\bm{C} \widehat{\bm{\tau}})}{p},
\end{align*}
which follows a F-distribution with degrees of freedom \((p,I-1)\), under \(\mathcal{H}_0\). The covariance matrix \(\widehat{\bm{V}}\) of the estimator \(\widehat{\bm{\tau}}\) is estimated via jackknife resampling. Let \(\widehat{\bm{\tau}}^{(-i)}\) denote the vector of estimands computed by omitting cluster \(i\), for \(i = 1, \dots, I\), and let \(\overline{\bm{\tau}} = \frac{1}{I} \sum_{i=1}^I \widehat{\bm{\tau}}^{(-i)}\) denote the jackknife mean. Then the jackknife variance estimator is given by
\[
\widehat{\bm{V}} = \frac{I-1}{I} \sum_{i=1}^I \left( \widehat{\bm{\tau}}^{(-i)} - \overline{\bm{\tau}} \right) \left( \widehat{\bm{\tau}}^{(-i)} - \overline{\bm{\tau}} \right)^\top.
\]
Each LOCO replicate \(\widehat{\bm{\tau}}^{(-i)}\) is obtained by recomputing all four estimators after removing the \(i\)th cluster entirely from the sample. The global test thus provides a unified assessment of any informative sizes by jointly evaluating all sources of discrepancy among the horizontal and vertical estimands. Rejection of the global null indicates some form of informative sizes in SW-CRTs. Beyond the global test, one can also specify reduced forms of the contrast matrix $\bm{C}$ to test informative cluster sizes, informative period sizes and informative cluster-period sizes (by comparing values among subsets of $\bm{\tau}$); we provide those additional details in Web Appendix H.


\section{Simulation studies} \label{simu}

\subsection{Simulation design with continuous outcomes} \label{simu_cont}

We conducted simulation studies under different data generating models for continuous outcomes to illustrate the operating characteristics of the model-robust standardization (MRS) methods. We focus on a cross-sectional design for simplicity. Each simulated dataset included \(I = 30\) clusters observed over \(J = 6\) time periods. We denote \(\overline{N} = \frac{1}{IJ} \sum_{i=1}^I \sum_{j=1}^J N_{ij}\) as the overall average cluster-period size. Two individual-level covariates were generated: a binary covariate \(X_{ijk,1} \sim \mathcal{B}(0.5)\), and a continuous covariate \(X_{ijk,2} \sim \mathcal{N}(0,0.01) \). Here \(\mathcal{B}(p)\) denotes the Bernoulli distribution with probability \(p\), and \(\mathcal{N}(\mu,\sigma^2)\) denotes the normal distribution with mean \(\mu\) and variance \(\sigma^2\). The potential outcomes were generated from the following linear mixed model:
\begin{align*}
Y_{ijk}(a) = \beta_{0j} + \beta_{1j} X_{ijk} + \beta_{2j} X_{ijk}^2 + \theta_{ijk}a + \alpha_i + \delta_{ij} + \epsilon_{ijk},~~~~a\in\{0,1\},
\end{align*}
where \(\beta_{0j}\) is the covariate-adjusted time trend that is linearly increasing from 0.25 to 0.27 across periods, and the remaining model coefficients are specified as \(\beta_{1j} = 3j/2\), \(\beta_{2j} = j/J\), $j=1,\ldots, J$. The random effects include cluster effects \(\alpha_i \sim \mathcal{N}(0,0.05)\), cluster-period effects \(\delta_{ij}\sim \mathcal{N}(0,0.05)\), and individual error \(\epsilon_{ijk} \sim \mathcal{N}(0,0.9)\), which are assumed mutually independent. This specification corresponds to within-period ICC of 0.10 and between-period ICC of 0.05, respectively. The individualized treatment effect \(\theta_{ijk}=Y_{ijk}(1)-Y_{ijk}(0)\) are specified differently across three different simulation scenarios. Let \(\mathcal{U}(a, b)\) denote a discrete uniform distribution over integers \(a, \dots, b\) with probability mass function \(1/(b-a+1)\).
We consider the following three Scenarios with increasing degrees of complexity:

\renewcommand{\theenumi}{C\arabic{enumi}}               
\renewcommand{\labelenumi}{(\textbf{\theenumi})}         
\begin{enumerate}
  \item\label{sc:C1}
    \[
    \theta_{ijk} = 1 + \sin(X_{ijk,1}) + \exp(-X_{ijk,2}), 
    \quad N_{ij} \sim \mathcal{U}(20,100).
    \]

  \item\label{sc:C2}
    \[
    \theta_{ijk} = \tfrac12 - \sin(X_{ijk,1})
      - 1.5\,\exp(-X_{ijk,2})
      + \frac{4\sqrt{N_{ij}}}{5\overline N}
      + \frac{3\log(N_{ij})\,N_{ij}^2}{2\,E[N_{ij}]^2},
    \quad N_{ij} \sim \mathcal{U}(20,100).
    \]

  \item\label{sc:C3}
    \[
    \theta_{ijk} = 1 + j\,\sin(X_{ijk,1})
      - j^2\,\exp(-X_{ijk,2})
      + \frac{\sqrt{N_{ij}}}{\overline N}
      + \frac{3\log(N_{ij})\,N_{ij}^2}{E[N_{i1}]^2},
    \quad N_{ij} \sim \mathcal{U}(10 + 10j,\,90 + 10j).
    \]
\end{enumerate}

Scenario \ref{sc:C1} represents a simple setting where the treatment effect depends solely on covariates and is independent of both cluster size and calendar time. Since \(\theta_{ijk}\) is structurally homogeneous across clusters and periods, all four estimands, h-iATE, h-cATE, v-iATE, and v-cATE, coincide. This scenario serves as a baseline for benchmarking estimator performance under non-informative sizes. In contrast, Scenario \ref{sc:C2} introduces informative cluster-period sizes by allowing the individualized treatment effect to depend on both covariates and the cluster-period size \(N_{ij}\), in a nonlinear fashion. 
Finally, Scenario \ref{sc:C3} further allows the cluster size and covariate effects to vary as a function of calendar time. 
Specifically, cluster-period sizes systematically vary over time with \(N_{ij} \sim \mathcal{U}(10 + 10j, 90 + 10j)\) for \(j = 1, \dots, J\), but are unrelated to covariates or outcomes to rule out selection bias. Scenario \ref{sc:C3} is the most complex scenario where the individualized treatment effects can even evolve over calendar time.

\begin{table}[htbp]
\centering
\caption{A summary of working outcome models compared in the simulations for continuous and binary outcomes. Note that the unadjusted estimator and ANCOVA III estimator in \citet{chen2024model} are summarized separately in the text.}
\label{tab:working-models}
\renewcommand{\arraystretch}{1}
\begin{tabular}{>{\bfseries}c p{5.5cm} p{6.5cm}}
\toprule
\multicolumn{3}{c}{\bfseries Simulation for continuous outcome} \\
\midrule
Abbreviation & \multicolumn{1}{c}{\bfseries Working model} & \multicolumn{1}{c}{\bfseries Specification} \\
\midrule
W1 & Independence GEE assuming a constant treatment effect (Section \ref{example4}) & 
\( E(Y_{ijk}) = \alpha_j + \tau Z_{ij} + \alpha_{x1} X_{ijk1} + \alpha_{x2} X_{ijk2} + \alpha_N N_{ij} \) \\

W2 & Independence GEE assuming a period-specific treatment effect  (Section \ref{example4}) & 
\( E(Y_{ijk}) = \beta_j + \tau_j Z_{ij} + \alpha_{x1} X_{ijk1} + \alpha_{x2} X_{ijk2} + \alpha_N N_{ij} \) \\

W3 & Linear mixed model assuming a constant treatment effect with cluster random effect (Section \ref{example1}) & 
\( Y_{ijk} = \beta_j + \tau Z_{ij} + \beta_{x1} X_{ijk1} + \beta_{x2} X_{ijk2} + \beta_N N_{ij} + \alpha_i + \epsilon_{ijk} \) \\

W4 & Linear mixed model assuming a period-specific treatment effect  with cluster random effect  (Section \ref{example1}) & 
\( Y_{ijk} = \beta_j + \tau_j Z_{ij} + \beta_{x1} X_{ijk1} + \beta_{x2} X_{ijk2} + \beta_N N_{ij} + \alpha_i + \epsilon_{ijk} \) \\

W5 & Linear mixed model assuming a constant treatment effect with cluster and cluster-period random effects (Section \ref{example1}) & 
\( Y_{ijk} = \beta_j + \tau Z_{ij} + \beta_{x1} X_{ijk1} + \beta_{x2} X_{ijk2} + \beta_N N_{ij} + \alpha_i + \delta_{ij} +  \epsilon_{ijk} \) \\

W6 & Linear mixed model assuming a period-specific treatment effect with cluster and cluster-period random effects  (Section \ref{example1}) & 
\( Y_{ijk} = \beta_j + \tau_j Z_{ij} + \beta_{x1} X_{ijk1} + \beta_{x2} X_{ijk2} + \beta_N N_{ij} + \alpha_i + \delta_{ij} + \epsilon_{ijk} \) \\

\midrule
\addlinespace
\multicolumn{3}{c}{\bfseries Simulation for binary outcome} \\
\midrule
Abbreviation & \multicolumn{1}{c}{\bfseries Working model} & \multicolumn{1}{c}{\bfseries Specification} \\
\midrule
W7 & Independence GEE with logit link assuming a constant treatment effect  (Section \ref{example4}) & 
\(\text{logit}\{P(Y_{ijk} = 1)\} = \beta_j + \tau Z_{ij} + \beta_{x1} X_{ijk1} + \beta_{x2} X_{ijk2} + \beta_N N_{ij}\) \\

W8 & Independence GEE with logit link assuming a period-specific treatment effect  (Section \ref{example4})  & 
\(\text{logit}\{P(Y_{ijk} = 1)\} = \beta_j + \tau_j Z_{ij} + \beta_{x1} X_{ijk1} + \beta_{x2} X_{ijk2} + \beta_N N_{ij}\) \\

W9 & Generalized linear mixed-effects model with logit link assuming a constant treatment effect, including cluster random effect (Section \ref{example3}) & 
\(\text{logit}\{P(Y_{ijk} = 1)\} = \beta_j + \tau Z_{ij} + \beta_{x1} X_{ijk1} + \beta_{x2} X_{ijk2} + \beta_N N_{ij} + \alpha_i\) \\

W10 & Generalized linear mixed-effects model with logit link assuming a period-specific treatment effect , including cluster random effect (Section \ref{example3}) & 
\(\text{logit}\{P(Y_{ijk} = 1)\} = \beta_j + \tau_j Z_{ij} + \beta_{x1} X_{ijk1} + \beta_{x2} X_{ijk2} + \beta_N N_{ij} + \alpha_i\) \\

W11 & Generalized linear mixed-effects model with logit link assuming a  constant treatment effect, including cluster and cluster-period random effects (Section \ref{example3}) & 
\(\text{logit}\{P(Y_{ijk} = 1)\} = \beta_j + \tau Z_{ij} + \beta_{x1} X_{ijk1} + \beta_{x2} X_{ijk2} + \beta_N N_{ij} + \alpha_i + \delta_{ij}\) \\

W12 & Generalized linear mixed-effects model with logit link assuming a period-specific treatment effect, including cluster and cluster-period random effects (Section \ref{example3}) & 
\(\text{logit}\{P(Y_{ijk} = 1)\} = \beta_j + \tau_j Z_{ij} + \beta_{x1} X_{ijk1} + \beta_{x2} X_{ijk2} + \beta_N N_{ij} + \alpha_i + \delta_{ij}\) \\

\bottomrule
\end{tabular}
\end{table}

Under each simulation configuration, we estimate h-iATE, h-cATE, v-iATE, and v-cATE, using different working outcome regression models based on all available data from period $1$ to $J$ (see Remark \ref{rem1}); each outcome model is listed in Table \ref{tab:working-models}. Outcome models (W1) and (W2) are independence GEE (see Section \ref{example4}). They differ in that (W1) assumes a constant treatment effect $\tau$ whereas (W2) considers a period-specific treatment effect through the coefficient $\tau_j$. Outcome models (W3) and (W4) are the counterparts of (W1) and (W2), except that individual-level linear mixed models (see Section \ref{example1}) with a random intercept are considered; these models imply the exchangeable working correlation structure. Outcome models (W5) and (W6) further introduce an additional cluster-period random effect, and imply a working nested exchangeable correlation structure. Each model adjusts for the main effect of the baseline covariate and cluster-period size. It is important to note that each model corresponds to a certain degree of misspecification of the true data generating model under Scenario \ref{sc:C1} - \ref{sc:C3}, regarding the treatment effect structure, functional form of covariates, and/or random-effects structure. Hence, the purpose of the simulations is to illustrate, when a working model is misspecified, whether MRS can still successfully recover different potential outcomes estimands defined in Section \ref{sec:estimands_def}. 

Under each working outcome model, we first apply the proposed MRS approach, where the predicted cluster-period means are computed using the steps described in Section \ref{example1} (for linear mixed models) and Section \ref{example4} (for GEE). As a comparison, we also examine the model-based coefficient (Coef) estimator. For (W1), (W3) and (W5), the Coef estimator is simply $\widehat{\tau}$, where for (W2), (W4) and (W6), the Coef estimator is based on a weighted average $\sum_{j=2}^{J-1} \omega_j \widehat{\tau}_j / \sum_{j=2}^{J-1} \omega_j$; the weights $\omega_j$ are chosen based on the definition of each estimand. We also compare two additional estimators. The first one is the unadjusted estimator given in \eqref{non_adj}, which is consistent but could be less efficient due to ignoring baseline covariates. The second one is an existing analysis of covariance estimator (ANCOVA III, which adjusts for both main covariate effect and treatment-by-covariate interactions) fitted by weighted least squares using weight \(\omega_{ijk}\) specific to each estimand. This estimator is one of the recommended estimators in \citet{chen2024model} and has been proved to be consistent with the mean difference estimands even under model misspecification. The variance of the unadjusted estimator and each MRS estimator is estimated by cluster jackknife. The variance of each Coef estimator from LMM and the ANCOVA III estimator is given by the cluster-robust sandwich variance estimator. We implement this using the \texttt{clubSandwich} package in R with \texttt{type = "CR3"} as this is the recommended small-sample correction  \citep{ouyang2024maintaining}. The variance of Coef estimator from GEE is estimated using a modified variance estimator proposed by \citet{kauermann2001note}. This is implemented via the \texttt{geesmv} package in R. All confidence intervals are constructed based on the $t(I-1)$ distribution to ensure a fair comparison.

\subsection{Simulation results with continuous outcomes} \label{simu_cont:res}
Under each of the $3$ simulation configurations, we will compare in total $14$ different estimators over $1000$ simulated stepped wedge trials. Performance metrics include absolute relative bias (RBias, in percentage), Monte Carlo standard deviation (MCSD), average of the estimated standard error (AESE), and empirical coverage probability (CP) of the 95\% confidence interval. We describe the simulation findings for each Scenario. The true estimands are calculated from a simulated superpopulation of $m=10^7$ clusters.

Table \ref{tab:sim_C3_results} presents the simulation results under Scenario \ref{sc:C3} with additional calendar time-varying heterogeneity in treatment effects. 
The proposed MRS estimator continues to show unbiased performance across all estimands. It achieves low relative bias and empirical coverage near the nominal \(95\%\) level, across all working models (W1)-(W6), despite the dual complexity of informative cluster-period size and calendar time treatment effect heterogeneity. 
In contrast, the Coef estimators show a varying degree of bias. For horizontal estimands, Coef estimators give moderate to substantial bias for h-iATE and h-cATE with several working models, particularly under (W5) and (W6), which include both cluster and cluster-period random effects, showing that a more complex random-effects structure can lead to even more biased inference under informative cluster-period sizes, a somewhat unconventional finding that has been previously identified in 2-period cluster randomized crossover designs \citep{lee2025estimated}. In comparison, the bias under simple exchangeable random-effects models becomes much smaller. The independence GEE models (W1) and (W2) tend to also perform well for both h-iATE and v-iATE under informative cluster-period sizes.  For vertical estimands, performance deteriorates further, especially for v-cATE, where Coef estimators suffer from high bias and severe undercoverage across most working models (W1)-(W4). This is consistent with the findings in parallel CRTs \citep{li2025model}. In contrast, the Coef estimator under (W5) and (W6) appears to offer unbiased estimation of v-cATE, with nominal coverage. The UNADJ and ANCOVA estimators retain small bias and near nominal coverage, but are less efficient than MRS. Across the four estimands and all working models in this scenario, the Monte Carlo variances of MRS are approximately 40--50\% lower than those of UNADJ. These findings show that model specification not only in terms of functional form but also in terms of correlation and variance structure plays a role in shaping the behavior of Coef estimators under informative cluster-period size. In practical settings, these distinctions directly impact which estimands are being targeted with Coef estimators. However, the MRS estimator offers a robust alternative regardless of the working outcome model specification. The simulation results for Scenario \ref{sc:C1} and \ref{sc:C2} are included in the Web Appendix I. In the absence of informative sizes, all four estimands coincide, and the MRS estimators produce results similar to the Coef estimators. In the presence of informative cluster-period sizes but constant covariate effect, the results for Scenario \ref{sc:C2} closely resemble those of Scenario \ref{sc:C3}. Finally, we examined the performance of estimators based on working models excluding covariate and cluster-period size adjustment; these results are presented in Web Appendix J (Web Table 3 - 5). For both MRS and Coef, ignoring baseline covariates in the outcome models leads to some efficiency loss, confirming the necessity of baseline adjustment.

We further investigated the impact of a smaller number of clusters by repeating Scenarios \ref{sc:C1}--\ref{sc:C3} with $I = 10$ clusters under two designs: \(J=6\) periods with two clusters per sequence and \(J=11\) periods with one cluster per sequence. For the one cluster per sequence design, adjustments to the jackknife variance estimator are necessary to account for cases in which deletion of specific clusters yields a rank-deficient design matrix, making some treatment effect parameters unestimable during jackknifing. The implementation details are presented in Web Appendix K. Across both designs, MRS estimates remain approximately unbiased with near-nominal coverage and achieves appreciable efficiency gains, with MCSDs roughly \(20\%\)-\(30\%\) smaller than those of UNADJ; these gains appear larger in the \(J=11\) setting. In contrast, the Coef estimators, especially under mixed-effects working models (W3--W6), exhibit larger bias and more pronounced undercoverage.

\begin{table}[ht!]
\centering
\caption{
Simulation results in Scenario C3 for estimating four estimands under a continuous outcome with $I=30$ clusters and $J=6$ periods using covariate-adjusted working models (W1)-(W6) with the Coef, MRS and ANCOVA estimators. MRS: proposed augmented estimator; Coef: treatment-effect coefficients from covariate-adjusted working model; ANCOVA: model-assisted ANCOVA estimator; UNADJ: nonparametric estimator. RBias (\%): absolute relative bias in percent; AESE: average estimated standard error; MCSD: Monte Carlo standard deviation; CP: 95\% confidence interval coverage.
}
\label{tab:sim_C3_results}
\begin{tabular}{cclrrrrrrrr}
\toprule
\multirow{2}{*}{Direction} & \multirow{2}{*}{Working model}
  & \multirow{2}{*}{Method}
  & \multicolumn{4}{c}{h-iATE}
  & \multicolumn{4}{c}{h-cATE} \\
\cmidrule(lr){4-7}\cmidrule(lr){8-11}
 & & 
 & RBias & AESE & MCSD & CP 
 & RBias & AESE & MCSD & CP \\
  & & 
 & \multicolumn{4}{c}{\(\tau_{\mathrm{I}}^{\mathrm{h}}=8.395\)} 
 & \multicolumn{4}{c}{\(\tau_{\mathrm{C}}^{\mathrm{h}}=7.857\)} \\
\midrule
\multirow{18}{*}{Horizontal}
  & \textbackslash & UNADJ 
    & 1.007 & 1.827 & 1.784 & 0.928 
    & 0.615 & 1.832 & 1.798 & 0.927 \\
  & \textbackslash & ANCOVA  
    & 1.009 & 1.866 & 1.781 & 0.932 
    & 0.620 & 1.869 & 1.797 & 0.930 \\
  & \multirow{2}{*}{W1}
    & Coef   
      & 0.793 & 1.233 & 1.251 & 0.935 
      & 4.742 & 1.233 & 1.251 & 0.889 \\
  &               & MRS    
      & 1.180 & 1.368 & 1.296 & 0.946 
      & 0.705 & 1.374 & 1.305 & 0.952 \\
  & \multirow{2}{*}{W2}
    & Coef   
      & 1.199 & 1.268 & 1.299 & 0.930 
      & 4.299 & 1.266 & 1.300 & 0.895 \\
  &               & MRS    
      & 1.199 & 1.373 & 1.299 & 0.946 
      & 0.724 & 1.378 & 1.308 & 0.954 \\
  & \multirow{2}{*}{W3}
    & Coef   
      & 3.830 & 1.670 & 1.601 & 0.936 
      & 1.536 & 1.670 & 1.601 & 0.955 \\
  &               & MRS    
      & 1.200 & 1.370 & 1.300 & 0.946 
      & 0.755 & 1.375 & 1.310 & 0.951 \\
  & \multirow{2}{*}{W4}
    & Coef   
      & 4.130 & 1.756 & 1.642 & 0.936 
      & 1.211 & 1.755 & 1.643 & 0.959 \\
  &               & MRS    
      & 1.220 & 1.376 & 1.304 & 0.945 
      & 0.775 & 1.380 & 1.313 & 0.950 \\
  & \multirow{2}{*}{W5}
    & Coef   
      & 21.376 & 1.215 & 1.178 & 0.084 
      & 16.989 & 1.215 & 1.178 & 0.318 \\
  &               & MRS    
      & 1.194 & 1.371 & 1.302 & 0.946 
      & 0.731 & 1.375 & 1.309 & 0.950 \\
  & \multirow{2}{*}{W6}
    & Coef   
      & 20.884 & 1.287 & 1.222 & 0.129 
      & 16.466 & 1.285 & 1.223 & 0.405 \\
  &               & MRS    
      & 1.201 & 1.375 & 1.304 & 0.945 
      & 0.737 & 1.378 & 1.311 & 0.949 \\
\toprule
& & 
  & \multicolumn{4}{c}{v-iATE}
  & \multicolumn{4}{c}{v-cATE} \\
  \cmidrule(lr){4-7}\cmidrule(lr){8-11}
 & & 
 & RBias & AESE & MCSD & CP 
 & RBias & AESE & MCSD & CP \\
  & & 
 & \multicolumn{4}{c}{\(\tau_{\mathrm{I}}^{\mathrm{v}}=8.821\)} 
 & \multicolumn{4}{c}{\(\tau_{\mathrm{C}}^{\mathrm{v}}=6.707\)} \\
 \midrule
\multirow{18}{*}{Vertical}
  & \textbackslash & UNADJ 
    & 1.503 & 1.841 & 1.787 & 0.925 
    & 0.331 & 1.688 & 1.671 & 0.935 \\
  & \textbackslash & ANCOVA  
    & 1.506 & 1.872 & 1.784 & 0.928 
    & 0.330 & 1.715 & 1.673 & 0.934 \\
  & \multirow{2}{*}{W1}
    & Coef   
      & 0.786 & 1.233 & 1.251 & 0.935 
      & 26.685 & 1.233 & 1.251 & 0.124 \\
  &               & MRS    
      & 1.667 & 1.378 & 1.298 & 0.939 
      & 0.353 & 1.268 & 1.211 & 0.952 \\
  & \multirow{2}{*}{W2}
    & Coef   
      & 1.690 & 1.271 & 1.301 & 0.925 
      & 25.531 & 1.271 & 1.301 & 0.167 \\
  &               & MRS    
      & 1.690 & 1.383 & 1.301 & 0.938 
      & 0.333 & 1.272 & 1.214 & 0.953 \\
  & \multirow{2}{*}{W3}
    & Coef   
      & 3.823 & 1.670 & 1.601 & 0.936 
      & 22.807 & 1.670 & 1.601 & 0.447 \\
  &               & MRS    
      & 1.688 & 1.380 & 1.302 & 0.939 
      & 0.328 & 1.269 & 1.215 & 0.952 \\
  & \multirow{2}{*}{W4}
    & Coef   
      & 4.631 & 1.758 & 1.643 & 0.925 
      & 21.776 & 1.758 & 1.643 & 0.531 \\
  &               & MRS    
      & 1.711 & 1.385 & 1.306 & 0.939 
      & 0.308 & 1.273 & 1.217 & 0.952 \\
  & \multirow{2}{*}{W5}
    & Coef   
      & 21.371 & 1.215 & 1.178 & 0.084 
      & 0.401 & 1.215 & 1.178 & 0.947 \\
  &               & MRS    
      & 1.682 & 1.380 & 1.303 & 0.937 
      & 0.332 & 1.262 & 1.208 & 0.949 \\
  & \multirow{2}{*}{W6}
    & Coef   
      & 21.407 & 1.281 & 1.216 & 0.111 
      & 0.354 & 1.281 & 1.216 & 0.946 \\
  &               & MRS    
      & 1.692 & 1.385 & 1.306 & 0.937 
      & 0.327 & 1.266 & 1.210 & 0.948 \\
\bottomrule
\end{tabular}
\end{table}

\subsection{Simulation design with binary outcomes} \label{simu_bin}

Next, we present the simulation studies for binary outcomes. The setup largely parallels that of the continuous outcome case. Each dataset consisted of \(I = 30\) clusters observed over \(J = 4\) time periods. We generated two covariates: a binary variable \(X_{ijk,1} \sim \text{Bernoulli}(0.1)\) and a continuous variable \(X_{ijk,2} = \mathcal{N}(0, 0.01)\). The potential outcome under treatment \(a \in \{0,1\}\) is generated from a generalized linear mixed model,
\[
\mu_{ijk}(a) = \beta_{0j} + \beta_{1j} X_{ijk1} + \beta_{2j} X_{ijk2}^2  + \theta_{ijk} a + \alpha_i + \delta_{ij},
\]
with \(Y_{ijk}(a) \sim \text{Bernoulli}(\text{expit}\{\mu_{ijk}(a)\})\). Time-varying coefficients were specified as \(\beta_{0j}\) increasing linearly from 0.05 to 0.25, \(\beta_{1j} = j\), and \(\beta_{2j} = j/J\). The random effect includes cluster effect \(\alpha_i \sim \mathcal{N}(0, 0.3)\), \(\delta_{ij} \sim \mathcal{N}(0,0.3)\). This specification corresponds to the within-period ICC of 0.154, and between-period ICC of 0.083, defined on the latent response scale \citep{ouyang2023estimating}. The individualized treatment effect \(\theta_{ijk} = \text{logit}\{\mu_{ijk}(1)\} - \text{logit}\{\mu_{ijk}(0)\}\) was specified differently across three simulation scenarios.

\renewcommand{\theenumi}{B\arabic{enumi}}               
\renewcommand{\labelenumi}{(\textbf{\theenumi})}   

\begin{enumerate}
    \item \label{sc:B1}
    \[
    \theta_{ijk} = 1 + 0.5\sin(\pi X_{ijk,1}) + \log(1+X_{ijk,1} + 2X_{ijk,2}^2) , \quad N_{ij} \sim \mathcal{U}(5,50).
    \]
    
    \item \label{sc:B2}
    \[
    \theta_{ijk} = 1 + \sin(\pi X_{ijk,1}) + 0.5 \log(1+X_{ijk,1}+X_{ijk,2}^2) + \frac{\log(N_{ij})N_{ij}^2}{E[N_{ij}]^2}, \quad N_{ij} \sim \mathcal{U}(5,50).
    \]
    
    \item \label{sc:B3}
    \begin{align*}
       & \theta_{ijk} = \frac{1}{10} +  \frac{j}{2J}\sin(\pi X_{ijk,1}) +  \frac{j^2}{J} \log(1+0.5X_{ijk,1}+0.2X_{ijk,2}^2) +\frac{\log(N_{ij})N_{ij}^2}{2E[N_{ij}]^2}\\
    &\quad N_{ij} \sim \mathcal{U}(5 + 5j, 45 + 10j).
    \end{align*}

\end{enumerate}

These simulation scenarios are constructed to evaluate the robustness of the proposed estimators under increasingly complex forms of heterogeneity and informative cluster-period sizes. Scenario \ref{sc:B1} serves as a baseline, where the treatment effect varies nonlinearly with individual-level covariates but otherwise remains constant across clusters and calendar time, and is independent of cluster-period size. Scenario \ref{sc:B2} introduces informative cluster-period sizes by allowing the treatment effect to depend not only on covariates but also directly on the cluster-period size. 
Scenario \ref{sc:B3} extends the complexity further by incorporating temporal variation. Both covariate effects and cluster-period sizes are now allowed to evolve over calendar time, mimicking designs with shifted population composition and exogenous time-varying associations. The individualized treatment effect in this last data generating model thus varies across individuals, clusters, and also calendar time.

Under each simulation scenario, we fit different working models to estimate four estimands, h-iATE, h-cATE, v-iATE, and v-cATE (on the log odds ratio scale), defined as 
$$\tau_\omega = \log\left[{\mu_\omega(1)}/\{1 - \mu_\omega(1)\} \right]  -\log\left[{\mu_\omega(0)}/\{1 - \mu_\omega(0)\} \right].$$
Estimation is conducted using both the MRS and Coef methods. All outcome regression models are fit to the complete dataset across periods \(j = 1, \ldots, J\), and are summarized in Table \ref{tab:working-models} (W7-W12), paralleling the models considered for continuous outcomes.  
For all models, predicted cluster-period means required for MRS augmentation are computed following the procedures outlined in Section \ref{example3} and Section \ref{example4}. All working models are intentionally misspecified relative to the true data generating mechanism, allowing evaluation of the robustness of MRS method under model misspecification. When the working model assumes a constant treatment effect (W7, W9, W11), the Coef estimator is taken to be the estimated coefficient on the treatment indicator. For models with period-specific effects (W8, W10, W12), the Coef estimator is computed as a weighted average: \(\sum_{j=2}^{J-1} \omega_j \widehat{\tau}_j/\sum_{j=2}^{J-1} \omega_j,\) where \(\omega_j\) is the corresponding weight for each target estimand. For comparison, we also include the unadjusted estimator defined in \eqref{non_adj}, which does not rely on any outcome model. Standard errors for MRS estimates are obtained via the cluster jackknife, while those for Coef are computed using sandwich variance estimators for GEE models (W7-W8), and model-based variance for GLMMs (W9-W12). The confidence intervals are constructed based on \(t(I-1)\) distribution. 

\begin{table}[htbp]
\centering
\caption{
Simulation results in Scenario B3 for estimating four estimands under a binary outcome with $I=30$ clusters and $J=4$ periods using covariate-adjusted working models (W7)--(W12) with the Coef, MRS and nonparametric (UNADJ) estimators. MRS: proposed augmented estimator; Coef: treatment‐effect coefficients from covariate‐adjusted working model; UNADJ: nonparametric estimator. RBias: absolute bias; AESE: average estimated standard error; MCSD: Monte Carlo standard deviation; CP: 95\% confidence interval coverage.
}
\label{tab:sim_B3_results}
\begin{tabular}{cclrrrrrrrr}
\toprule
\multirow{2}{*}{Direction} & \multirow{2}{*}{Working model}
  & \multirow{2}{*}{Method}
  & \multicolumn{4}{c}{h-iATE}
  & \multicolumn{4}{c}{h-cATE} \\
\cmidrule(lr){4-7}\cmidrule(lr){8-11}
 & & 
 & RBias & AESE & MCSD & CP 
 & RBias & AESE & MCSD & CP \\
 & & 
 & \multicolumn{4}{c}{\(\Delta_{\mathrm{I}}^{\mathrm{h}}=3.195\)} 
 & \multicolumn{4}{c}{\(\Delta_{\mathrm{C}}^{\mathrm{h}}=2.879\)} \\
\midrule
\multirow{13}{*}{Horizontal}
  & \textbackslash & UNADJ 
    & 1.099 & 0.508 & 0.494 & 0.949 
    & 2.158 & 0.516 & 0.502 & 0.947 \\
  & \multirow{2}{*}{W7}
    & Coef   
      & 6.308 & 0.417 & 0.482 & 0.918 
      & 17.971 & 0.417 & 0.482 & 0.805 \\
  &               & MRS    
      & 1.122 & 0.492 & 0.480 & 0.954 
      & 2.217 & 0.500 & 0.487 & 0.941 \\
  & \multirow{2}{*}{W8}
    & Coef   
      & 7.947 & 0.413 & 0.511 & 0.892 
      & 19.827 & 0.414 & 0.512 & 0.758 \\
  &               & MRS    
      & 1.237 & 0.490 & 0.477 & 0.955 
      & 2.337 & 0.499 & 0.485 & 0.942 \\
  & \multirow{2}{*}{W9}
    & Coef   
      & 11.245 & 0.219 & 0.487 & 0.570 
      & 23.450 & 0.219 & 0.487 & 0.307 \\
  &               & MRS    
      & 1.105 & 0.486 & 0.475 & 0.953 
      & 2.177 & 0.495 & 0.481 & 0.941 \\
  & \multirow{2}{*}{W10}
    & Coef   
      & 12.549 & 0.233 & 0.518 & 0.568 
      & 24.937 & 0.233 & 0.519 & 0.314 \\
  &               & MRS    
      & 1.249 & 0.484 & 0.471 & 0.956 
      & 2.328 & 0.494 & 0.479 & 0.942 \\
  & \multirow{2}{*}{W11}
    & Coef   
      & 24.154 & 0.404 & 0.449 & 0.564 
      & 37.775 & 0.404 & 0.449 & 0.272 \\
  &               & MRS    
      & 1.260 & 0.472 & 0.459 & 0.952 
      & 2.341 & 0.480 & 0.465 & 0.945 \\
  & \multirow{2}{*}{W12}
    & Coef   
      & 25.516 & 0.406 & 0.465 & 0.532 
      & 39.326 & 0.406 & 0.466 & 0.244 \\
  &               & MRS    
      & 1.415 & 0.468 & 0.454 & 0.957 
      & 2.505 & 0.479 & 0.461 & 0.945 \\
\toprule
& & 
  & \multicolumn{4}{c}{v-iATE}
  & \multicolumn{4}{c}{v-cATE} \\
  \cmidrule(lr){4-7}\cmidrule(lr){8-11}
 & & 
 & RBias & AESE & MCSD & CP 
 & RBias & AESE & MCSD & CP \\
& & 
 & \multicolumn{4}{c}{\(\Delta_{\mathrm{I}}^{\mathrm{v}}=3.172\)} 
 & \multicolumn{4}{c}{\(\Delta_{\mathrm{C}}^{\mathrm{v}}=2.524\)} \\
 \midrule
\multirow{13}{*}{Vertical}
  & \textbackslash & UNADJ 
    & 0.922 & 0.519 & 0.504 & 0.948 
    & 2.902 & 0.511 & 0.503 & 0.940 \\
  & \multirow{2}{*}{W7}
    & Coef   
      & 7.069 & 0.417 & 0.482 & 0.916 
      & 34.543 & 0.417 & 0.482 & 0.518 \\
  &               & MRS    
      & 0.949 & 0.503 & 0.488 & 0.947 
      & 3.004 & 0.497 & 0.488 & 0.938 \\
  & \multirow{2}{*}{W8}
    & Coef   
      & 7.690 & 0.412 & 0.509 & 0.892 
      & 35.323 & 0.412 & 0.509 & 0.478 \\
  &               & MRS    
      & 1.067 & 0.500 & 0.485 & 0.950 
      & 3.134 & 0.496 & 0.486 & 0.940 \\
  & \multirow{2}{*}{W9}
    & Coef   
      & 12.042 & 0.219 & 0.487 & 0.551 
      & 40.792 & 0.219 & 0.487 & 0.070 \\
  &               & MRS    
      & 0.933 & 0.497 & 0.483 & 0.949 
      & 3.003 & 0.491 & 0.483 & 0.939 \\
  & \multirow{2}{*}{W10}
    & Coef   
      & 12.270 & 0.233 & 0.516 & 0.564 
      & 41.078 & 0.233 & 0.516 & 0.081 \\
  &               & MRS    
      & 1.078 & 0.494 & 0.479 & 0.952 
      & 3.169 & 0.491 & 0.481 & 0.940 \\
  & \multirow{2}{*}{W11}
    & Coef   
      & 25.043 & 0.404 & 0.449 & 0.544 
      & 57.129 & 0.404 & 0.449 & 0.068 \\
  &               & MRS    
      & 1.085 & 0.480 & 0.466 & 0.950 
      & 3.170 & 0.476 & 0.467 & 0.941 \\
  & \multirow{2}{*}{W12}
    & Coef   
      & 25.146 & 0.405 & 0.465 & 0.547 
      & 57.259 & 0.405 & 0.465 & 0.063 \\
  &               & MRS    
      & 1.243 & 0.477 & 0.461 & 0.950 
      & 3.346 & 0.476 & 0.463 & 0.938 \\
\bottomrule
\end{tabular}
\end{table}

\subsection{Simulation results with binary outcomes} \label{simu_bin_res}

Table \ref{tab:sim_B3_results} presents simulation results under Scenario \ref{sc:B3}, which involves both informative cluster-period sizes and time-varying treatment effects. Despite these added complexities, the MRS estimator remains robust, consistently achieving low bias and empirical coverage close to the nominal level across all estimands and working models. 
In contrast, the Coef estimators show more inflated bias and under-coverage, particularly for cluster-average estimands and GLMM (W9-W12). In addition, the bias for Coef estimators appears more pronounced even for individual-average estimands; this suggests that Coef estimators from independence GEE and GLMM are generally not preferred for estimating potential outcomes estimands on the odds ratio scale. In GLMMs with only a cluster-level random effect (W9-W10), the model-based variance estimator also dramatically underestimates the true variability. For binary outcomes, similarly divergent performance of Coef estimators for cluster-average versus individual-average estimands were discussed in \citet{li2025model} in simpler parallel-arm CRTs. Interestingly, the GEE Coef estimators (W7-W8) lead to small bias, with generally closer to nominal coverage for h-iATE estimands.
Finally, the UNADJ estimator gives unbiased estimates with near-nominal coverage but remains less efficient than MRS throughout. Across the four estimands and all working models, the MRS estimator shows uniformly smaller Monte Carlo standard deviation than the UNADJ estimator, with approximately \(6\)-\(16 \%\) efficiency gain. The larger efficiency improvements are observed under the GLMM working models (W9-W12). The simulation results for Scenarios \ref{sc:B1} and \ref{sc:B2} are included in the Web Appendix L. Overall, the results under Scenario \ref{sc:B2} are similar to those under Scenario \ref{sc:B3}. In the absence of informative cluster-period size (Scenario \ref{sc:B1}), all four estimands coincide, and the MRS estimators give results similar to the Coef estimators from the GEE working model (W7-W8), whereas the Coef estimator from the GLMM exhibits clear bias. 
Additionally, we examined the performance of Coef estimators based on working models (W7-W12) excluding covariate adjustment; these results are presented in Web Appendix M (Web Table 14 - 16) and are similar to those with covariate adjustment (but with some efficiency loss). Finally, we examine the impact of fewer clusters by repeating Scenarios \ref{sc:B1}-\ref{sc:B3} with $I=9$ clusters. The findings are generally similar to those in Section \ref{simu_cont:res} and presented in Web Appendix N.

\subsection{Simulations with the tests for informative sizes}
In Web Appendix O, we conducted additional simulations to verify the properties of the proposed tests for informative sizes in Section \ref{info_test}. For both continuous and binary outcomes, data were generated based on Scenarios \ref{sc:C1} and \ref{sc:B1}, but included an additional informative-size component whose strength was governed by the parameter $\delta$ (details are provided in Web Appendix O). Under the null case of $\delta = 0$ (non-informative sizes), the tests appear conservative, with empirical rejection rates of approximately 0.03 for continuous outcomes and 0.02 for binary outcomes. As $\delta$ increases, corresponding to higher degree of informative sizes, the rejection rates rise sharply, demonstrating the ability of the proposed test to detect this phenomenon when present.


\section{Applications to two stepped wedge designs} \label{real_data}

\subsection{The TSOS Trial}

Post-traumatic stress disorder (PTSD) and related comorbidities are common following traumatic physical injury. Prior studies have demonstrated that early psychotherapeutic and pharmacological interventions targeting PTSD symptoms can be effective in improving patient outcomes. The Trauma Survivors Outcomes and Support (TSOS) trial was a pragmatic, stepped-wedge, cluster-randomized, effectiveness-implementation hybrid trial conducted at 25 Level I trauma centers across the United States to evaluate the effectiveness of implementing high-quality PTSD screening and intervention services in routine trauma care \citep{zatzick2021stepped}. The trial employed a stepped wedge design in which sites were randomized to initiate the intervention at staggered time points. Patients in the intervention arm received stepped collaborative care, while those in the control arm received enhanced usual care. The primary clinical outcome was PTSD symptom severity, measured using the PTSD Checklist-Civilian Version (PCL-C). A total of 491 injured patients were included in our analysis. Across the five study periods, the number of patients per cluster-period varies between 1 and 16. The primary estimands of interest were the ATEs on PTSD symptoms, defined as h-iATE, h-cATE, v-iATE, and v-cATE. These estimands represent the difference in mean PCL-C scores between the intervention and control conditions at different aggregation levels. To estimate these effects, we used the proposed MRS approach under six working models (W1) through (W6), as described in Table \ref{tab:working-models} and consistent with the simulation in Section \ref{simu_cont}. All models were adjusted for gender and age as baseline covariates. We report results for the MRS estimators and their corresponding Coef estimators from working models.

\begin{figure}[htbp]
    \centering
    \includegraphics[width=0.65\linewidth]{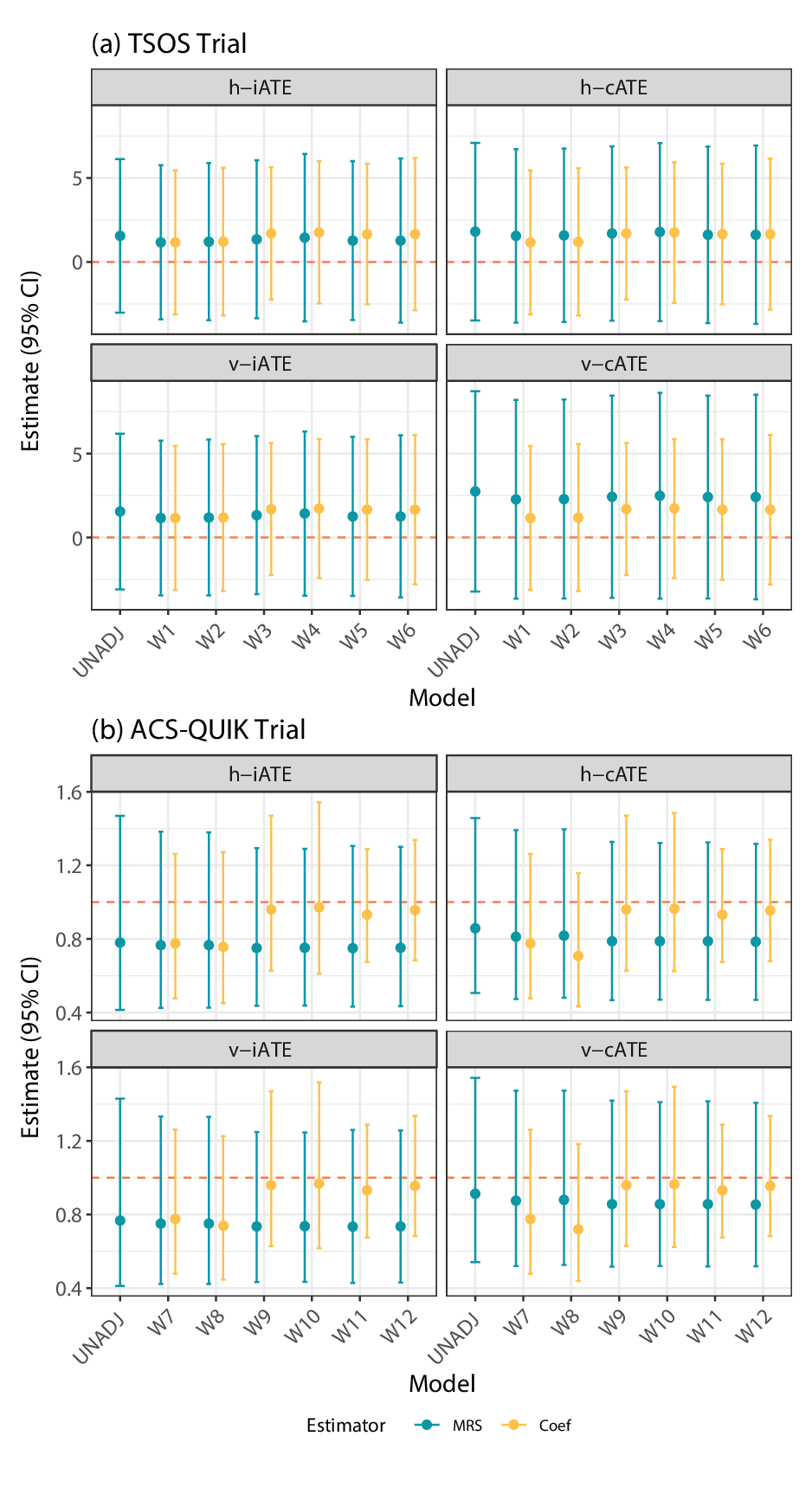}
    \caption{Panel (a): Estimated treatment effects in difference on PTSD symptom severity from the TSOS data using both MRS and Coef estimators under working models (W1) through (W6), as well as an unadjusted estimator. Each panel corresponds to one of the four estimands: h-iATE, h-cATE, v-iATE, and v-cATE. Panel (b): Estimated odds ratios of 30-day mortality from the ACS-QUIK trial under MRS and Coef estimators using working models (W7) through (W12), as well as an unadjusted estimator. Each panel corresponds to one of the four estimands: h-iATE, h-cATE, v-iATE, and v-cATE. For each model, point estimates and 95\% confidence intervals are displayed for both the MRS and Coef estimators.}
    \label{fig:data_fig}
\end{figure}

Figure \ref{fig:data_fig} (a) presents the estimated treatment effects on PTSD symptom scores using both MRS and Coef estimators across working models (W1) through (W6), along with an unadjusted method (UNADJ). Each panel corresponds to one of the four estimands: h-iATE, h-cATE, v-iATE, and v-cATE. Point estimates and 95\% confidence intervals are shown for each estimator. Across all working models, MRS estimates carry similar magnitude for each estimand, showing robustness to working model specification. In contrast, Coef estimates vary more substantially across models. Under GEE (W1-W1), Coef estimators align closely with h-iATE which is consistent with simulation results in Section \ref{simu_cont:res}. Across panels, the individual-average MRS estimators (h-iATE and v-iATE) show similar results, as do the cluster-average MRS estimators (h-cATE and v-cATE). However, the MRS estimators differ between individual-average and cluster-average, suggesting the presence of informative sizes. Although the tests reported in Web Appendix P are not statistically significant at the \(0.05\) level, this study may not also be designed to have adequate power for this test (and hence the test could at most serve as partial evidence). Among linear mixed models (W3-W6), Coef estimates shows larger bias for the v-cATE than for the other estimands. As an illustrative example, under working model (W4), a linear mixed model with a cluster-level random effect and a period-specific treatment effect, the MRS estimator produced estimates of 1.448 for h-iATE,  1.371 for h-cATE, 1.425 for v-iATE, and 2.486 for v-cATE. The Coef estimator under the same model produced different estimates: 1.767 for h-iATE, 1.745 for h-cATE, and 1.728 for both v-iATE and v-cATE. 
Consistent with simulation findings in Section \ref{simu_cont:res}, the UNADJ estimator is close to MRS but is associated with wider confidence intervals.

\subsection{The ACS-QUIK Trial}

The Acute Coronary Syndrome Quality Improvement in Kerala (ACS-QUIK) trial was a pragmatic, cluster-randomized, stepped-wedge clinical trial designed to evaluate the effectiveness of a locally adapted quality improvement toolkit in reducing 30-day mortality among patients presenting with acute myocardial infarction in hospitals across Kerala, India. Between November 2014 and November 2016, 63 hospitals were enrolled and sequentially randomized to implement the intervention at one of five predefined four-month steps. At each step, 12 or 13 hospitals transitioned to the intervention phase, beginning approximately two weeks before the scheduled implementation date \citep{huffman2018effect}. The dataset consists of 21,079 patients from these hospitals. The number of patients per cluster varied over time, ranging from 2 to 1,977, with an average of 345.  
The estimands of interest were defined as the odds ratio comparing 30-day mortality between intervention and control periods. Four estimands were considered: h-iATE, h-cATE, v-iATE, and v-cATE. To estimate these causal effects, we applied the proposed MRS method and Coef of treatment effect using six working models, (W7)-(W12), as summarized in Table \ref{tab:working-models}. All models were adjusted for gender, age, and cluster size as baseline covariates.

Figure \ref{fig:data_fig} (b) presents the estimated treatment effects on 30-day mortality odds ratio using both MRS and Coef estimators across working models (W7) through (W12), as well as the unadjusted (UNADJ) estimator. Each panel corresponds to one of the four estimands: h-iATE, h-cATE, v-iATE, and v-cATE. Across all models, MRS estimates continue to carry similar magnitude across working models for each estimand, though not statistically significant at the \(0.05\) level. In contrast, the Coef estimates exhibit greater variability depending on the model choice. Patterns across estimands show that MRS estimates for h-iATE and v-iATE give comparable results, as do h-cATE and v-cATE. However, the individual-average and cluster-average estimates differ, which indicate likely informative sizes. Although the formal test reported in Web Appendix P did not reach the \(0.05\) significance level, the trial may also lack sufficient power for such a test.  Moreover, Coef estimates from the independence GEE (W7-W8) are more aligned with the individual-average estimands, particularly h-iATE, consistent with the patterns observed in the simulation study. However, under GLMM (W9-W12), Coef estimates were consistently smaller in magnitude and closer to 1. For example, under working model (W10), a GLMM with period-specific treatment effects and a cluster-level random effect, the MRS estimator gives the estimates of odds ratio: $0.752$ for h-iATE, $0.855$ for h-cATE, $0.736$  for v-iATE, and $0.856$ for v-cATE. The Coef estimator under the same model gives the estimates closer to null: $0.957$ for h-iATE, $0.961$ for h-cATE, $0.959$ for v-iATE, and $0.961$ for v-cATE.

\section{Discussion} \label{discuss}

In this article, we clarify four distinct average treatment effect estimands in SW-CRTs, distinguished by the direction and sequence of aggregation across individuals, clusters, and time periods. These estimands correspond to explicit and interpretable weighting schemes and provide a unifying taxonomy for understanding how treatment effects are summarized under informative sizes. We further propose a model-robust standardization approach that enables consistent estimation of each estimand using familiar regression working models, even under outcome model misspecification. Importantly, our contribution is directly relevant to the ICH E9(R1) estimands framework \citep{EMA}. In particular, while the original ICH E9(R1) was not specifically developed to address CRTs, the ongoing CRT-Estimands Framework extension \citep{kahan2025development,Bi2026CRTEstimandsScoping} emphasizes the need to explicitly specify the population-level summary measure, including how individuals and clusters are weighted. Correspondingly, the four estimands studied here provide a systematic characterization of population-level summary measures for SW-CRTs, and facilitate the implementation of this key component by providing robust estimation strategies.

Our effort in developing a model-robust standardization strategy is connected to but different from prior work on causal inference for SW-CRTs. \citet{chen2024model} defined a class of weighted average treatment effect estimands under informative cluster size. Our estimands should be viewed as the super-population counterparts of those in \citet{chen2024model}. They have also pioneered the discussion on model-assisted estimation of potential outcomes estimands via linear analysis of covariance models. To connect to this prior work, we prove an interesting result in Web Appendix G that each model-assisted estimator proposed by \citet{chen2024model} is a special case of our proposed model-robust standardization estimator. In comparison, however, the advantage of using model-robust standardization is that one does not need to restrict to linear regression and the working independence assumption. Second, \citet{wang2024achieve} defined four interpretable potential outcomes estimands for SW-CRTs by allowing the treatment effect to potentially vary by exposure time and/or calendar time. However, they only focused on the cluster-average treatment effect defined based on an underlying cohort within each cluster. Our estimands are different in a sense that we went beyond the cluster-average definition and addressed an entire class of weighted average treatment effect estimands. However, we focused at most on the period-specific treatment structure, and did not address additional treatment effect estimands that can vary by exposure time or duration of treatment \citep{kenny2022analysis,maleyeff2022assessing}. \citet{wang2024achieve} also proved a set of central limit theorems to show that one can use a specially designed g-computation formula to consistently estimate potential outcomes estimands under LMM and GEE (with some constraints on mean and correlation model specification). Our model-robust standardization approach bears a similar spirit, but we do not require stringent working model specifications to achieve consistency. 

Although we primarily focus on SW-CRTs as a context, the estimands and model-robust standardization approach extend more broadly to longitudinal CRTs. That is, the same principles apply to longitudinal parallel-arm CRTs and cluster randomized crossover trials with little modifications, and we provide further elaborations on the details in Web Appendix Q. We have also developed an R package, \texttt{MRStdLCRT}, to implement our model-robust standardization procedure under all three longitudinal CRT designs, with example syntax provided in Web Appendix R. Second, while our primary exposition assumes a cross-sectional SW-CRT, closed-cohort designs, in which the same individuals are followed across periods, are still possible. The corresponding modifications to the estimand definitions and to the specification of working outcome models under closed-cohort designs are described in Web Appendix A. By contrast, open-cohort SW-CRTs, where individuals may enter or leave clusters over time, raise additional challenges for defining super-population estimands and for standardization. A full treatment of such designs is beyond the scope of this paper and represents an important direction for future research. Third, SW-CRTs with an implementation period, as a form of an incomplete design, require explicit clarification of the target population of interest. In practice, data observations from the implementation periods are often excluded, which may modify the set of eligible cluster-period cells contributing to the estimand. Removing the implementation periods requires redefining the estimands in Table \ref{tab:estimands} by excluding certain ineligible cluster-period cells in the summation. Such modifications do not affect the validity of the model-robust standardization procedure but instead reflect transparent alignment between the estimand definition and the subset of data deemed relevant for statistical inference. A technical presentation of this extension is provided in Web Appendix T.

In conventional practice, the analyses of SW-CRTs typically pre-specify a primary working outcome regression model for estimating the treatment effect. As a further step, implementation of the model-robust standardization approach requires a two-tier pre-specification of both the target potential outcomes estimand (operationalized through the weighting scheme in Table 1) and the working outcome regression model. That is, in contrast to conventional practice, which focus primarily on model choice, our framework makes clear that the estimand should be defined first, as it reflects the scientific question and determines how outcomes are aggregated across individuals, clusters, and periods. Conditional on this estimand specification, a working outcome model can then be selected, typically mirroring the model that would otherwise be used in a standard pre-specified SW-CRT analysis \citep{li2021mixed}. For example, we would prespecify as the primary working model a generalized linear mixed model with fixed effects for period, treatment, and possible prognostic covariates, together with a nested exchangeable random-effects structure, using a link function appropriate for the outcome type. While we demonstrate that model-robust standardization substantially reduces sensitivity of the estimated treatment effect to outcome model misspecification, the choice of working model remains consequential for precision, with greater efficiency generally achieved when the working model more closely approximates the underlying data generating process. Given these considerations, we recommend pre-specifying a primary working model in the protocol or statistical analysis plan alongside a clear estimand definition, while reserving additional, scientifically plausible working models for secondary or sensitivity analyses implemented through the same model-robust standardization procedure.

While pre-specification of a working regression model is standard practice, explicit pre-specification of the target estimand may be less familiar to many practitioners. We offer several preliminary considerations below. First, the choice between cluster-average and individual-average estimands generally depend on the unit of inference \citep{li2025model}. The cluster-average estimand is a natural candidate when studying a cluster-level intervention targeting improvements for entire clusters, whereas the individual-average estimand is a natural candidate when the intent is to mimic an individually-randomized trial but when cluster randomization is used for practical or administrative concerns. Second, a key difference in estimands consideration for SW-CRTs is the layer of horizontal (across clusters) and vertical (across time) aggregation strategies. Horizontal estimands collapse potential outcomes across periods within each cluster non-differentially, and are more akin to the estimands definition in parallel-arm CRTs. Hence, these estimands do not fully address potential effect heterogeneity across calendar time. In contrast, vertical estimands treat SW-CRTs as a series of parallel-arm CRTs (defined by distinct periods), and therefore are averages of period-specific treatment effects. Therefore, it is easy to draw inference about more granular, period-specific cluster-average and individual-average treatment effect estimands. Thus, the choice between the direction of aggregation could depend on considerations for potential calendar time treatment effect heterogeneity. In the TSOS trial, the intervention involved implementing PTSD screening and intervention services within routine trauma care and targets cluster-level improvements. As a result, we believe the cluster-average treatment effect may be of greater interest. In contrast, the ACS-QUIK trial introduced a quality improvement toolkit designed for individual patients to improve survival, despite cluster randomization. It is therefore reasonable to think that the trial mimics an individually-randomized trial and the individual-average treatment effect appears to be a more suitable target. In both trials, we do not see any major differences between the effect estimates for horizontal and vertical estimands, and the vertical estimands may be preferred should the study team anticipate calendar time treatment effect heterogeneity \emph{a priori}. We emphasize that these are example considerations rather than fixed rules. In practice, we encourage study teams to carefully think through estimands of interest in stepped wedge designs, and envision that this work promotes a conversation in that direction. 

As a closing remark, we note that inference with a small number of clusters poses consistent challenges for all CRTs, particularly for variance and interval estimation. We conducted additional simulations in Section \ref{simu} (Web Appendices K and N) with $I=10$ and $I=9$ clusters, respectively. We find that the proposed model-robust standardization estimators remain approximately unbiased with near-nominal coverage. This suggests that the model-robust standardization approach continues to be useful in settings with fewer clusters. However, for extremely small numbers of clusters (e.g., fewer than 5), reliable model-robust inference in SW-CRTs remains an open methodological challenge. While the jackknife variance estimator used here is closely related to bias-corrected variance estimators previously studied for SW-CRTs \citep{ouyang2024maintaining}, further work is needed to develop its extensions that address an extremely small number of clusters under model-robust standardization. We also note that although SW-CRTs with very few clusters do occur in practice, their ability to support strong causal conclusions has been debated \citep{taljaard2016substantial}. We anticipate more future work to combat small-sample challenges under model-robust standardization.

\section*{Acknowledgements}
Research in this article was supported by a Patient-Centered Outcomes Research Institute Award\textsuperscript{\textregistered} (PCORI\textsuperscript{\textregistered} Award ME-2022C2-27676) and the National Institute Of Allergy And Infectious Diseases of the National Institutes of Health under Award Number R00AI173395. This work is also partially supported by the National Institute on Aging (NIA) of the National Institutes of Health (NIH) under Award Number U54AG063546, which funds NIA Imbedded Pragmatic Alzheimer’s Disease (AD) and AD-Related Dementias Clinical Trials Collaboratory (NIA IMPACT Collaboratory). The statements presented in this article are solely the responsibility of the authors and do not necessarily represent the official views of the National Institutes of Health or PCORI\textsuperscript{\textregistered}, its Board of Governors, or the Methodology Committee. The authors also would like to thank Dr Douglas Zatzick for sharing the data from the TSOS study.







\clearpage
\newpage
\printbibliography

\newpage
\appendix

\section{Modifications to accommodate closed-cohort SW-CRTs} \label{close-sw}

In a closed-cohort SW-CRT, a fixed number of individuals will be enrolled at baseline and followed across all periods \citep{gasparini2025analysis}. In contrast to the cross-sectional design, where different subjects are observed in each cluster-period, the closed-cohort design maintains the same set of individuals \( k = 1,\dots, N_i \) within each cluster \( i = 1,\ldots, I \) across all periods \( j = 1,\ldots, J \); hence, in the absence of attrition, \( N_{ij} = N_i \) for all \( j \). The potential outcome \( Y_{ijk}(z) \) denotes the response of individual \( k \) in cluster \( i \) and period \( j \), were the cluster-period assigned treatment status \( Z_{ij} = z \). We assume temporal consistency---namely, the outcome at a given time depends only on the contemporaneous treatment status and not on the timing of treatment initiation. 

Following the definition of the weighted average treatment effects in \eqref{weight_causal_effect}, the fixed individual set and constant cluster size in fact simplify the estimands defined for cross-sectional designs. In particular, the \(\mathrm{h\text{-}iATE}\) and \(\mathrm{v\text{-}iATE}\) estimands coincide, and the \(\mathrm{h\text{-}cATE}\) and \(\mathrm{v\text{-}cATE}\) estimands also coincide. Specifically,
\[
\text{iATE}=\mu_\text{I}(z)=\mu_{\text{I}}^\text{h}(z) = \mu_{\text{I}}^\text{v}(z)= \frac{1}{J-2}\sum_{j=2}^{J-1} \frac{\mathbb{E}\left[\sum_{k=1}^{N_i} Y_{ijk}(z)\right]}{\mathbb{E}[N_i]},
\]
and
\[
\text{cATE}= \mu_\text{C}(z)=\mu_{\text{C}}^\text{h}(z)= \mu_{\text{C}}^\text{v}(z) = \frac{1}{J-2}\sum_{j=2}^{J-1} \mathbb{E}\left[\frac{1}{N_i} \sum_{k=1}^{N_i} Y_{ijk}(z)\right].
\]
That is, the individual-average and cluster-average estimands directly follow the counterparts defined for parallel-arm CRTs \citep{kahan2023estimands,kahan2024demystifying}, and are invariant to horizontal or vertical aggregation. In a closed-cohort SW-CRT, both estimands can be estimated using the proposed model-robust standardization estimator \eqref{aug_mu}, with possible modifications for the working outcome model tailored to features of repeated outcome assessment within the same individuals. Specifically, the individual-average estimand \(\mu_{\text{I}}(z)\) can be estimated as
\[
\widehat{\mu}_{\text{I}}^{\text{aug}}(z) = \frac{1}{J-2} \sum_{j=2}^{J-1} \widehat{\mu}_{\text{I},j}^{\text{aug}}(z),
\]
where
\[
\widehat{\mu}_{\text{I},j}^{\text{aug}}(z) =  \frac{\sum_{i=1}^{I} N_i\, m_{zj}(\underline{\bm{X}}_{ij}, N_i)}{\sum_{i=1}^{I} N_i}+\frac{\sum_{i=1}^{I} N_i\, \mathbb{I}(Z_{ij}=z)\, \left\{\overline{Y}_{ij} - m_{zj}(\underline{\bm{X}}_{ij}, N_i)\right\}}{\sum_{i=1}^{I} \mathbb{I}(Z_{ij}=z)\, N_i}.
\]
Similarly, the cluster-average estimand \(\mu_\text{C}(z)\) is estimated as
\[
\widehat{\mu}_{\text{C}}^{\text{aug}}(z) = \frac{1}{J-2} \sum_{j=2}^{J-1} \widehat{\mu}_{\text{C},j}^{\text{aug}}(z),
\]
where
\[
\widehat{\mu}_{\text{C},j}^{\text{aug}}(z) = \frac{1}{I}\sum_{i=1}^{I} m_{zj}(\underline{\bm{X}}_{ij}, N_i)+\frac{\sum_{i=1}^{I} \mathbb{I}(Z_{ij}=z)\, \left\{\overline{Y}_{ij} - m_{zj}(\underline{\bm{X}}_{ij}, N_i)\right\}}{\sum_{i=1}^{I} \mathbb{I}(Z_{ij}=z)} .
\]

To implement the model-robust standardization estimator in closed-cohort SW-CRTs, the working model \( m_{zj}(\underline{\bm{X}}_{ij}, N_i) \) can be specified using regression approaches introduced in main paper Section \ref{example1}-\ref{example4} and Web Appendix Section \ref{example2} - \ref{example5}, with appropriate modifications that additionally account for the within-individual ICC for repeated outcome measures over time. For example, one could consider the following linear mixed model suitable for closed-cohort designs:
\begin{align}
    Y_{ijk} = \beta_j + Z_{ij} \tau_j + g(\underline{\bm{X}}_{ij}, N_i, Z_{ij}) + RE_{ik} + \epsilon_{ijk},
\end{align}
where an additional random effect for the repeated measures on the same individual \(k\) in cluster \(i\) may be added to \(RE_{ij}\). One concrete specification, proposed by Girling and Hemming \citep{girling2016statistical}, takes the form \(RE_{ik} = \alpha_i + \gamma_{ij} + \phi_{ik}\), where \(\phi_{ij} \sim \mathcal{N}(0,\tau_{\phi}^2)\) is independent of both \(\alpha_i\) and \(\gamma_{ij}\) as we defined in Section \ref{example1}. This structure defines the block exchangeable correlation model, where the correlations are exchangeable both within and across periods. Under this structure, the correlations can be summarized through the following ICCs: the within individual-ICC for repeated measure is \(\rho_a = (\tau_\alpha^2+\tau_\phi^2)/(\tau_\alpha^2+\tau_\gamma^2+\tau_\phi^2+\sigma_\epsilon^2) \), the within-period-ICC is \(\phi_w = (\tau_\alpha^2 + \tau_\gamma^2)/(\tau_\alpha^2 + \tau_\gamma^2 + \tau_\phi^2 + \sigma_\epsilon^2) \), and the between-period ICC is \(\rho_b = \tau_\alpha^2/(\tau_\alpha^2+\tau_\gamma^2 + \tau_\phi^2 + \sigma_\epsilon^2) \). As an alternative that allows for between-period correlation decay, \citet{li2020design} introduced the proportional decay model, in which \( RE_{ik} = \gamma_{ik} \) with \( \gamma_{ik} \sim \mathcal{N}(0, \tau_\gamma^2 \widetilde{\bm{M}}) \). The  \(\widetilde{\bm{M}}\) follows an antoregressive structure with decay parameter \( r \). This leads to an exponential decreasing correlation structure over time, with within-individual ICC as \(\rho_{a,|j-l|} = r^{|j-l|} \), within-period and between period ICC as \(\rho_w = \tau_\gamma^2 /(\tau_\gamma^2 + \sigma_\epsilon^2) \) and \(\rho_{b,|j-l|} = \tau_\gamma^2 r^{|j-l|} /(\tau_\gamma^2 + \sigma_\epsilon^2) \). These correlations can also be integrated into the GLMM model described in main paper Section \ref{example3}. 
For marginal models such as those in main paper Section \ref{example4} and Section \ref{example5} in this Web Appendix, the working correlation matrix should often be adapted to account for repeated measurements inherent in closed-cohort SW-CRT designs. Example block exchangeable working correlation models and proportional decay working correlation models are given by prior work that studies GEE in closed-cohort SW-CRTs \citep{li2018,li2020design,zhang2023general}.

In a closed-cohort SW-CRT, the fixed composition of individuals over time introduces a natural risk of attrition due to intermittent nonresponse or dropout. While all baseline covariates should be collected at baseline when study participants are enrolled, the outcomes could be missing for some participants when they miss a visit or leave the study. That is, two possible patterns of missingness can arise. A monotone missing pattern occurs when an individual, once missing, is never observed again, typically referred to as dropout. In contrast, a nonmonotone missing pattern allows for reappearance of values after a missing observation. Multiple imputation (MI) provides a framework to handle both monotone and nonmonotone missingness under the assumptions of missing at random (MAR) or missing not at random (MNAR). MI is typically conducted in two stages. First, missing values are imputed multiple times by sampling from an approximation to the posterior predictive distribution of the missing data given the observed data. Second, each completed dataset is analyzed using the method of interest, and the results are combined using Rubin's rules \citep{rubin1978multiple}. Closed-cohort SW-CRTs involve three sources of correlation: repeated measures within individuals, within-period correlation, and within-cluster correlation and it is recommended to use multilevel multiple imputation (MMI) methods that are congenial with the substantive analysis model \citep{drechsler2015multiple,van2012flexible}. There are two major model-based strategies for conducting MM: joint modeling (JM) and fully conditional specification (FCS). In JM, all incomplete variables are modeled jointly, typically assuming multivariate normality, treating incomplete variables as multivariate responses and fully observed variables as predictor. In contrast, FCS iteratively imputes each variable using a series of univariate models. The extension of JM method for MMI uses a multivariate linear mixed model (LMM) as the imputation model \citep{schafer2002computational}. Similarly, the multilevel extension to the FCS imputes missing values using a series of univariate LMMs \citep{van2012flexible}. As mentioned by \citet{huque2020multiple} and \citet{wijesuriya2022multiple}, extending MI models for such three-layer data with dummy indicators for hierarchical levels can produce approximately unbiased estimates. Bayesian multiple imputation methods offer an alternative strategy, often using predictive mean matching, which imputes only from observed values. \citet{luo2016bayesian} extended this idea to develop a Bayesian MMI framework for multivariate longitudinal data with both monotone and non-monotone missingness \citet{gasparini2025analysis}. further proposed a joint longitudinal-survival model to address informative dropout by combining a nested exchangeable model for the outcome with a proportional hazards model for time-to-dropout, specifically for closed-cohort SW-CRT settings. Regarding the choice of \(M\), the number of imputations, it is generally accepted that using 2 to 10 imputations is sufficient to give efficient point estimates. However, if users wish to ensure stable standard errors that do not vary substantially across different numbers of imputations, a larger \(M\) may be required. \citet{white2011multiple} recommend, as a rule of thumb, choosing the number of imputations to be at least as large as the percentage of subjects with any missing data. More recently, \citet{von2020many} proposed a more formal, variance-based approach for determining the optimal number of imputations required to achieve a desired level of precision.

In practice, several \texttt{R} packages support MMI. The \texttt{pan} and \texttt{jomo} packages implement MI under multivariate joint modeling, \citet{grund2016multiple} while 
the \texttt{ml.lmer} function in package \texttt{miceadds} provides FCS-based implementation for three-level imputation. \citet{robitzsch2020miceadds} Once multiple imputations are completed, estimates across the \(M\) imputed datasets are combined using Rubin's rules. Suppose \(\widehat{\mu}_{\omega}^{\text{aug},(m)}(z)\) and \(\widehat{V}^{\text{aug},(m)}\) denote the point estimate and its Jackknife variance from the \(m\)-th imputed dataset. The combined estimate is
\[
\overline{\mu}_{\omega}(z) = \frac{1}{M} \sum_{m=1}^{M} \widehat{\mu}_{\omega}^{\text{aug},(m)}(z),
\]
with total variance given by
\(
T_{\omega}(z) = \overline{V}_{\omega}(z) + \left(1 + \frac{1}{M} \right) B_{\omega}(z),
\)
where the average within-imputation variance is
\(
\overline{V}_{\omega}(z) = \frac{1}{M} \sum_{m=1}^{M} \widehat{V}^{\text{aug},(m)},
\)
and the between-imputation variance is
\(
B_{\omega}(z) = \frac{1}{M-1} \sum_{m=1}^{M} \left(\widehat{\mu}_{\omega}^{\text{aug},(m)}(z) - \overline{\mu}_{\omega}(z) \right)^2.
\) The inference based on MMI uses the \(t\)-distribution, with degrees of freedom given as \citet{hossain2017missing}
\[
df = (M-1) \left(1 + \frac{M}{M+1} \frac{\overline{V}_\omega(z)}{B_\omega(z)} \right)^2.
\]
A modified degree of freedom, proposed by \citet{barnard1999miscellanea} adjusts for finite-sample and is given by:
\[
df_{\text{adj}} = \left( df^{-1} + \widehat{df}^{-1} \right)^{-1},
\]
where
\[
\widehat{df} = \left( \frac{df_{\text{com}} + 1}{df_{\text{com}} + 3} \cdot df_{\text{com}} \right) \left( 1 + \frac{M+1}{M} \frac{B_\omega(z)}{\overline{V}_\omega(z)} \right)^{-1},
\]
and \(df_{\text{com}}\) denotes the degrees of freedom under the assumption of a complete dataset, typically approximated as \(I - 1\).

\section{Proof of the consistency of nonparametric estimators} \label{non_adjust_proof}
In this section, we provide proof of the consistency of \(\widehat{\mu}_\omega^{\text{unadj}}(z)\). 

By the law of large numbers, and under the assumption of SUTVA and randomization, as \(I \rightarrow \infty\), 
\[ \frac{1}{I} \sum_{i=1}^{I} \omega_{ij} \mathbb{I}(Z_{ij} = z)  \overline{Y}_{ij} \stackrel{p}{\rightarrow} \mathbb{E} \left[ \omega_{ij} \mathbb{I}(Z_{ij}=z) \overline{Y}_{ij} \right], \]
and 

\[ \frac{1}{I} \sum_{i=1}^{I} \omega_{ij} \mathbb{I}(Z_{ij} = z)  \stackrel{p}{\rightarrow} \mathbb{E} \left[ \omega_{ij} \mathbb{I}(Z_{ij}=z) \right]. \]
Hence, by the continuous mapping  theorem, 

\begin{align*}
   & \widehat{\mu}_\omega^{\text{unadj}}(z)\\
    & \stackrel{p}{\rightarrow} \frac{1}{\sum_{j=2}^{J-1} \omega_j} \sum_{j=2}^{J-1} \omega_j \frac{\mathbb{E} \left[ \omega_{ij} \mathbb{I}(Z_{ij}=z) \overline{Y}_{ij} \right]  }{\mathbb{E} \left[ \omega_{ij} \mathbb{I}(Z_{ij}=z) \right]} \\
    & = \frac{\mathbb{E} \left[ \omega_{ij} \mathbb{I}(Z_{ij}=z) \overline{Y}_{ij} \right]}{\mathbb{E} \left[ \omega_{ij} \mathbb{I}(Z_{ij}=z) \right]} \\
    & =\mu_\omega(z).
\end{align*}
The proof is complete.

\section{Proof of the consistency of the augmented estimators} \label{aug_proof}

In this section we prove the consistency of \(\widehat{\mu}_\omega^{\text{aug}}\). 

By definition, we have:
\begin{align}
\widehat{\mu}_{\omega,j}^{\text{aug}}(z)  & = \frac{\sum_{i=1}^{I} \omega_{ij} \mathbb{I}(Z_{ij}=z) \left[ \overline{Y}_{ij} - m_{zj}(\underline{\bm{X}}_{ij},N_{ij} ) \right]  }{\sum_{i=1}^{I}\omega_{ij} \mathbb{I}(Z_{ij}=z) } \label{term1}\\
& + \frac{\sum_{i=1}^{I}\omega_{ij}  m_{zj}(\underline{\bm{X}}_{ij},N_{ij} ) }{\sum_{i=1}^{I} \omega_{ij}}.  \label{term2} 
\end{align}
By the law of large numbers and continuous mapping theorem, we can show that 
\begin{align*}
    \eqref{term1} \stackrel{p}{\rightarrow} \frac{\mathbb{E}\left[ \omega_{ij} \mathbb{I}(Z_{ij}=z) \left\{ \overline{Y}_{ij} - m_{zj}(\underline{\bm{X}}_{ij},N_{ij} ) \right\} \right]}{\mathbb{E}\left[ \omega_{ij} \mathbb{I}(Z_{ij}=z) \right]}. 
\end{align*}
Since randomization is conditionally independent with \(\overline{Y}_{ij}(z)\), we have
\begin{align*}
  \eqref{term1} + \eqref{term2} & \stackrel{p}{\rightarrow} \frac{\mathbb{E}\left[\omega_{ij}\mathbb{I}(Z_{ij}=z) \overline{Y}_{ij}   \right]  }{E[\omega_{ij} \mathbb{I}(Z_{ij}=z)] } \\
  & - \frac{E\left[\omega_{ij} \mathbb{I}(Z_{ij}=z) m_{zj}(\underline{\bm{X}}_{ij},N_{ij})  \right] }{\mathbb{E}[\omega_{ij} \mathbb{I}(Z_{ij}=z)] } + \frac{\mathbb{E}\left[\omega_{ij} m_{zj}(\underline{\bm{X}}_{ij},N_{ij})  \right]  }{E[\omega_{ij}]}. 
\end{align*}
If the working model is incorrect:
\begin{align*}
     \eqref{term1} + \eqref{term2} & \stackrel{p}{\rightarrow} \frac{\mathbb{E}\left[\omega_{ij}\mathbb{I}(Z_{ij}=z) \overline{Y}_{ij}   \right]  }{E[\omega_{ij} \mathbb{I}(Z_{ij}=z)] } \\
  & - \frac{E[\mathbb{I}(Z_{ij}=z)]\mathbb{E}\left[\omega_{ij} m_{zj}(\underline{\bm{X}}_{ij},N_{ij})  \right] }{E[\mathbb{I}(Z_{ij}=z)]\mathbb{E}[\omega_{ij}] } + \frac{\mathbb{E}\left[\omega_{ij} m_{zj}(\underline{\bm{X}}_{ij},N_{ij})  \right]  }{\mathbb{E}[\omega_{ij}]} =  \frac{\mathbb{E}\left[\omega_{ij}\mathbb{I}(Z_{ij}=z) \overline{Y}_{ij}   \right]  }{\mathbb{E}[\omega_{ij} \mathbb{I}(Z_{ij}=z)] }. 
\end{align*}
If the working model is correct, \(m_{zj}(\underline{\bm{X}}_{ij},N_{ij}) = \mathbb{E}\left[\overline{Y}_{ij} \mid Z_{ij}=z,\underline{\bm{X}}_{ij},N_{ij} \right] \), which gives us \(\eqref{term1} = 0 \), and 

\begin{align*}
     \eqref{term1} + \eqref{term2} & \stackrel{p}{\rightarrow} \frac{\mathbb{E}\left[\omega_{ij} m_{zj}(\underline{\bm{X}}_{ij},N_{ij})  \right]  }{E[\omega_{ij}]} = \frac{\mathbb{E}\left[\omega_{ij} \mathbb{I}(Z_{ij}=z) m_{zj}(\underline{\bm{X}}_{ij},N_{ij})  \right]  }{E[\mathbb{I}(Z_{ij}=z)\omega_{ij}]}  =\frac{\mathbb{E}\left[\omega_{ij}\mathbb{I}(Z_{ij}=z) \overline{Y}_{ij}   \right]  }{\mathbb{E}[\omega_{ij} \mathbb{I}(Z_{ij}=z)] }. 
\end{align*}
Since \(\widehat{\mu}_{\omega,j}^{\text{aug}}\) is a consistent estimator of \[\frac{\mathbb{E}\left[\omega_{ij}\mathbb{I}(Z_{ij}=z) \overline{Y}_{ij}   \right]  }{E[\omega_{ij} \mathbb{I}(Z_{ij}=z)] },\] we can show that 

\[
\widehat{\mu}_\omega^{\text{aug}}(z) = \frac{\sum_{j=2}^{J-1} \omega_j \widehat{\mu}_{\omega,j}^{\text{aug}}(z)  }{ \sum_{j=2}^{J-1} \omega_j } \stackrel{p}{\rightarrow}     \mu_{\omega}(z) = \frac{\mathbb{E}\left[ \sum_{j=2}^{J-1} \omega_{ij} \overline{Y}_{ij}(z) \right] }{\mathbb{E}\left[\sum_{j=2}^{J-1} \omega_{ij} \right]}.
\]

\section{The optimal choice of working model} \label{optimum_proof}

Define
\[
p_{z,\omega,j} = \mathbb{E} \left[ \frac{ \sum_{i=1}^{I} \omega_{ij} \mathbb{I}(Z_{ij} = z)}{ \sum_{i=1}^{I} \omega_{ij} } \right]
\]
as the weighted proportion of clusters receiving treatment \(Z_{ij} = z\) in period \(j\). The influence function for period \(j\) is then defined as
\[
h(\underline{\bm{X}}) = \frac{Z \omega \overline{Y}}{p_{1,\omega}} - \frac{(1 - Z) \omega \overline{Y}}{1 - p_{1,\omega}} - \tau_{\omega},
\]
and any influence function can be written as
\[
h(\underline{\bm{X}}) + (Z - p_{1,\omega}) m(\underline{\bm{X}}, N),
\]
where \(m(\underline{\bm{X}}, N)\) is a user-specified augmentation function.

According to \citet{tsiatis2006semiparametric}, the optimal influence function is obtained by choosing the augmentation term \((Z - p_{1,\omega}) m(\underline{\bm{X}}, N)\) such that it minimizes the projection of \(h(\underline{\bm{X}})\) onto \((Z - p_{1,\omega}) m(\underline{\bm{X}}, N)\). Due to the linearity of projections, this is equivalent to finding a function \(m_1(\underline{\bm{X}}, N)\) such that
\[
\mathbb{E} \left[ \left\{ Z \overline{Y}(1) - (Z - p_{1,\omega}) m_1(\underline{\bm{X}}, N) \right\} (Z - p_{1,\omega}) h(\underline{\bm{X}}) \right] = 0.
\]
Therefore,
\begin{align}
    & \mathbb{E} \left[ \left\{ Z \overline{Y}(1) - (Z - p_{1,\omega}) m_1(\underline{\bm{X}}, N) \right\} (Z - p_{1,\omega}) h(\underline{\bm{X}}) \right] = 0 \nonumber \\
    \Rightarrow \; & \mathbb{E} \left[ Z (Z - p_{1,\omega}) h(\underline{\bm{X}}) \overline{Y}(1) - (Z - p_{1,\omega})^2 m_1(\underline{\bm{X}}, N) \right] = 0 \nonumber \\
    \Rightarrow \; & \mathbb{E} \left[ \mathbb{E} \left\{ Z (Z - p_{1,\omega}) h(\underline{\bm{X}}) \overline{Y}(1) \mid Z, \underline{\bm{X}}, N \right\} \right] - \mathbb{E} \left[ p_{1,\omega} (1 - p_{1,\omega}) m_1(\underline{\bm{X}}, N) h(\underline{\bm{X}}) \right] = 0 \nonumber \\
    \Rightarrow \; & \mathbb{E} \left( \mathbb{E} \left[ Z (Z - p_{1,\omega}) h(\underline{\bm{X}}) \mathbb{E}\{\overline{Y}(1) \mid Z = 1, \underline{\bm{X}}, N\} \mid \underline{\bm{X}}, N \right] \right) - \mathbb{E} \left[ p_{1,\omega} (1 - p_{1,\omega}) m_1(\underline{\bm{X}}, N) h(\underline{\bm{X}}) \right] = 0 \nonumber \\
    \Rightarrow \; & \mathbb{E} \left[ p_{1,\omega} \{1 - p_{1,\omega}\} \left( \mathbb{E}\{\overline{Y}(1) \mid Z = 1, \underline{\bm{X}}, N\} - m_1(\underline{\bm{X}}, N) \right) h(\underline{\bm{X}}) \right] = 0. \label{aug_expect}
\end{align}

Since \(p_{1,\omega}\) is bounded away from zero almost surely and \(\mathbb{E}\{\overline{Y}(1) \mid Z = 1, \underline{\bm{X}}, N\} - m_1(\underline{\bm{X}}, N)\) is a function of \(\underline{\bm{X}}, N\), the expectation in \eqref{aug_expect} can only be zero for all \(h(\underline{\bm{X}})\) if and only if
\[
\mathbb{E}\{\overline{Y}(1) \mid Z = 1, \underline{\bm{X}}, N\} = m_1(\underline{\bm{X}}, N).
\]

A similar result holds for \(Z = 0\). Therefore, for each period \(j\), the optimal augmentation function that minimizes the asymptotic variance of \(\tau_\omega^{\text{aug}}\) is
\(
m_{z,j}(\underline{\bm{X}}_{ij}, N_{ij}) = \mathbb{E} \left[ \overline{Y}_{ij} \mid Z_{ij} = z, \underline{\bm{X}}_{ij}, N_{ij} \right].
\)

\section{Proof of Remark \ref{rem:NP_wo_cov}} \label{np_aug_proof}
\begin{proof}
Fix a period \(j\) and treatment level \(z\).  Suppose there exists a constant \(\mu_{zj} \in \mathbb{R}\) such that 
\[ m_{zj}(\underline{\bm{X}}_{ij},N_{ij}) = \mu_{zj} \quad\text{for all } i=1,\dots,I.\]
Then the first term in the augmentation component is
\[\frac{\sum_{i=1}^{I} \omega_{ij} m_{zj}(\underline{\bm{X}}_{ij},N_{ij})}{\omega_j}=\frac{\sum_{i=1}^{I} \omega_{ij} c}{\sum_{i=1}^{I} \omega_{ij}}=c,
\]
and the second term is
\[\frac{\sum_{i=1}^{I} \omega_{ij} \mathbb{I}(Z_{ij}=z)\, m_{zj}(\underline{\bm{X}}_{ij},N_{ij})}{\sum_{i=1}^{I} \omega_{ij} \mathbb{I}(Z_{ij}=z)}=\frac{\sum_{i=1}^{I} \omega_{ij} \mathbb{I}(Z_{ij}=z) c}{\sum_{i=1}^{I} \omega_{ij} \mathbb{I}(Z_{ij}=z)}= c.
\]
Thus,
\[\frac{\sum_{i=1}^{I} \omega_{ij} m_{zj}(\underline{\bm{X}}_{ij},N_{ij})}{\omega_j} - \frac{\sum_{i=1}^{I} \omega_{ij} \mathbb{I}(Z_{ij}=z)\, m_{zj}(\underline{\bm{X}}_{ij},N_{ij})}{\sum_{i=1}^{I} \omega_{ij} \mathbb{I}(Z_{ij}=z)}= 0,
\]
where the augmentation term vanishes.
As a result,
\[\widehat{\mu}_{\omega,j}^{\text{aug}}(z)=\frac{\sum_{i=1}^{I} \omega_{ij} \mathbb{I}(Z_{ij}=z)\, \overline{Y}_{ij}}{\sum_{i=1}^{I} \omega_{ij} \mathbb{I}(Z_{ij}=z)}=\widehat{\mu}_{\omega,j}^{\text{unadj}}(z).
\]
Conditional on \(\underline{\bm{X}}_{ij},N_{ij}\), for \(j=1,\dots, J\), and \(i=1,\dots, I\), and cluster randomization, we have
 \begin{align*}
0&=\frac{\sum_{i=1}^{I} \omega_{ij} m_{zj}(\underline{\bm{X}}_{ij},N_{ij})}{\omega_j}
-\frac{\sum_{i=1}^{I} \omega_{ij} \mathbb{I}(Z_{ij}=z)\, m_{zj}(\underline{\bm{X}}_{ij},N_{ij})}
       {\sum_{i=1}^{I} \omega_{ij} \mathbb{I}(Z_{ij}=z)} \\
&\Longleftrightarrow \frac{\sum_{i=1}^{I} \omega_{ij} m_{zj}(\underline{\bm{X}}_{ij},N_{ij})}{\omega_j}= \frac{\sum_{i=1}^{I} \omega_{ij} \mathbb{I}(Z_{ij}=z)\, m_{zj}(\underline{\bm{X}}_{ij},N_{ij})}{\sum_{i=1}^{I} \omega_{ij} \mathbb{I}(Z_{ij}=z)} \\
&\Longleftrightarrow \frac{\sum_{i=1}^{I} \omega_{ij} m_{zj}(\underline{\bm{X}}_{ij},N_{ij})}{\omega_j} = \frac{\sum_{i=1}^{I} \omega_{ij} \mathbb{I}(Z_{ij}=z)\, m_{zj}(\underline{\bm{X}}_{ij},N_{ij})}{\sum_{i=1}^{I} \omega_{ij} \mathbb{I}(Z_{ij}=z)}=: \mu_{zj} \\
&\Longleftrightarrow \sum_{i=1}^{I} \omega_{ij} m_{zj}(\underline{\bm{X}}_{ij},N_{ij})=\mu_{zj}\,\sum_{i=1}^{I} \omega_{ij} \quad\text{and}\quad \sum_{i=1}^{I} \omega_{ij} \mathbb{I}(Z_{ij}=z)\, m_{zj}(\underline{\bm{X}}_{ij},N_{ij}) = \mu_{zj}\,\sum_{i=1}^{I} \omega_{ij} \mathbb{I}(Z_{ij}=z) \\
&\Longleftrightarrow
\sum_{i=1}^{I} \omega_{ij} \bigl\{m_{zj}(\underline{\bm{X}}_{ij},N_{ij}) - \mu_{zj}\bigr\} = 0
\quad\text{and}\quad
\sum_{i=1}^{I} \omega_{ij} \mathbb{I}(Z_{ij}=z)\, \bigl\{m_{zj}(\underline{\bm{X}}_{ij},N_{ij}) - \mu_{zj}\bigr\}
= 0.
\end{align*}
This leads to \(\mu_{zj}(\underline{\bm{X}}_{ij},N_{ij})=\mu_{zj}\) for any \(j=2\dots, J-1\) and \(i=1,\dots, I\).

\end{proof}

\section{Specifying the working outcome regression model with cluster-period means} \label{clus_period_model}

\subsection{Linear mixed model with cluster-period-level means}\label{example2}

An alternative modeling strategy specifies a linear mixed-effects model directly for the cluster-period mean outcome \(\overline{Y}_{ij}\), aggregated from the individual-level data:
\begin{align} \label{lmm_cp}
\overline{Y}_{ij} = \beta_j^{*} + Z_{ij} \tau_j^{*} + g^{*}(\overline{\bm{X}}_{ij}, N_{ij}, Z_{ij}) + \alpha_i^{*} + \gamma_{ij}^{*},
\end{align}
where \(\beta_j^{*}\) and \(\tau_j^{*}\) denote period-specific fixed and treatment effects, and again, \(\tau_j^{*}\) may be simplified under a constant treatment effect assumption by setting \(\tau_j^{*} = \tau^{*}\). The function \(g^{*}(\overline{\bm{X}}_{ij}, N_{ij}, Z_{ij})\) represents a general function of cluster-period-level covariates (summary of individual-level covariates using the form \(\overline{\bm{X}}_{ij} = \sum_{k=1}^{N_{ij}} \bm{X}_{ijk} / N_{ij}\)), cluster-period size, and potential treatment-by-covariate interactions. Because each cluster now contains at most $J$ cluster-period summary outcomes, we can at most consider a working random effect \(\alpha_i^{*} \sim \mathcal{N}(0, \sigma_{\alpha^*}^2)\) that accounts for between-cluster variability. In principle, assuming that the outcome data are generated from a linear mixed model at the individual-level, Web Appendix A of \citet{li2022marginal} have shown that the residual error \(\epsilon_{ijk}\) contributes a heteroskedastic error term with variance \(\sigma_\epsilon^2 / N_{ij}\). However, a convenient modeling practice is to continue assuming \(\gamma_{ij}^{*} \sim \mathcal{N}(0, \sigma_{\gamma^*}^2)\), so that standard model fitting routine for linear mixed models can be used. In fact, even if such a cluster-period-level linear mixed model potentially misspecifies the true individual-level data generating process when cluster-period sizes vary, (and could potentially yield biased estimates of individual-level ICC parameters \citep{li2022clarifying}) our model-robost standardization procedure can still lead to consistent estimator for the target estimands when such a working outcome model is considered to predict the cluster-period-level mean potential outcomes. That is, the implied conditional mean from model \eqref{lmm_cp}, which serves as the augmentation term in the proposed estimator, is given by
\[
\widehat{m}_{zj}(\overline{\bm{X}}_{ij}, N_{ij}) = \widehat{\mathbb{E}}[\overline{Y}_{ij} \mid Z_{ij} = z, \overline{\bm{X}}_{ij}, N_{ij}] =  \widehat{\beta}_j^{*} + z \widehat{\tau}_j^{*} + \widehat{g}^{*}(\overline{\bm{X}}_{ij}, N_{ij}, z),
\]
where \(\widehat{g}^{*}(\cdot)\) is the estimated contribution of cluster-period-level summary covariates and cluster size to the expected cluster-period mean outcome. 

Parameters in cluster-period-level linear mixed models can be estimated via maximum likelihood or restricted maximum likelihood using standard R packages, including \texttt{lme4}, \texttt{nlme}, \texttt{lmerTest}, and \texttt{glmmTMB}. Both \texttt{lme4} and \texttt{nlme} support modeling nested exchangeable correlation structures via random intercepts, while \texttt{lmerTest} extends \texttt{lme4} can also fit the model with nested exchangeable correlation structure. In all cases, the data must first be aggregated to the cluster-period level, where the outcome is the cluster-period mean and covariates are summarized (e.g., using averages of individual-level covariates). These summary datasets are then used as inputs to model-fitting functions (e.g., \texttt{lmer()}) with appropriate fixed and random effect specifications.

\subsection{Marginal model with cluster-period-level observations}\label{example5}
As a final example, one could also consider GEE approaches based on cluster-period summary data. While individual-level GEE provides a flexible and asymptotically consistent framework, its application can be computationally intensive in large-scale SW-CRTs due to the size of the estimating equations and repeated inversion of large working correlation matrices. To address these limitations, \citet{li2022marginal} recently proposed a cluster-period GEE formulation, in which the outcome is aggregated to the cluster-period level. Let \(\mu_{ij}  = \mathbb{E}[\overline{Y}_{ij} \mid \overline{\bm{X}}_{ij}, Z_{ij} ,N_{ij} ]\) and the cluster-period marginal model can be written as
\[
\eta\{\mu_{ij} \} =  \alpha^{*}_j + Z_{ij}\nu^{*}_j + g^{*}(\overline{\bm{X}}_{ij}, N_{ij},Z_{ij}),
\]
where \(g^{*}(.)\) captures the contribution of cluster-period level covariates, cluster size, and possible interactions.
In the generalized estimation equation, a working covariance matrix need to be specified for the vector of cluster-period means \(\overline{\bm{Y}}_i = ( \overline{Y}_{i1},\dots, \overline{Y}_{iJ} )\) to account for both within period and between period correlation for each pair of outcomes. \citet{li2022marginal} introduced two covariance structures: nested exchangeable and exponential decay structures. In both structures, the diagonal elements of the covariance matrix take the common form, which accounts for variance of cluster-period means, while the off-diagonal elements correspond to a first-order auto-regressive structure.  The cluster-period GEE can largely reduce computational burden and improve scalability for large trials. \citet{li2022marginal} showed that in cross-sectional SW-CRTs, both individual-level and cluster-period GEE estimators are asymptotically equivalent, and the realized efficiency largely depends on the accuracy of the working correlation specification. The conditional mean can be computed as 
\[
\widehat{m}_{zj}(\overline{\bm{X}}_{ij}, N_{ij}) = \widehat{\mathbb{E}}[\overline{Y}_{ij} \mid Z_{ij} = z, \overline{\bm{X}}_{ij},N_{ij}] = \eta^{-1} \left\{ \widehat{\alpha}^{*}_j + z\widehat{\nu}^{*}_j +  \widehat{g}^{*}(\overline{\bm{X}}_{ij}, N_{ij}, z)\right\}.
\]
This method can be implemented through \texttt{geeCRT} package in \texttt{R} (currently only available for binary outcomes).

\begin{remark}\label{rmk:parallel}
Consider a parallel-arm CRT with periods \(j\in\{1,\dots,J\}\) and a cluster-level treatment \(A_i\in\{0,1\}\) that is constant over periods, so that \(A_{ij}=A_i\) for all \(j\). Following the cluster-period mean defined in (2) of main paper, a general weighted parallel-arm estimand can be written as
\[ \tau^{\mathrm{par}}_{\omega} = f\bigl\{\mu^{\mathrm{par}}_{\omega}(1),\,\mu^{\mathrm{par}}_{\omega}(0)\bigr\}, \qquad \mu^{\mathrm{par}}_{\omega}(a) = \frac{\mathbb{E}\bigl[\sum_{j=1}^{J}\omega_{ij}\,\overline Y_{ij}(a)\bigr]}{\mathbb{E}\bigl[\sum_{j=1}^{J}\omega_{ij}\bigr]}.
\]
When \(\omega_{ijk} = 1\), this leads to an individual-average estimand, and when \(\omega_{ijk}=1/N_{ij}\), this gives a cluster-period estimand. The augmented model-robust standardization estimator to the weighted parallel-arm estimand is
\[ \widehat\mu^{\mathrm{par}}_{\omega,j}(a) = \frac{\sum_{i=1}^{I}\omega_{ij}\,m_{aj}(X_{ij},N_{ij})}{\sum_{i=1}^{I}\omega_{ij}} + \frac{\sum_{i=1}^{I}\omega_{ij}\,\mathbb{I}(A_i=a)\,\{\overline Y_{ij}-m_{aj}(X_{ij},N_{ij})\}} {\sum_{i=1}^{I}\omega_{ij}\,\mathbb{I}(A_i=a)},
\]
and aggregating over periods gives
\[
\widehat\mu^{\mathrm{par}}_{\omega}(a) = \frac{\sum_{j=1}^{J}\bigl(\sum_{i=1}^{I}\omega_{ij}\bigr)\,\widehat\mu^{\mathrm{par}}_{\omega,j}(a)} {\sum_{j=1}^{J}\sum_{i=1}^{I}\omega_{ij}},
\qquad \widehat\tau^{\mathrm{par}}_{\omega,\mathrm{aug}} = f\bigl\{\widehat\mu^{\mathrm{par}}_{\omega}(1),\,\widehat\mu^{\mathrm{par}}_{\omega}(0)\bigr\}.
\]
\end{remark}

\section{Proof of the equivalence between the proposed estimator and the ANCOVA estimator} \label{chen_eq_proof}

In this section, we show that the proposed augmented estimator is a generalization of the ANCOVA estimator introduced by \citet{chen2024model}. For illustration, we focus on the ANCOVA-I estimator; the arguments for other ANCOVA variants follow analogously.

We consider the ANCOVA model defined by \citet{chen2024model} as 
\[Y_{ijk} = \beta_{j} + \tau_{j} Z_{ij} + \bm{\widetilde{X}}_{ijk} \gamma + \epsilon_{ijk}, \]
where \(\widetilde{\bm{X}}_{ijk} = \bm{X}_{ijk} -\overline{\bm{X}}_j^{\omega}  \) denotes the period-mean centered baseline covariate vector, \(\overline{\bm{X}}^{\omega}_j = \sum_{i=1}^{I}\omega_{ij}\overline{\bm{X}}_{ij}^{\omega} /\sum_{i=1}^{I} \omega_{ij} \) and \(\overline{\bm{X}}_{ij}^{\omega} = \sum_{k=1}^{N_{ij}}\omega_{ijk} \bm{X}_{ijk} /\sum_{k=1}^{N_{ij}}\omega_{ijk} \).  As shown in Proposition 1 by \citet{chen2024model}, the estimator for causal effect can be written as

\[\widehat{\tau}_{\omega,j}^{\text{ancova}} = \frac{1}{\sum_{i=1}^{I}\mathbb{I}(Z_{ij}=1) \omega_{ij} } \left( \overline{Y}_{ij}(1) - \widetilde{\bm{X}}_{ij}^{\omega}\widehat{\bm{\gamma}} \right) - \frac{1}{\sum_{i=1}^{I}\mathbb{I}(Z_{ij}=0) \omega_{ij} } \left( \overline{Y}_{ij}(0) - \widetilde{\bm{X}}_{ij}^{\omega}\widehat{\bm{\gamma}} \right). \]
The model-robust estimator is 
\begin{align*}
    \widehat{\tau}_{\omega,j}^{\text{aug}} = & \frac{\sum_{i=1}^{I} \omega_{ij} \mathbb{I}(Z_{ij} =1) \{\overline{Y}_{ij} - \widehat{m}_{1j} (\underline{\bm{X}}_{ij},N_{ij})  \}  }{\sum_{i=1}^{I}\mathbb{I}(Z_{ij}=1) \omega_{ij} } - \frac{\sum_{i=1}^{I} \omega_{ij} \mathbb{I}(Z_{ij} =0) \{\overline{Y}_{ij} - \widehat{m}_{1j} (\underline{\bm{X}}_{ij},N_{ij})  \}  }{\sum_{i=1}^{I}\mathbb{I}(Z_{ij}=0) \omega_{ij} }  \nonumber\\
    & + \frac{\sum_{i=1}^{I }\omega_{ij} \widehat{m}_{1j}(\underline{\bm{X}}_{ij},N_{ij})  }{\sum_{i=1}^{I} \omega_{ij}} - \frac{\sum_{i=1}^{I }\omega_{ij} \widehat{m}_{0j}(\underline{\bm{X}}_{ij},N_{ij})  }{\sum_{i=1}^{I} \omega_{ij}}. 
\end{align*}

By setting \(m_{zj}(\underline{\bm{X}}_{ij},N_{ij}) = \widetilde{\bm{X}}_{ij}^{\omega}\bm{\gamma}\), 
we have

\begin{align*}
    \widehat{\tau}_{\omega,j}^{\text{aug}} = & \frac{\sum_{i=1}^{I} \omega_{ij} \mathbb{I}(Z_{ij} =1) \{\overline{Y}_{ij} - \widetilde{\bm{X}}_{ij}^{\omega}\widehat{\bm{\gamma}} \}  }{\sum_{i=1}^{I}\mathbb{I}(Z_{ij}=1) \omega_{ij} } - \frac{\sum_{i=1}^{I} \omega_{ij} \mathbb{I}(Z_{ij} =0) \{\overline{Y}_{ij} - \widetilde{\bm{X}}_{ij}^{\omega}\widehat{\bm{\gamma}} \}  }{\sum_{i=1}^{I}\mathbb{I}(Z_{ij}=0) \omega_{ij} }  \nonumber\\
    & + \frac{\sum_{i=1}^{I }\omega_{ij} \widetilde{\bm{X}}_{ij}^{\omega}\widehat{\bm{\gamma}}  }{\sum_{i=1}^{I} \omega_{ij}} - \frac{\sum_{i=1}^{I }\omega_{ij}\widetilde{\bm{X}}_{ij}^{\omega}\widehat{\bm{\gamma}}  }{\sum_{i=1}^{I} \omega_{ij}} \\
    & = \frac{\overline{Y}_{ij}(1) - \widetilde{\bm{X}}_{ij}^{\omega}\widehat{\bm{\gamma}}}{\sum_{i=1}^{I}\mathbb{I}(Z_{ij}=1) \omega_{ij} } - \frac{ \overline{Y}_{ij}(0) - \widetilde{\bm{X}}_{ij}^{\omega}\widehat{\bm{\gamma}}}{\sum_{i=1}^{I}\mathbb{I}(Z_{ij}=0) \omega_{ij} } \\
    & = \widehat{\tau}_{\omega,j}^{\text{ancova}}.
\end{align*}

\section{Hypothesis testing procedures for informative size} \label{ics_test}

\subsection{Comparing horizontal individual-average and cluster-average estimands}\label{sec:test1}
The first test compares the individual-level and cluster-level estimands at the horizontal direction. The null hypothesis is defined as 
\begin{equation} \label{h-test}
    \mathcal{H}_0: \ \tau_{\text{I}}^{\text{h}} = \tau_{\text{C}}^{\text{h}},
\end{equation}
which tests that the ATE remains the same whether calculated by pooling all individuals equally or by averaging cluster-specific means across time. The test statistics can be defined as a contrast between $\psi(\widehat{\tau}_{\text{I}}^{\text{h}})$ and $\psi(\widehat{\tau}_{\text{C}}^{\text{h}})$. Here, we consider a possible monotonic transformation $\psi$ to stabilize the test statistic. Since both \(\tau_{\text{I}}^{\text{h}}\) and \(\tau_{\text{C}}^{\text{h}}\) are constructed using the contrast function \(f(a,b)\), the choice of the transformation \(\psi\) depends of the effect scale. Under the difference scale \(f(a, b) = a - b\), we typically set $\psi(.)$ to be the identity function. But under the ratio scale where \(f(a, b) = a / b\) (risk ratio) or \(f(a, b) = (a(1-b))/((1-a)b)\) (odds ratio), we often set $\psi(.)$ to be the log function to obtain a linear structure on the log scale; see \citet{ZhouRJ-2022-011} for examples of constructing such statistics for observational studies. For simplicity of exposition and better illustrating insights, we focus on the difference scale in the following development, but generalization to ratio estimands is straightforward. Interestingly, under the difference scale, the null hypothesis is algebraically equivalent to
\[
\mathbb{E} \left[ \left( \sum_{j=2}^{J-1} \sum_{k=1}^{N_{ij}} \{ Y_{ijk}(1) - Y_{ijk}(0) \} \right) \left( 1 - \frac{\mathbb{E} \left( \sum_{j=2}^{J-1} N_{ij} \right)}{\sum_{j=2}^{J-1} N_{ij}} \right) \right] = 0,
\]
and after some simplification, can be written as
\[
\text{Cov} \left( \sum_{j=2}^{J-1} N_{ij}, \ \frac{\sum_{j=2}^{J-1} \sum_{k=1}^{N_{ij}} \{ Y_{ijk}(1) - Y_{ijk}(0) \} }{\sum_{j=2}^{J-1} N_{ij}} \right) = 0.
\]
The details of derivation are provided below.
\begin{proof}
    For the null hypothesis of \( \mathcal{H}_0: \tau_{\text{I}}^{\text{h}} = \tau_{\text{C}}^{\text{h}} \), under different scale it is equivalent to
\begin{align*}
    & \frac{\sum_{j=2}^{J-1} \mathbb{E}\left[\sum_{k=1}^{N_{ij}} \{Y_{ijk}(1) - Y_{ijk}(0) \}\right] }{ \sum_{j=2}^{J-1} \mathbb{E}(N_{ij}) } - \mathbb{E}\left[ \frac{\sum_{j=2}^{J-1} \sum_{k=1}^{N_{ij}} \{ Y_{ijk}(1) - Y_{ijk}(0) \} }{\sum_{j=2}^{J-1}N_{ij} } \right] = 0 \\
    & \Rightarrow \mathbb{E} \left[ \left\{ \frac{1}{ \sum_{j=2}^{J-1} \mathbb{E}(N_{ij})} - \frac{1}{\sum_{j=2}^{J-1}N_{ij} } \right\} \sum_{j=2}^{J-1} \sum_{k=1}^{N_{ij}} \left\{ Y_{ijk}(1) - Y_{ijk}(0) \right\} \right] = 0 \\
    & \Rightarrow \mathbb{E}\left[ \left\{ 1- \frac{ \sum_{j=2}^{J-1} \mathbb{E}(N_{ij})}{\sum_{j=2}^{J-1}N_{ij} } \right\} \sum_{j=2}^{J-1} \sum_{k=1}^{N_{ij}} \left\{ Y_{ijk}(1) - Y_{ijk}(0) \right\} \right] = 0 \\
    & \Rightarrow \mathbb{E} \left[\left\{ \frac{ \sum_{j=2}^{J-1} \sum_{k=1}^{N_{ij}} \left\{ Y_{ijk}(1) - Y_{ijk}(0)  \right\}}{\sum_{j=2}^{J-1}N_{ij}} \right\}\sum_{j=2}^{J-1}N_{ij}   \right] - \mathbb{E} \left[ \frac{ \sum_{j=2}^{J-1} \sum_{k=1}^{N_{ij}} \left\{ Y_{ijk}(1) - Y_{ijk}(0)  \right\}}{\sum_{j=2}^{J-1}N_{ij}} \right] \mathbb{E} \left[ \sum_{j=2}^{J-1}N_{ij}\right] = 0 \\
    & \Rightarrow \text{Cov} \left[\sum_{j=2}^{J-1}N_{ij} , \frac{ \sum_{j=2}^{J-1} \sum_{k=1}^{N_{ij}} \left\{ Y_{ijk}(1) - Y_{ijk}(0)  \right\}}{\sum_{j=2}^{J-1}N_{ij}}  \right] = 0.
\end{align*}
\end{proof}
That is, the test aims to assess, within the roll-out periods, whether the cluster size (total cluster-period size from period $2$ to $J-1$) is correlated with the horizontally aggregated cluster-specific treatment effect. Rejection of this null hypothesis suggests that different cluster sizes could marginally associate with different treatment effects, implying that estimands that treat individuals and clusters differently may not agree due to underlying cluster size heterogeneity. 
The variance of the test statistic can be estimated using the Jackknife variance estimator as discussed in Section \ref{sec:var_jack}. Specifically, let  
\[
D_i = \psi(\widehat{\tau}_{\text{I}, -i}^{\text{h}}) - \psi(\widehat{\tau}_{\text{C}, -i}^{\text{h}})
\]  
denote the contrast-based statistic computed with the \(i\)th cluster removed, where \(\widehat{\tau}_{\text{I}, -i}^{\text{h}}\) and \(\widehat{\tau}_{\text{C}, -i}^{\text{h}}\) are the individual- and cluster-level horizontal estimands calculated after omitting cluster \(i\). The Jackknife estimate of the variance is then given by
\[
\widehat{V}_{D}= \frac{I - 1}{I} \sum_{i=1}^{I} \left( D_i - \overline{D} \right)^2,
\quad \text{where } \overline{D} = \frac{1}{I} \sum_{i=1}^I D_i.
\]
The test statistic is defined as
\(W = \left\{ \psi(\widehat{\tau}_{\text{I}}^{\text{h}}) - \psi(\widehat{\tau}_{\text{C}}^{\text{h}}) \right\}/\sqrt{ \widehat{V}_{D} }\),
which approximately follows a t-distribution considering finite sample size with \( I - 1 \) degrees of freedom under the null hypothesis. A two-sided Wald-type test rejects the null hypothesis at level \(\alpha\) if
\(|W| > t_{I-1, 1 - \alpha/2}\).

\subsection{Comparing vertical individual-average and cluster-average estimands} \label{sec:test2}

The second test contrasts individual-average and cluster-average treatment effect estimands based on vertical aggregations. The null hypothesis is
\begin{equation} \label{v-test}
    \mathcal{H}_0: \ \tau_{\text{I}}^{\text{v}} = \tau_{\text{C}}^{\text{v}},
\end{equation}
with the test statistic given by $\psi(\widehat{\tau}_{\text{I}}^{\text{v}}) - \psi(\widehat{\tau}_{\text{C}}^{\text{v}})  $. Under the difference scale with \(f(a, b) = a - b\), this null hypothesis is equivalent to
\[
\frac{1}{J-2}\sum_{j=2}^{J-1} \frac{1}{\mathbb{E}(N_{ij})} \mathbb{E} \left[ \sum_{k=1}^{N_{ij}} \{ Y_{ijk}(1) - Y_{ijk}(0) \} \left( 1 - \frac{\mathbb{E}(N_{ij})}{N_{ij}} \right) \right] = 0,
\]
which, after some algebraic manipulation, becomes
\[
\sum_{j=2}^{J-1}\frac{1}{\mathbb{E}(N_{ij})}\text{Cov} \left( N_{ij}, \ \frac{1}{N_{ij}} \sum_{k=1}^{N_{ij}} \{ Y_{ijk}(1) - Y_{ijk}(0) \} \right) = 0.
\]
The details of the derivation are provided below.
\begin{proof}
    For the hypothesis: \( \mathcal{H}_0: \tau_{\text{I}}^{\text{v}} = \tau_{\text{C}}^{\text{v}} \), under different scale it is equivalent to
\begin{align*}
    & \frac{1}{J-2} \sum_{j=2}^{J-1} \frac{\mathbb{E}\left[ \sum_{k=1}^{N_{ij}} \{Y_{ijk}(1) - Y_{ijk}(0) \}\right]}{\mathbb{E}(N_{ij})} - \frac{1}{J-2} \sum_{j=2}^{J-1} \mathbb{E}\left[ \frac{\sum_{k=1}^{N_{ij}} \{Y_{ijk}(1) - Y_{ijk}(0) \}}{N_{ij}}\right] = 0 \\
    & \Rightarrow \frac{1}{J-2} \sum_{j=2}^{J-1} \mathbb{E}\left[  \left\{ \sum_{k=1}^{N_{ijk}} \{ Y_{ijk}(1) - Y_{ijk}(0)  \}\right\} \left\{ \frac{1}{\mathbb{E}(N_{ij})}  - \frac{1}{N_{ij}}  \right\}  \right] = 0 \\
    & \Rightarrow \frac{1}{J-2} \sum_{j=2}^{J-1} \mathbb{E} \left[  \frac{\sum_{k=1}^{N_{ijk}} \{Y_{ijk}(1) - Y_{ijk}(0) \} }{N_{ij}} \frac{N_{ij}}{\mathbb{E}(N_{ij})} \right] - \mathbb{E} \left[  \frac{\sum_{k=1}^{N_{ijk}} \{ Y_{ijk}(1) - Y_{ijk}(0) \} }{N_{ij}}  \right] \mathbb{E} \left[\frac{N_{ij}}{\mathbb{E}(N_{ij})}\right] = 0 \\
    & \Rightarrow \sum_{j=2}^{J-1} \text{Cov} \left[\frac{N_{ij}}{\mathbb{E}(N_{ij})}, \frac{\sum_{k=1}^{N_{ijk}} \{Y_{ijk}(1) - Y_{ijk}(0) \} }{N_{ij}}  \right] = 0.
\end{align*}
\end{proof}
Because in SW-CRT, the expected cluster-period sizes \(\mathbb{E}(N_{ij})\) can vary across periods, the standardized cluster-period size \(N_{ij}/\mathbb{E}(N_{ij})\) remove the period-to-period difference. A closer examination of this testable implication says that such a test aims to assess whether, on average, the cluster-period size is marginally associated with the cluster-average treatment effect at each roll-out period. Since the testable implication is a summation of covariance terms, it differs in a subtle way from the testable implication in Section \ref{sec:test1}. 
Rejection of the null implies that cluster-period-specific treatment effects vary locally with the number of individuals observed within each cluster-period. The test statistic and testing procedure are analogous to those in Section \ref{sec:test1}.

\subsection{Comparing vertical cluster-average and horizontal cluster-average estimands} \label{sec:test3}
The third test contrasts two cluster-average treatment effect estimands based on vertical versus horizontal aggregation. The null hypothesis is
\begin{equation} \label{hv-ctest}
    \mathcal{H}_0: \ \tau_{\text{C}}^{\text{v}} = \tau_{\text{C}}^{\text{h}},
\end{equation}
with test statistic $\psi(\widehat{\tau}_{\text{C}}^{\text{v}}) - \psi(\widehat{\tau}_{\text{C}}^{\text{h}})$. Under the difference scale with $f(a,b) = a - b$, this null is equivalent to
\[
\mathbb{E}\left[\sum_{j=2}^{J-1}\left\{\frac{N_{ij}}{\sum_{s=2}^{J-1} N_{is}} - \frac{1}{J-2}\right\}\left\{\frac{\sum_{k=1}^{N_{ij}} \{Y_{ijk}(1) - Y_{ijk}(0)\}}{N_{ij}}\right\}\right] = 0.
\]
In the special case where the expected cluster-period sizes do not vary across roll-out periods, thus, \(\mathbb{E}\left\{\frac{N_{ij}}{\sum_{s=2}^{J-1} N_{is}}\right\}=\frac{1}{J-2}\) for all \(j \in \{2,\dots J-1\}\) the above test is algebraically equivalent
\[\sum_{j=2}^{J-1}\mathrm{Cov}\left(\frac{N_{ij}}{\sum_{s=2}^{J-1} N_{is}},\;\frac{\sum_{k=1}^{N_{ij}} \{Y_{ijk}(1) - Y_{ijk}(0)\}}{N_{ij}}\right) = 0,
\]
which implies a test for informative cluster-period size. It assesses whether, on average across the roll-out periods, the cluster-period size (or the share of a cluster’s total roll-out size allocated to a given period) is marginally associated with the cluster-period average treatment effect. In SW-CRTs, however, the expected cluster-period sizes $\mathbb{E}(N_{ij})$ may vary across periods. In such cases, the condition $\mathbb{E}\left(\frac{N_{ij}}{\sum_{s=2}^{J-1} N_{is}}\right) = \frac{1}{J-2}$ need not hold, and the equality of $\tau_{\text{C}}^{\text{v}}$ and $\tau_{\text{C}}^{\text{h}}$ reflects two intertwined features: (i) whether the cluster-period size is correlated with the cluster-period average treatment effect; and (ii) how the expected cluster-period sizes over calendar time are correlated with the cluster-period average treatment effect. In other words, when $\mathbb{E}(N_{ij})$ varies across $j$, the test no longer isolates a ``pure'' informative size effect. Rejection of $\mathcal{H}_0$ therefore indicates that either cluster-period average treatment effects vary with the cluster-period size, or the expected roll-out profile of cluster-period sizes is itself aligned with the pattern of treatment effect heterogeneity over time, or both. The test statistic and testing procedure are analogous to those in Section \ref{sec:test1}.

\subsection{Comparing vertical individual-average and horizontal individual-average estimands} 
\label{sec:test4}
The fourth test contrasts two individual-average treatment effect estimands based on vertical versus horizontal aggregation. The null hypothesis is
\begin{equation} \label{vh-itest}
    \mathcal{H}_0: \ \tau_{\text{I}}^{\text{v}} = \tau_{\text{I}}^{\text{h}},
\end{equation}
with test statistic $\psi(\widehat{\tau}_{\text{I}}^{\text{v}}) - \psi(\widehat{\tau}_{\text{I}}^{\text{h}})$. Under the difference scale with $f(a,b) = a - b$, and using the definitions of $\tau_{\text{I}}^{\text{v}}$ and $\tau_{\text{I}}^{\text{h}}$, the null hypothesis in \eqref{vh-itest} is equivalent to
\[\mathbb{E} \left[ \sum_{j=2}^{J-1} \left\{ \frac{1}{J-2} - \frac{\mathbb{E}(N_{ij})}{ \sum_{s=2}^{J-1} \mathbb{E}(N_{is}) } \right\} \frac{\sum_{k=1}^{N_{ij}} \{Y_{ijk}(1) - Y_{ijk}(0) \} }{\mathbb{E}(N_{ij})}\right] = 0.\]
This is equivalent to 
\[\mathrm{Cov}_j \left\{\mathbb{E}(N_{ij}),\,\frac{\mathbb{E}\left[\sum_{k=1}^{N_{ij}} \{Y_{ijk}(1) - Y_{ijk}(0)\} \right]}{\mathbb{E}(N_{ij})} \right\} = 0,
\]
where $\mathrm{Cov}_j(\cdot,\cdot)$ denotes covariance across periods $j = 2,\dots, J-1$ under the uniform distribution. This representation implies a test for informative period size (or, equivalently, informative expected cluster-period size over calendar time). If $\mathbb{E}(N_{ij})$ is constant over periods $j=2,\dots, J-1$, then $\mathbb{E}(N_{ij})$ has no variation across $j$ and the covariance above is identically zero. In this case, $\tau_{\text{I}}^{\text{v}} = \tau_{\text{I}}^{\text{h}}$ holds by construction. This behavior aligns with our simulation Scenarios C1-C2 and B1-B2, where the expected cluster-period sizes are constant across the roll-out and the vertical and horizontal individual-average estimands coincide. In contrast, when $\mathbb{E}(N_{ij})$ varies across periods $j=2,\dots,J-1$, the equality $\tau_{\text{I}}^{\text{v}} = \tau_{\text{I}}^{\text{h}}$ holds if and only if the temporal pattern of expected cluster-period sizes is uncorrelated with the individual-average treatment effects. This is related to our Scenario C3 and B3 in simulation studies. Rejection of $\mathcal{H}_0$ therefore indicates that periods contributing more observations in expectation tend to exhibit either larger or smaller individual-average causal effects than periods contributing fewer observations. Similarly, the test statistic and testing procedure are analogous to those in Section \ref{sec:test1}.

\section{Simulation results for continuous outcome in Scenario C1 and C2} \label{sim:cont_adj}

Table \ref{tab:sim_C1_results} summarizes the simulation results under Scenario \ref{sc:C1}, when there is no informative cluster size and all four estimands coincide. Across all working models, the proposed MRS estimator consistently gives nearly unbiased estimates, with nominal coverage.
The robustness of MRS holds regardless of the choice between GEE and linear mixed models, and does not depend on the specified random-effects structure. On the other hand, the Coef estimators also perform well in terms of bias and coverage. In the absence of informative cluster size, the Monte Carlo standard deviation of the Coef estimator tends to be smaller than that of the MRS estimator. Finally, the unadjusted estimator, although unbiased, is much less efficient due to ignoring baseline covariates. The model-assisted ANCOVA estimator is unbiased and offers improved efficiency over UNADJ, but remains less efficient than MRS estimators. 

Table \ref{tab:sim_C2_results} summarizes the results under Scenario \ref{sc:C2}. Even in the presence of informative cluster size, the MRS estimator continues to deliver unbiased estimates and nominal coverage regardless of working outcome model specification. In contrast, the performance of Coef estimators critically depends on the model specification. For individual-average estimands, the Coef estimators generally correspond to small bias and nominal coverage. The exceptions emerge under models (W5) and (W6), which include both cluster and cluster-period random effects, showing that a more complex random-effects structure can lead to even more biased inference under informative cluster size, a finding that has been previously identified in 2-period cluster randomized crossover designs \citep{lee2025estimated}. In comparison, the bias under simple exchangeable random-effects models becomes much smaller \citep{lee2025estimated}. The independence GEE models (W1) and (W2) tend to also perform well for i-ATE under informative cluster size. However, for cluster-average estimands, particularly in the vertical direction, the Coef estimators can lead to substantial bias and severe undercoverage, especially under working models (W1)-(W4). This is aligned with the findings in parallel CRT \citep{li2025model}. Interestingly, under working model (W5) and (W6), the Coef estimator gives improved empirically unbiased inference for v-cATE.  From Scenario \ref{sc:C1}, it is more likely that MRS leads to a smaller MCSD than Coef under informative cluster size. In this Scenario, the UNADJ and ANCOVA estimators also produce approximately unbiased estimates with coverage rates close to nominal across estimands. Again, although the ANCOVA estimator improves the efficiency over UNADJ, it still remains less efficient than the MRS estimators.

\begin{table}[ht!]
\centering
\caption{
Simulation results in Scenario C1 for estimating four estimands under a continuous outcome with $I=30$ clusters and $J=6$ periods using covariate-adjusted working models (W1)-(W6) with the Coef, MRS and ANCOVA estimators. MRS: proposed augmented estimator; Coef: treatment-effect coefficients from covariate-adjusted working model; ANCOVA: model-assisted ANCOVA estimator; UNADJ: nonparametric estimator. RBias (\%): absolute relative bias in percent; AESE: average estimated standard error; MCSD: Monte Carlo standard deviation; CP: 95\% confidence interval coverage.
}
\label{tab:sim_C1_results}
\begin{tabular}{cclrrrrrrrr}
\toprule
\multirow{2}{*}{Direction} & \multirow{2}{*}{Working model}
  & \multirow{2}{*}{Method}
  & \multicolumn{4}{c}{h-iATE}
  & \multicolumn{4}{c}{h-cATE} \\
\cmidrule(lr){4-7}\cmidrule(lr){8-11}
 & & 
 & RBias & AESE & MCSD & CP 
 & RBias & AESE & MCSD & CP \\
  & & 
 & \multicolumn{4}{c}{\(\tau_{\mathrm{I}}^{\mathrm{h}}=2.472\)} 
 & \multicolumn{4}{c}{\(\tau_{\mathrm{C}}^{\mathrm{h}}=2.472\)} \\
\midrule
\multirow{18}{*}{Horizontal}
  & \textbackslash & UNADJ 
    & 0.192 & 0.127 & 0.124 & 0.945 
    & 0.220 & 0.127 & 0.125 & 0.947 \\
  & \textbackslash & ANCOVA  
    & 0.167 & 0.107 & 0.101 & 0.964 
    & 0.170 & 0.106 & 0.101 & 0.958 \\
  & \multirow{2}{*}{W1}
    & Coef   
      & 0.204 & 0.092 & 0.094 & 0.947 
      & 0.204 & 0.092 & 0.094 & 0.947 \\
  &               & MRS    
      & 0.185 & 0.098 & 0.096 & 0.963 
      & 0.182 & 0.097 & 0.096 & 0.961 \\
  & \multirow{2}{*}{W2}
    & Coef   
      & 0.186 & 0.092 & 0.096 & 0.941 
      & 0.187 & 0.092 & 0.096 & 0.942 \\
  &               & MRS    
      & 0.186 & 0.098 & 0.096 & 0.963 
      & 0.183 & 0.097 & 0.096 & 0.961 \\
  & \multirow{2}{*}{W3}
    & Coef   
      & 0.096 & 0.080 & 0.076 & 0.956 
      & 0.096 & 0.080 & 0.076 & 0.956 \\
  &               & MRS    
      & 0.179 & 0.098 & 0.096 & 0.961 
      & 0.178 & 0.097 & 0.096 & 0.961 \\
  & \multirow{2}{*}{W4}
    & Coef   
      & 0.081 & 0.084 & 0.077 & 0.960 
      & 0.083 & 0.084 & 0.077 & 0.961 \\
  &               & MRS    
      & 0.180 & 0.098 & 0.096 & 0.963 
      & 0.179 & 0.097 & 0.096 & 0.960 \\
  & \multirow{2}{*}{W5}
    & Coef   
      & 0.113 & 0.076 & 0.072 & 0.953 
      & 0.112 & 0.076 & 0.072 & 0.953 \\
  &               & MRS    
      & 0.180 & 0.098 & 0.096 & 0.962 
      & 0.178 & 0.097 & 0.096 & 0.960 \\
  & \multirow{2}{*}{W6}
    & Coef   
      & 0.105 & 0.079 & 0.074 & 0.962 
      & 0.106 & 0.079 & 0.074 & 0.962 \\
  &               & MRS    
      & 0.181 & 0.098 & 0.096 & 0.963 
      & 0.179 & 0.097 & 0.096 & 0.960 \\
\toprule
& & 
  & \multicolumn{4}{c}{v-iATE}
  & \multicolumn{4}{c}{v-cATE} \\
  \cmidrule(lr){4-7}\cmidrule(lr){8-11}
 & & 
 & RBias & AESE & MCSD & CP 
 & RBias & AESE & MCSD & CP \\
  & & 
 & \multicolumn{4}{c}{\(\tau_{\mathrm{I}}^{\mathrm{v}}=2.472\)} 
 & \multicolumn{4}{c}{\(\tau_{\mathrm{C}}^{\mathrm{v}}=2.472\)} \\
 \midrule
\multirow{18}{*}{Vertical}
  & \textbackslash & UNADJ 
    & 0.192 & 0.128 & 0.124 & 0.943 
    & 0.204 & 0.130 & 0.126 & 0.954 \\
  & \textbackslash & ANCOVA  
    & 0.164 & 0.107 & 0.102 & 0.962 
    & 0.152 & 0.106 & 0.101 & 0.959 \\
  & \multirow{2}{*}{W1}
    & Coef   
      & 0.204 & 0.092 & 0.094 & 0.947 
      & 0.205 & 0.092 & 0.094 & 0.947 \\
  &               & MRS    
      & 0.182 & 0.098 & 0.097 & 0.960 
      & 0.151 & 0.096 & 0.095 & 0.953 \\
  & \multirow{2}{*}{W2}
    & Coef   
      & 0.183 & 0.092 & 0.097 & 0.942 
      & 0.184 & 0.092 & 0.097 & 0.942 \\
  &               & MRS    
      & 0.183 & 0.098 & 0.097 & 0.959 
      & 0.151 & 0.096 & 0.095 & 0.954 \\
  & \multirow{2}{*}{W3}
    & Coef   
      & 0.096 & 0.080 & 0.076 & 0.956 
      & 0.097 & 0.080 & 0.076 & 0.956 \\
  &               & MRS    
      & 0.176 & 0.098 & 0.097 & 0.959 
      & 0.149 & 0.096 & 0.095 & 0.953 \\
  & \multirow{2}{*}{W4}
    & Coef   
      & 0.076 & 0.084 & 0.077 & 0.961 
      & 0.077 & 0.084 & 0.077 & 0.961 \\
  &               & MRS    
      & 0.178 & 0.098 & 0.097 & 0.958 
      & 0.149 & 0.096 & 0.095 & 0.953 \\
  & \multirow{2}{*}{W5}
    & Coef   
      & 0.113 & 0.076 & 0.072 & 0.953 
      & 0.113 & 0.076 & 0.072 & 0.953 \\
  &               & MRS    
      & 0.177 & 0.098 & 0.097 & 0.959 
      & 0.149 & 0.096 & 0.095 & 0.953 \\
  & \multirow{2}{*}{W6}
    & Coef   
      & 0.100 & 0.079 & 0.074 & 0.963 
      & 0.100 & 0.079 & 0.074 & 0.963 \\
  &               & MRS    
      & 0.178 & 0.098 & 0.097 & 0.959 
      & 0.149 & 0.096 & 0.095 & 0.953 \\
\bottomrule
\end{tabular}
\end{table}

\begin{table}[ht!]
\centering
\caption{
Simulation results in Scenario C2 for estimating four estimands under a continuous outcome with $I=30$ clusters and $J=6$ periods using covariate-adjusted working models (W1)-(W6) with the Coef, MRS and ANCOVA estimators. MRS: proposed augmented estimator; Coef: treatment-effect coefficients from covariate-adjusted working model; ANCOVA: model-assisted ANCOVA estimator; UNADJ: nonparametric estimator. RBias (\%): absolute relative bias in percent; AESE: average estimated standard error; MCSD: Monte Carlo standard deviation; CP: 95\% confidence interval coverage.
}
\label{tab:sim_C2_results}
\begin{tabular}{cclrrrrrrrr}
\toprule
\multirow{2}{*}{Direction} & \multirow{2}{*}{Working model}
  & \multirow{2}{*}{Method}
  & \multicolumn{4}{c}{h-iATE}
  & \multicolumn{4}{c}{h-cATE} \\
\cmidrule(lr){4-7}\cmidrule(lr){8-11}
 & & 
 & RBias & AESE & MCSD & CP 
 & RBias & AESE & MCSD & CP \\
  & & 
 & \multicolumn{4}{c}{\(\tau_{\mathrm{I}}^{\mathrm{h}}=8.135\)} 
 & \multicolumn{4}{c}{\(\tau_{\mathrm{C}}^{\mathrm{h}}=7.617\)} \\
\midrule
\multirow{18}{*}{Horizontal}
  & \textbackslash & UNADJ 
    & 1.186 & 0.921 & 0.897 & 0.930 
    & 0.717 & 0.923 & 0.905 & 0.935 \\
  & \textbackslash & ANCOVA  
    & 1.190 & 0.939 & 0.894 & 0.931 
    & 0.729 & 0.940 & 0.902 & 0.937 \\
  & \multirow{2}{*}{W1}
    & Coef   
      & 0.926 & 0.622 & 0.629 & 0.939 
      & 5.802 & 0.622 & 0.629 & 0.891 \\
  &               & MRS    
      & 1.396 & 0.691 & 0.652 & 0.944 
      & 0.833 & 0.693 & 0.657 & 0.952 \\
  & \multirow{2}{*}{W2}
    & Coef   
      & 1.419 & 0.640 & 0.654 & 0.927 
      & 5.259 & 0.639 & 0.654 & 0.894 \\
  &               & MRS    
      & 1.419 & 0.693 & 0.654 & 0.945 
      & 0.856 & 0.695 & 0.658 & 0.954 \\
  & \multirow{2}{*}{W3}
    & Coef   
      & 4.595 & 0.838 & 0.801 & 0.933 
      & 1.884 & 0.838 & 0.801 & 0.953 \\
  &               & MRS    
      & 1.422 & 0.692 & 0.654 & 0.943 
      & 0.895 & 0.694 & 0.659 & 0.954 \\
  & \multirow{2}{*}{W4}
    & Coef   
      & 4.961 & 0.881 & 0.822 & 0.936 
      & 1.484 & 0.881 & 0.822 & 0.957 \\
  &               & MRS    
      & 1.445 & 0.694 & 0.656 & 0.943 
      & 0.919 & 0.696 & 0.660 & 0.953 \\
  & \multirow{2}{*}{W5}
    & Coef   
      & 25.598 & 0.615 & 0.593 & 0.090 
      & 20.544 & 0.615 & 0.593 & 0.329 \\
  &               & MRS    
      & 1.413 & 0.692 & 0.655 & 0.946 
      & 0.865 & 0.694 & 0.659 & 0.952 \\
  & \multirow{2}{*}{W6}
    & Coef   
      & 24.999 & 0.651 & 0.617 & 0.145 
      & 19.900 & 0.650 & 0.617 & 0.422 \\
  &               & MRS    
      & 1.422 & 0.694 & 0.656 & 0.945 
      & 0.873 & 0.696 & 0.660 & 0.952 \\
\toprule
& & 
  & \multicolumn{4}{c}{v-iATE}
  & \multicolumn{4}{c}{v-cATE} \\
  \cmidrule(lr){4-7}\cmidrule(lr){8-11}
 & & 
 & RBias & AESE & MCSD & CP 
 & RBias & AESE & MCSD & CP \\
  & & 
 & \multicolumn{4}{c}{\(\tau_{\mathrm{I}}^{\mathrm{v}}=8.134\)} 
 & \multicolumn{4}{c}{\(\tau_{\mathrm{C}}^{\mathrm{v}}=6.011\)} \\
 \midrule
\multirow{18}{*}{Vertical}
  & \textbackslash & UNADJ 
    & 1.784 & 0.927 & 0.899 & 0.927 
    & 0.457 & 0.852 & 0.840 & 0.939 \\
  & \textbackslash & ANCOVA  
    & 1.789 & 0.942 & 0.896 & 0.928 
    & 0.448 & 0.864 & 0.840 & 0.939 \\
  & \multirow{2}{*}{W1}
    & Coef   
      & 0.918 & 0.622 & 0.629 & 0.939 
      & 34.072 & 0.622 & 0.629 & 0.121 \\
  &               & MRS    
      & 1.983 & 0.695 & 0.653 & 0.939 
      & 0.477 & 0.641 & 0.610 & 0.951 \\
  & \multirow{2}{*}{W2}
    & Coef   
      & 2.010 & 0.642 & 0.655 & 0.924 
      & 32.595 & 0.642 & 0.655 & 0.172 \\
  &               & MRS    
      & 2.010 & 0.698 & 0.655 & 0.939 
      & 0.451 & 0.643 & 0.611 & 0.950 \\
  & \multirow{2}{*}{W3}
    & Coef   
      & 4.587 & 0.838 & 0.801 & 0.933 
      & 29.107 & 0.838 & 0.801 & 0.456 \\
  &               & MRS    
      & 2.009 & 0.696 & 0.655 & 0.937 
      & 0.444 & 0.641 & 0.612 & 0.950 \\
  & \multirow{2}{*}{W4}
    & Coef   
      & 5.564 & 0.882 & 0.822 & 0.926 
      & 27.785 & 0.882 & 0.822 & 0.535 \\
  &               & MRS    
      & 2.037 & 0.699 & 0.657 & 0.937 
      & 0.418 & 0.643 & 0.613 & 0.950 \\
  & \multirow{2}{*}{W5}
    & Coef   
      & 25.591 & 0.615 & 0.593 & 0.090 
      & 0.686 & 0.615 & 0.593 & 0.950 \\
  &               & MRS    
      & 2.001 & 0.697 & 0.656 & 0.940 
      & 0.449 & 0.638 & 0.609 & 0.946 \\
  & \multirow{2}{*}{W6}
    & Coef   
      & 25.629 & 0.648 & 0.614 & 0.121 
      & 0.634 & 0.648 & 0.614 & 0.949 \\
  &               & MRS    
      & 2.013 & 0.699 & 0.657 & 0.938 
      & 0.443 & 0.640 & 0.610 & 0.946 \\
\bottomrule
\end{tabular}
\end{table}


\clearpage

\section{Simulation results for continuous outcome with non-covariate-adjusted model} \label{sim:cont_unadj}

Tables \ref{tab:sim_C1_unadj}-\ref{tab:sim_C3_unadj} present the results for the Coef estimators under Scenarios C1 to C3 using six working models without covariate adjustment. Similar to the adjusted models, when informative cluster size exists, models (W1)-(W4) give less biased estimates for individual-average estimands but exhibit substantial bias for cluster-average estimands, particularly for v-cATE. In contrast, estimators from models (W5)-(W6) are biased for individual-average estimands but show minimal bias for v-iATE. Overall, the efficiency of the unadjusted models is generally lower than that of the adjusted models.

\begin{table}[ht!]
\centering
\caption{
Simulation results in Scenario C1 for estimating four estimands under a continuous outcome with $I=30$ clusters and $J=6$ periods using non-covariate-adjusted working models (W1)-(W6) with the Coef estimators. RBias (\%): absolute relative bias in percent; AESE: average estimated standard error; MCSD: Monte Carlo standard deviation; CP: 95\% confidence interval coverage.
}

\label{tab:sim_C1_unadj}
\begin{tabular}{cclrrrrrrrr}
\toprule
\multirow{2}{*}{Direction} & \multirow{2}{*}{Working model}
  & \multicolumn{4}{c}{h-iATE}
  & \multicolumn{4}{c}{h-cATE} \\
\cmidrule(lr){3-6}\cmidrule(lr){7-10}
 & & RBias & AESE & MCSD & CP 
     & RBias & AESE & MCSD & CP \\
  & & \multicolumn{4}{c}{\(\tau_{\text{I}}^{\text{h}}=2.472\)}
        & \multicolumn{4}{c}{\(\tau_{\text{C}}^{\text{h}}=2.472\)} \\
\midrule
\multirow{6}{*}{Horizontal}
  & UNADJ-W1 & 0.201 & 0.119 & 0.121 & 0.946 
             & 0.201 & 0.119 & 0.121 & 0.946 \\
  & UNADJ-W2 & 0.192 & 0.121 & 0.124 & 0.937 
             & 0.194 & 0.121 & 0.124 & 0.937 \\
  & UNADJ-W3 & 0.110 & 0.114 & 0.111 & 0.968 
             & 0.110 & 0.114 & 0.111 & 0.968 \\
  & UNADJ-W4 & 0.099 & 0.120 & 0.114 & 0.964 
             & 0.101 & 0.120 & 0.114 & 0.962 \\
  & UNADJ-W5 & 0.126 & 0.113 & 0.111 & 0.959 
             & 0.126 & 0.113 & 0.111 & 0.959 \\
  & UNADJ-W6 & 0.113 & 0.119 & 0.114 & 0.963 
             & 0.115 & 0.119 & 0.114 & 0.964 \\
\toprule
& 
  & \multicolumn{4}{c}{v-iATE}
  & \multicolumn{4}{c}{v-cATE} \\
  \cmidrule(lr){3-6}\cmidrule(lr){7-10}
 & 
 & RBias & AESE & MCSD & CP 
 & RBias & AESE & MCSD & CP \\
 
  & & \multicolumn{4}{c}{\(\tau_{\text{I}}^{\text{v}}=2.472\)}
        & \multicolumn{4}{c}{\(\tau_{\text{C}}^{\text{v}}=2.472\)} \\
        \midrule
\multirow{6}{*}{Vertical}

  & UNADJ-W1 & 0.201 & 0.119 & 0.121 & 0.946 
             & 0.201 & 0.119 & 0.121 & 0.946 \\
  & UNADJ-W2 & 0.192 & 0.121 & 0.124 & 0.937 
             & 0.193 & 0.121 & 0.124 & 0.937 \\
  & UNADJ-W3 & 0.110 & 0.114 & 0.111 & 0.968 
             & 0.110 & 0.114 & 0.111 & 0.968 \\
  & UNADJ-W4 & 0.098 & 0.120 & 0.114 & 0.963 
             & 0.098 & 0.120 & 0.114 & 0.963 \\
  & UNADJ-W5 & 0.126 & 0.113 & 0.111 & 0.959 
             & 0.126 & 0.113 & 0.111 & 0.959 \\
  & UNADJ-W6 & 0.113 & 0.119 & 0.113 & 0.962 
             & 0.113 & 0.119 & 0.113 & 0.962 \\
\bottomrule
\end{tabular}
\end{table}

\begin{table}[ht!]
\centering
\caption{
Simulation results in Scenario C2 for estimating four estimands under a continuous outcome with $I=30$ clusters and $J=6$ periods using non-covariate-adjusted working models (W1)-(W6) with the Coef estimators. RBias (\%): absolute relative bias in percent; AESE: average estimated standard error; MCSD: Monte Carlo standard deviation; CP: 95\% confidence interval coverage.
}
\label{tab:sim_C2_unadj}
\begin{tabular}{cclrrrrrrrr}
\toprule
\multirow{2}{*}{Direction} & \multirow{2}{*}{Working model}
  & \multicolumn{4}{c}{h-iATE}
  & \multicolumn{4}{c}{h-cATE} \\
\cmidrule(lr){3-6}\cmidrule(lr){7-10}
 & & RBias & AESE & MCSD & CP 
     & RBias & AESE & MCSD & CP \\
  & & \multicolumn{4}{c}{\(\tau_{\text{I}}^{\text{h}}=8.135\)}
        & \multicolumn{4}{c}{\(\tau_{\text{C}}^{\text{h}}=7.617\)} \\
\midrule
\multirow{6}{*}{Horizontal}
  & UNADJ-W1 & 0.743 & 0.842 & 0.870 & 0.935 
             & 5.998 & 0.842 & 0.870 & 0.907 \\
  & UNADJ-W2 & 1.186 & 0.862 & 0.897 & 0.921 
             & 5.502 & 0.861 & 0.897 & 0.899 \\
  & UNADJ-W3 & 4.452 & 1.138 & 1.114 & 0.944 
             & 2.038 & 1.138 & 1.114 & 0.955 \\
  & UNADJ-W4 & 4.764 & 1.187 & 1.130 & 0.939 
             & 1.690 & 1.186 & 1.130 & 0.958 \\
  & UNADJ-W5 & 25.485 & 0.826 & 0.829 & 0.347 
             & 20.424 & 0.826 & 0.829 & 0.558 \\
  & UNADJ-W6 & 24.879 & 0.866 & 0.850 & 0.415 
             & 19.779 & 0.864 & 0.849 & 0.617 \\
\toprule
& 
  & \multicolumn{4}{c}{v-iATE}
  & \multicolumn{4}{c}{v-cATE} \\
  \cmidrule(lr){3-6}\cmidrule(lr){7-10}
 & 
 & RBias & AESE & MCSD & CP 
 & RBias & AESE & MCSD & CP \\
    & & \multicolumn{4}{c}{\(\tau_{\text{I}}^{\text{v}}=8.134\)}
        & \multicolumn{4}{c}{\(\tau_{\text{C}}^{\text{v}}=6.011\)} \\
        \midrule
\multirow{6}{*}{Vertical}

  & UNADJ-W1 & 0.735 & 0.842 & 0.870 & 0.935 
             & 34.321 & 0.842 & 0.870 & 0.339 \\
  & UNADJ-W2 & 1.784 & 0.864 & 0.899 & 0.914 
             & 32.901 & 0.864 & 0.899 & 0.409 \\
  & UNADJ-W3 & 4.444 & 1.138 & 1.114 & 0.944 
             & 29.302 & 1.138 & 1.114 & 0.684 \\
  & UNADJ-W4 & 5.381 & 1.188 & 1.129 & 0.934 
             & 28.034 & 1.188 & 1.129 & 0.730 \\
  & UNADJ-W5 & 25.479 & 0.826 & 0.829 & 0.347 
             & 0.838 & 0.826 & 0.829 & 0.941 \\
  & UNADJ-W6 & 25.526 & 0.862 & 0.846 & 0.392 
             & 0.774 & 0.862 & 0.846 & 0.937 \\
\bottomrule
\end{tabular}
\end{table}

\begin{table}[ht!]
\centering
\caption{
Simulation results in Scenario C3 for estimating four estimands under a continuous outcome with $I=30$ clusters and $J=6$ periods using non-covariate-adjusted working models (W1)-(W6) with the Coef estimators. RBias (\%): absolute relative bias in percent; AESE: average estimated standard error; MCSD: Monte Carlo standard deviation; CP: 95\% confidence interval coverage.
}
\label{tab:sim_C3_unadj}
\begin{tabular}{cclrrrrrrrr}
\toprule
\multirow{2}{*}{Direction} & \multirow{2}{*}{Working model}
  & \multicolumn{4}{c}{h-iATE}
  & \multicolumn{4}{c}{h-cATE} \\
\cmidrule(lr){3-6}\cmidrule(lr){7-10}
 & & RBias & AESE & MCSD & CP 
     & RBias & AESE & MCSD & CP \\
  & & \multicolumn{4}{c}{\(\tau_{\text{I}}^{\text{h}}=8.395\)}
        & \multicolumn{4}{c}{\(\tau_{\text{C}}^{\text{h}}=7.857\)} \\
\midrule
\multirow{6}{*}{Horizontal}
  & UNADJ-W1 & 0.641 & 1.671 & 1.728 & 0.936 
             & 4.903 & 1.671 & 1.728 & 0.904 \\
  & UNADJ-W2 & 1.007 & 1.710 & 1.784 & 0.916 
             & 4.498 & 1.708 & 1.783 & 0.907 \\
  & UNADJ-W3 & 3.717 & 2.273 & 2.227 & 0.947 
             & 1.655 & 2.273 & 2.227 & 0.955 \\
  & UNADJ-W4 & 3.973 & 2.369 & 2.260 & 0.939 
             & 1.373 & 2.368 & 2.260 & 0.956 \\
  & UNADJ-W5 & 21.360 & 1.636 & 1.647 & 0.334 
             & 16.972 & 1.636 & 1.647 & 0.550 \\
  & UNADJ-W6 & 20.839 & 1.715 & 1.691 & 0.401 
             & 16.423 & 1.712 & 1.690 & 0.609 \\
\toprule
& 
  & \multicolumn{4}{c}{v-iATE}
  & \multicolumn{4}{c}{v-cATE} \\
  \cmidrule(lr){3-6}\cmidrule(lr){7-10}
 & 
 & RBias & AESE & MCSD & CP 
 & RBias & AESE & MCSD & CP \\
   & & \multicolumn{4}{c}{\(\tau_{\text{I}}^{\text{v}}=8.821\)}
        & \multicolumn{4}{c}{\(\tau_{\text{C}}^{\text{v}}=6.707\)} \\
        \midrule
\multirow{6}{*}{Vertical}

  & UNADJ-W1 & 0.634 & 1.671 & 1.728 & 0.936 
             & 26.880 & 1.671 & 1.728 & 0.331 \\
  & UNADJ-W2 & 1.503 & 1.715 & 1.787 & 0.914 
             & 25.770 & 1.715 & 1.787 & 0.398 \\
  & UNADJ-W3 & 3.711 & 2.273 & 2.227 & 0.947 
             & 22.951 & 2.273 & 2.227 & 0.687 \\
  & UNADJ-W4 & 4.485 & 2.372 & 2.257 & 0.933 
             & 21.962 & 2.372 & 2.257 & 0.730 \\
  & UNADJ-W5 & 21.354 & 1.636 & 1.647 & 0.334 
             & 0.422 & 1.636 & 1.647 & 0.942 \\
  & UNADJ-W6 & 21.375 & 1.707 & 1.683 & 0.370 
             & 0.395 & 1.707 & 1.683 & 0.933 \\
\bottomrule
\end{tabular}
\end{table}


\section{Simulation results for continuous outcome in Scenario C1-C3 with smaller number of clusters \(I=10\)}
\label{cont_small}

In the one-cluster-per-sequence SW-CRT, each cluster is assigned a unique rollout period at which it switches from control ($Z_{ij}=0$) to treatment ($Z_{ij}=1$) and remains on treatment thereafter. In this setting, the leave-one-cluster-out jackknife procedure described in Section \ref{sec:var_jack} requires a minor modification. For some clusters $g$ and some rollout periods $j \in \{2,\dots,J-1\}$, cluster $g$ may be the only treated cluster (immediately after its switch time) or the only control cluster (just before its switch time). After removing such a cluster $g$, the leave-one-cluster-out dataset based on clusters $i \neq g$ can therefore contain periods in which all remaining clusters are treated or all remaining clusters are control. In these periods the period-specific treatment effect estimators are not identifiable, because one of the arm-specific weights satisfies $\sum_{i \neq g} \omega_{ij} \mathbb{I}(Z_{ij}=z) = 0$, and the associated design matrix in the working outcome regression with period-by-treatment effects is rank deficient; as a result, the fitted values $\widehat{m}_{zj}^{-g}(\underline{\bm{X}}_{ij},N_{ij})$ needed to construct $\widehat{\mu}_{\omega,j}^{-g}(z)$ are not well-defined. To ensure that each leave-one-cluster-out estimator $\widehat{\mu}_\omega^{-g}(z)$ is estimable, we therefore implement the jackknife by, for each deleted cluster $g$, refitting the same outcome regression model used in the full-sample analysis on the reduced dataset of clusters $i \neq g$ and restricting to rollout periods $j \in \{2,\dots,J-1\}$ that continue to contain at least one treated and one control cluster after removing cluster $g$, that is, periods satisfying
\[
\sum_{i \neq g} \omega_{ij} \mathbb{I}(Z_{ij}=1) > 0 
\quad \text{and} \quad
\sum_{i \neq g} \omega_{ij} \mathbb{I}(Z_{ij}=0) > 0.
\]
For these periods we compute $\widehat{m}_{zj}^{-g}(\underline{\bm{X}}_{ij},N_{ij})$ and define 
\[\widehat{\mu}_{\omega,j}^{-g}(z)=\frac{\sum_{i \neq g} \omega_{ij} \widehat{m}_{zj}^{-g}(\underline{\bm{X}}_{ij},N_{ij})}{\omega_j^{-g}}+\frac{\sum_{i \neq g} \omega_{ij} \mathbb{I}(Z_{ij}=z)\bigl\{\overline{Y}_{ij} - \widehat{m}_{zj}^{-g}(\underline{\bm{X}}_{ij},N_{ij})\bigr\}}{\sum_{i \neq g} \omega_{ij} \mathbb{I}(Z_{ij}=z)},
\]
with $\omega_j^{-g} = \sum_{i \neq g} \omega_{ij}$. We obtain the corresponding leave-one-cluster-out estimator of $\mu_\omega(z)$ by averaging only over these informative periods,
\[\widehat{\mu}_\omega^{-g}(z)=\left(\sum_{j \in \mathcal{J}_g} \omega_j^{-g}\right)^{-1}\sum_{j \in \mathcal{J}_g} \omega_j^{-g} \widehat{\mu}_{\omega,j}^{-g}(z),
\]
where $\mathcal{J}_g \subset \{2,\dots,J-1\}$ denotes the subset of periods that satisfy the above ``both arms present'' condition after deleting cluster $g$. We then compute the jackknife covariance matrix $\widehat{\Sigma}$ using the same formula as in Section \ref{sec:var_jack}, and construct $t$-based inference for $\tau_\omega$ with $I-1$ degrees of freedom.

Tables \ref{tab:sim_C1_small}--\ref{tab:sim_C3_small} report the continuous-outcome results for Scenarios C1-C3 with $I = 10$ clusters and $J = 6$ periods. Across all three scenarios, all four estimands, and all working models, the proposed MRS estimator continues to exhibit small bias (RBias generally below 8\%) and coverage probabilities mostly in the 0.91-0.95\%, while achieving appreciable efficiency gains. The Monte Carlo standard deviations are about 20-25\% smaller than those of the UNADJ estimator in each scenario. In contrast, the Coef estimators display marked sensitivity to the working model, with substantial bias and severe undercoverage for some GLMM specifications (especially for cluster-average estimands in Scenarios C2--C3), whereas UNADJ and ANCOVA maintain small bias but show more pronounced undercoverage and larger variability than MRS. Tables \ref{tab:sim_C1_small_seq}-\ref{tab:sim_C3_small_seq} show a very similar pattern for the case where each sequence includes only one cluster switch.

\begin{table}[ht!]
\centering
\caption{
Simulation results in Scenario C1 for estimating four estimands under a continuous outcome with $I=10$ clusters and $J=6$ periods using covariate-adjusted working models (W1)-(W6) with the Coef, MRS and ANCOVA estimators. MRS: proposed augmented estimator; Coef: treatment-effect coefficients from covariate-adjusted working model; ANCOVA: model-assisted ANCOVA estimator; UNADJ: nonparametric estimator. RBias (\%): absolute relative bias in percent; AESE: average estimated standard error; MCSD: Monte Carlo standard deviation; CP: 95\% confidence interval coverage.
}
\label{tab:sim_C1_small}
\begin{tabular}{cclrrrrrrrr}
\toprule
\multirow{2}{*}{Direction} & \multirow{2}{*}{Working model}
  & \multirow{2}{*}{Method}
  & \multicolumn{4}{c}{h-iATE}
  & \multicolumn{4}{c}{h-cATE} \\
\cmidrule(lr){4-7}\cmidrule(lr){8-11}
 & & 
 & RBias & AESE & MCSD & CP 
 & RBias & AESE & MCSD & CP \\
  & & 
 & \multicolumn{4}{c}{\(\tau_{\mathrm{I}}^{\mathrm{h}}=2.472\)} 
 & \multicolumn{4}{c}{\(\tau_{\mathrm{C}}^{\mathrm{h}}=2.472\)} \\
\midrule
\multirow{18}{*}{Horizontal}
  & \textbackslash & UNADJ 
    & 0.064 & 0.242 & 0.216 & 0.960 
    & 0.099 & 0.241 & 0.218 & 0.955 \\
  & \textbackslash & ANCOVA  
    & 0.148 & 0.214 & 0.172 & 0.953 
    & 0.151 & 0.213 & 0.172 & 0.952 \\
  & \multirow{2}{*}{W1}
    & Coef   
      & 0.071 & 0.154 & 0.163 & 0.924 
      & 0.071 & 0.154 & 0.163 & 0.924 \\
  &               & MRS    
      & 0.092 & 0.184 & 0.165 & 0.945 
      & 0.093 & 0.183 & 0.165 & 0.943 \\
  & \multirow{2}{*}{W2}
    & Coef   
      & 0.096 & 0.149 & 0.165 & 0.910 
      & 0.096 & 0.149 & 0.165 & 0.910 \\
  &               & MRS    
      & 0.096 & 0.185 & 0.165 & 0.946 
      & 0.097 & 0.184 & 0.165 & 0.942 \\
  & \multirow{2}{*}{W3}
    & Coef   
      & 0.083 & 0.146 & 0.137 & 0.967 
      & 0.083 & 0.146 & 0.137 & 0.967 \\
  &               & MRS    
      & 0.096 & 0.183 & 0.165 & 0.943 
      & 0.101 & 0.182 & 0.165 & 0.942 \\
  & \multirow{2}{*}{W4}
    & Coef   
      & 0.095 & 0.164 & 0.140 & 0.974 
      & 0.094 & 0.164 & 0.140 & 0.976 \\
  &               & MRS    
      & 0.098 & 0.184 & 0.165 & 0.944 
      & 0.104 & 0.183 & 0.165 & 0.941 \\
  & \multirow{2}{*}{W5}
    & Coef   
      & 0.073 & 0.138 & 0.132 & 0.962 
      & 0.073 & 0.138 & 0.132 & 0.962 \\
  &               & MRS    
      & 0.100 & 0.183 & 0.165 & 0.945 
      & 0.106 & 0.182 & 0.165 & 0.942 \\
  & \multirow{2}{*}{W6}
    & Coef   
      & 0.085 & 0.156 & 0.136 & 0.960 
      & 0.085 & 0.156 & 0.135 & 0.959 \\
  &               & MRS    
      & 0.103 & 0.184 & 0.165 & 0.945 
      & 0.110 & 0.183 & 0.165 & 0.941 \\
\toprule
& & 
  & \multicolumn{4}{c}{v-iATE}
  & \multicolumn{4}{c}{v-cATE} \\
  \cmidrule(lr){4-7}\cmidrule(lr){8-11}
 & & 
 & RBias & AESE & MCSD & CP 
 & RBias & AESE & MCSD & CP \\
  & & 
 & \multicolumn{4}{c}{\(\tau_{\mathrm{I}}^{\mathrm{v}}=2.472\)} 
 & \multicolumn{4}{c}{\(\tau_{\mathrm{C}}^{\mathrm{v}}=2.472\)} \\
 \midrule
\multirow{18}{*}{Vertical}
  & \textbackslash & UNADJ 
    & 0.094 & 0.244 & 0.217 & 0.961 
    & 0.070 & 0.241 & 0.223 & 0.952 \\
  & \textbackslash & ANCOVA  
    & 0.184 & 0.214 & 0.172 & 0.950 
    & 0.148 & 0.207 & 0.171 & 0.946 \\
  & \multirow{2}{*}{W1}
    & Coef   
      & 0.071 & 0.154 & 0.163 & 0.924 
      & 0.072 & 0.154 & 0.163 & 0.924 \\
  &               & MRS    
      & 0.124 & 0.185 & 0.164 & 0.947 
      & 0.089 & 0.180 & 0.163 & 0.945 \\
  & \multirow{2}{*}{W2}
    & Coef   
      & 0.128 & 0.149 & 0.165 & 0.907 
      & 0.128 & 0.149 & 0.165 & 0.907 \\
  &               & MRS    
      & 0.128 & 0.186 & 0.165 & 0.947 
      & 0.093 & 0.181 & 0.163 & 0.943 \\
  & \multirow{2}{*}{W3}
    & Coef   
      & 0.083 & 0.146 & 0.137 & 0.967 
      & 0.083 & 0.146 & 0.137 & 0.967 \\
  &               & MRS    
      & 0.128 & 0.183 & 0.164 & 0.944 
      & 0.093 & 0.179 & 0.163 & 0.946 \\
  & \multirow{2}{*}{W4}
    & Coef   
      & 0.119 & 0.165 & 0.139 & 0.979 
      & 0.120 & 0.165 & 0.139 & 0.979 \\
  &               & MRS    
      & 0.130 & 0.184 & 0.165 & 0.944 
      & 0.095 & 0.180 & 0.163 & 0.942 \\
  & \multirow{2}{*}{W5}
    & Coef   
      & 0.073 & 0.138 & 0.132 & 0.962 
      & 0.074 & 0.138 & 0.132 & 0.962 \\
  &               & MRS    
      & 0.132 & 0.184 & 0.164 & 0.946 
      & 0.098 & 0.179 & 0.163 & 0.945 \\
  & \multirow{2}{*}{W6}
    & Coef   
      & 0.112 & 0.156 & 0.135 & 0.962 
      & 0.112 & 0.156 & 0.135 & 0.962 \\
  &               & MRS    
      & 0.135 & 0.185 & 0.165 & 0.944 
      & 0.101 & 0.180 & 0.163 & 0.941 \\
\bottomrule
\end{tabular}
\end{table}

\begin{table}[ht!]
\centering
\caption{
Simulation results in Scenario C2 for estimating four estimands under a continuous outcome with $I=10$ clusters and $J=6$ periods using covariate-adjusted working models (W1)-(W6) with the Coef, MRS and ANCOVA estimators. MRS: proposed augmented estimator; Coef: treatment-effect coefficients from covariate-adjusted working model; ANCOVA: model-assisted ANCOVA estimator; UNADJ: nonparametric estimator. RBias (\%): absolute relative bias in percent; AESE: average estimated standard error; MCSD: Monte Carlo standard deviation; CP: 95\% confidence interval coverage.
}
\label{tab:sim_C2_small}
\begin{tabular}{cclrrrrrrrr}
\toprule
\multirow{2}{*}{Direction} & \multirow{2}{*}{Working model}
  & \multirow{2}{*}{Method}
  & \multicolumn{4}{c}{h-iATE}
  & \multicolumn{4}{c}{h-cATE} \\
\cmidrule(lr){4-7}\cmidrule(lr){8-11}
 & & 
 & RBias & AESE & MCSD & CP 
 & RBias & AESE & MCSD & CP \\
  & & 
 & \multicolumn{4}{c}{\(\tau_{\mathrm{I}}^{\mathrm{h}}=8.135\)} 
 & \multicolumn{4}{c}{\(\tau_{\mathrm{C}}^{\mathrm{h}}=7.617\)} \\
\midrule
\multirow{18}{*}{Horizontal}
  & \textbackslash & UNADJ 
    & 4.329 & 1.688 & 1.551 & 0.904 
    & 2.647 & 1.676 & 1.550 & 0.917 \\
  & \textbackslash & ANCOVA  
    & 4.300 & 1.814 & 1.540 & 0.912 
    & 2.625 & 1.789 & 1.540 & 0.928 \\
  & \multirow{2}{*}{W1}
    & Coef   
      & 3.714 & 1.080 & 1.144 & 0.919 
      & 2.826 & 1.080 & 1.144 & 0.926 \\
  &               & MRS    
      & 5.157 & 1.378 & 1.171 & 0.925 
      & 3.598 & 1.370 & 1.177 & 0.931 \\
  & \multirow{2}{*}{W2}
    & Coef   
      & 5.270 & 1.049 & 1.171 & 0.888 
      & 1.133 & 1.044 & 1.172 & 0.909 \\
  &               & MRS    
      & 5.270 & 1.396 & 1.171 & 0.927 
      & 3.703 & 1.387 & 1.178 & 0.930 \\
  & \multirow{2}{*}{W3}
    & Coef   
      & 7.270 & 1.616 & 1.412 & 0.957 
      & 0.972 & 1.616 & 1.412 & 0.973 \\
  &               & MRS    
      & 5.171 & 1.377 & 1.172 & 0.923 
      & 3.686 & 1.368 & 1.179 & 0.930 \\
  & \multirow{2}{*}{W4}
    & Coef   
      & 8.345 & 1.847 & 1.440 & 0.959 
      & 2.132 & 1.844 & 1.442 & 0.978 \\
  &               & MRS    
      & 5.277 & 1.400 & 1.173 & 0.930 
      & 3.785 & 1.391 & 1.179 & 0.933 \\
  & \multirow{2}{*}{W5}
    & Coef   
      & 26.050 & 1.178 & 1.057 & 0.644 
      & 21.028 & 1.178 & 1.057 & 0.771 \\
  &               & MRS    
      & 5.191 & 1.365 & 1.172 & 0.920 
      & 3.634 & 1.357 & 1.178 & 0.929 \\
  & \multirow{2}{*}{W6}
    & Coef   
      & 23.988 & 1.363 & 1.110 & 0.750 
      & 18.796 & 1.356 & 1.112 & 0.840 \\
  &               & MRS    
      & 5.224 & 1.382 & 1.173 & 0.923 
      & 3.651 & 1.373 & 1.179 & 0.929 \\
\toprule
& & 
  & \multicolumn{4}{c}{v-iATE}
  & \multicolumn{4}{c}{v-cATE} \\
  \cmidrule(lr){4-7}\cmidrule(lr){8-11}
 & & 
 & RBias & AESE & MCSD & CP 
 & RBias & AESE & MCSD & CP \\
  & & 
 & \multicolumn{4}{c}{\(\tau_{\mathrm{I}}^{\mathrm{v}}=8.134\)} 
 & \multicolumn{4}{c}{\(\tau_{\mathrm{C}}^{\mathrm{v}}=6.011\)} \\
 \midrule
\multirow{18}{*}{Vertical}
  & \textbackslash & UNADJ 
    & 6.386 & 1.694 & 1.551 & 0.889 
    & 0.514 & 1.546 & 1.433 & 0.914 \\
  & \textbackslash & ANCOVA  
    & 6.355 & 1.807 & 1.541 & 0.901 
    & 0.557 & 1.624 & 1.424 & 0.919 \\
  & \multirow{2}{*}{W1}
    & Coef   
      & 3.706 & 1.080 & 1.144 & 0.919 
      & 30.300 & 1.080 & 1.144 & 0.674 \\
  &               & MRS    
      & 7.183 & 1.378 & 1.167 & 0.915 
      & 0.137 & 1.282 & 1.071 & 0.954 \\
  & \multirow{2}{*}{W2}
    & Coef   
      & 7.337 & 1.042 & 1.169 & 0.876 
      & 25.387 & 1.042 & 1.169 & 0.752 \\
  &               & MRS    
      & 7.337 & 1.397 & 1.169 & 0.915 
      & 0.261 & 1.301 & 1.073 & 0.951 \\
  & \multirow{2}{*}{W3}
    & Coef   
      & 7.263 & 1.616 & 1.412 & 0.957 
      & 25.487 & 1.616 & 1.412 & 0.896 \\
  &               & MRS    
      & 7.194 & 1.377 & 1.168 & 0.912 
      & 0.136 & 1.281 & 1.072 & 0.951 \\
  & \multirow{2}{*}{W4}
    & Coef   
      & 10.410 & 1.836 & 1.433 & 0.948 
      & 21.228 & 1.836 & 1.433 & 0.954 \\
  &               & MRS    
      & 7.345 & 1.401 & 1.169 & 0.917 
      & 0.248 & 1.306 & 1.074 & 0.950 \\
  & \multirow{2}{*}{W5}
    & Coef   
      & 26.044 & 1.178 & 1.057 & 0.644 
      & 0.073 & 1.178 & 1.057 & 0.959 \\
  &               & MRS    
      & 7.218 & 1.365 & 1.168 & 0.909 
      & 0.160 & 1.268 & 1.069 & 0.950 \\
  & \multirow{2}{*}{W6}
    & Coef   
      & 26.103 & 1.342 & 1.085 & 0.703 
      & 0.007 & 1.342 & 1.085 & 0.966 \\
  &               & MRS    
      & 7.288 & 1.383 & 1.170 & 0.909 
      & 0.161 & 1.287 & 1.071 & 0.953 \\
\bottomrule
\end{tabular}
\end{table}

\begin{table}[ht!]
\centering
\caption{
Simulation results in Scenario C3 for estimating four estimands under a continuous outcome with $I=10$ clusters and $J=6$ periods using covariate-adjusted working models (W1)-(W6) with the Coef, MRS and ANCOVA estimators. MRS: proposed augmented estimator; Coef: treatment-effect coefficients from covariate-adjusted working model; ANCOVA: model-assisted ANCOVA estimator; UNADJ: nonparametric estimator. RBias (\%): absolute relative bias in percent; AESE: average estimated standard error; MCSD: Monte Carlo standard deviation; CP: 95\% confidence interval coverage.
}
\label{tab:sim_C3_small}
\begin{tabular}{cclrrrrrrrr}
\toprule
\multirow{2}{*}{Direction} & \multirow{2}{*}{Working model}
  & \multirow{2}{*}{Method}
  & \multicolumn{4}{c}{h-iATE}
  & \multicolumn{4}{c}{h-cATE} \\
\cmidrule(lr){4-7}\cmidrule(lr){8-11}
 & & 
 & RBias & AESE & MCSD & CP 
 & RBias & AESE & MCSD & CP \\
  & & 
 & \multicolumn{4}{c}{\(\tau_{\mathrm{I}}^{\mathrm{h}}=8.395\)} 
 & \multicolumn{4}{c}{\(\tau_{\mathrm{C}}^{\mathrm{h}}=7.857\)} \\
\midrule
\multirow{18}{*}{Horizontal}
  & \textbackslash & UNADJ 
    & 3.606 & 3.346 & 3.088 & 0.903 
    & 2.188 & 3.322 & 3.087 & 0.910 \\
  & \textbackslash & ANCOVA  
    & 3.592 & 3.600 & 3.076 & 0.908 
    & 2.177 & 3.551 & 3.076 & 0.923 \\
  & \multirow{2}{*}{W1}
    & Coef   
      & 3.100 & 2.141 & 2.278 & 0.921 
      & 2.307 & 2.141 & 2.278 & 0.923 \\
  &               & MRS    
      & 4.301 & 2.734 & 2.330 & 0.921 
      & 2.973 & 2.718 & 2.342 & 0.933 \\
  & \multirow{2}{*}{W2}
    & Coef   
      & 4.395 & 2.080 & 2.331 & 0.885 
      & 0.914 & 2.070 & 2.333 & 0.909 \\
  &               & MRS    
      & 4.395 & 2.771 & 2.331 & 0.921 
      & 3.059 & 2.753 & 2.343 & 0.930 \\
  & \multirow{2}{*}{W3}
    & Coef   
      & 6.057 & 3.224 & 2.820 & 0.958 
      & 0.815 & 3.224 & 2.820 & 0.970 \\
  &               & MRS    
      & 4.313 & 2.732 & 2.332 & 0.921 
      & 3.047 & 2.715 & 2.346 & 0.928 \\
  & \multirow{2}{*}{W4}
    & Coef   
      & 6.947 & 3.685 & 2.874 & 0.962 
      & 1.765 & 3.677 & 2.878 & 0.976 \\
  &               & MRS    
      & 4.401 & 2.780 & 2.334 & 0.927 
      & 3.128 & 2.761 & 2.347 & 0.931 \\
  & \multirow{2}{*}{W5}
    & Coef   
      & 21.732 & 2.335 & 2.103 & 0.635 
      & 17.364 & 2.335 & 2.103 & 0.761 \\
  &               & MRS    
      & 4.329 & 2.709 & 2.333 & 0.920 
      & 3.003 & 2.692 & 2.343 & 0.931 \\
  & \multirow{2}{*}{W6}
    & Coef   
      & 20.004 & 2.702 & 2.208 & 0.743 
      & 15.516 & 2.689 & 2.213 & 0.829 \\
  &               & MRS    
      & 4.356 & 2.743 & 2.335 & 0.919 
      & 3.016 & 2.725 & 2.345 & 0.932 \\
\toprule
& & 
  & \multicolumn{4}{c}{v-iATE}
  & \multicolumn{4}{c}{v-cATE} \\
  \cmidrule(lr){4-7}\cmidrule(lr){8-11}
 & & 
 & RBias & AESE & MCSD & CP 
 & RBias & AESE & MCSD & CP \\
  & & 
 & \multicolumn{4}{c}{\(\tau_{\mathrm{I}}^{\mathrm{v}}=8.821\)} 
 & \multicolumn{4}{c}{\(\tau_{\mathrm{C}}^{\mathrm{v}}=6.707\)} \\
 \midrule
\multirow{18}{*}{Vertical}
  & \textbackslash & UNADJ 
    & 5.320 & 3.356 & 3.087 & 0.879 
    & 0.387 & 3.060 & 2.852 & 0.912 \\
  & \textbackslash & ANCOVA  
    & 5.305 & 3.588 & 3.076 & 0.891 
    & 0.413 & 3.222 & 2.843 & 0.921 \\
  & \multirow{2}{*}{W1}
    & Coef   
      & 3.093 & 2.141 & 2.278 & 0.922 
      & 23.739 & 2.141 & 2.278 & 0.670 \\
  &               & MRS    
      & 5.989 & 2.734 & 2.322 & 0.912 
      & 0.129 & 2.542 & 2.131 & 0.948 \\
  & \multirow{2}{*}{W2}
    & Coef   
      & 6.117 & 2.066 & 2.325 & 0.877 
      & 19.878 & 2.066 & 2.325 & 0.754 \\
  &               & MRS    
      & 6.117 & 2.772 & 2.325 & 0.914 
      & 0.226 & 2.581 & 2.133 & 0.949 \\
  & \multirow{2}{*}{W3}
    & Coef   
      & 6.050 & 3.224 & 2.820 & 0.958 
      & 19.964 & 3.224 & 2.820 & 0.899 \\
  &               & MRS    
      & 5.999 & 2.732 & 2.324 & 0.910 
      & 0.129 & 2.540 & 2.133 & 0.947 \\
  & \multirow{2}{*}{W4}
    & Coef   
      & 8.667 & 3.662 & 2.860 & 0.948 
      & 16.622 & 3.662 & 2.860 & 0.955 \\
  &               & MRS    
      & 6.124 & 2.780 & 2.327 & 0.916 
      & 0.216 & 2.591 & 2.135 & 0.951 \\
  & \multirow{2}{*}{W5}
    & Coef   
      & 21.726 & 2.335 & 2.103 & 0.635 
      & 0.053 & 2.335 & 2.103 & 0.958 \\
  &               & MRS    
      & 6.019 & 2.709 & 2.324 & 0.909 
      & 0.148 & 2.515 & 2.126 & 0.948 \\
  & \multirow{2}{*}{W6}
    & Coef   
      & 21.765 & 2.660 & 2.157 & 0.700 
      & 0.102 & 2.660 & 2.157 & 0.960 \\
  &               & MRS    
      & 6.076 & 2.743 & 2.327 & 0.911 
      & 0.149 & 2.552 & 2.129 & 0.948 \\
\bottomrule
\end{tabular}
\end{table}

\begin{table}[ht!]
\centering
\caption{
Simulation results in Scenario C1 for estimating four estimands under a continuous outcome with $I=10$ clusters and $J=11$ periods (switching one cluster per sequence) using covariate-adjusted working models (W1)-(W6) with the Coef, MRS and ANCOVA estimators. MRS: proposed augmented estimator; Coef: treatment-effect coefficients from covariate-adjusted working model; ANCOVA: model-assisted ANCOVA estimator; UNADJ: nonparametric estimator. RBias (\%): absolute relative bias in percent; AESE: average estimated standard error; MCSD: Monte Carlo standard deviation; CP: 95\% confidence interval coverage.
}
\label{tab:sim_C1_small_seq}
\begin{tabular}{cclrrrrrrrr}
\toprule
\multirow{2}{*}{Direction} & \multirow{2}{*}{Working model}
  & \multirow{2}{*}{Method}
  & \multicolumn{4}{c}{h-iATE}
  & \multicolumn{4}{c}{h-cATE} \\
\cmidrule(lr){4-7}\cmidrule(lr){8-11}
 & & 
 & RBias & AESE & MCSD & CP 
 & RBias & AESE & MCSD & CP \\
  & & 
 & \multicolumn{4}{c}{\(\tau_{\mathrm{I}}^{\mathrm{h}}=2.472\)} 
 & \multicolumn{4}{c}{\(\tau_{\mathrm{C}}^{\mathrm{h}}=2.472\)} \\
\midrule
\multirow{18}{*}{Horizontal}
  & \textbackslash & UNADJ 
    & 0.296 & 0.265 & 0.244 & 0.941 
    & 0.287 & 0.266 & 0.243 & 0.946 \\
  & \textbackslash & ANCOVA  
    & 0.129 & 0.123 & 0.189 & 0.824 
    & 0.137 & 0.123 & 0.188 & 0.824 \\
  & \multirow{2}{*}{W1}
    & Coef   
      & 0.145 & 0.139 & 0.154 & 0.931 
      & 0.145 & 0.139 & 0.154 & 0.931 \\
  &               & MRS    
      & 0.148 & 0.176 & 0.167 & 0.956 
      & 0.151 & 0.176 & 0.166 & 0.951 \\
  & \multirow{2}{*}{W2}
    & Coef   
      & 0.148 & 0.112 & 0.167 & 0.836 
      & 0.148 & 0.112 & 0.167 & 0.836 \\
  &               & MRS    
      & 0.148 & 0.177 & 0.167 & 0.955 
      & 0.151 & 0.177 & 0.166 & 0.952 \\
  & \multirow{2}{*}{W3}
    & Coef   
      & 0.064 & 0.114 & 0.123 & 0.940 
      & 0.063 & 0.114 & 0.123 & 0.940 \\
  &               & MRS    
      & 0.150 & 0.176 & 0.167 & 0.954 
      & 0.155 & 0.176 & 0.166 & 0.953 \\
  & \multirow{2}{*}{W4}
    & Coef   
      & 0.067 & 0.104 & 0.140 & 0.880 
      & 0.066 & 0.104 & 0.140 & 0.880 \\
  &               & MRS    
      & 0.149 & 0.177 & 0.168 & 0.957 
      & 0.154 & 0.176 & 0.166 & 0.952 \\
  & \multirow{2}{*}{W5}
    & Coef   
      & 0.038 & 0.111 & 0.122 & 0.937 
      & 0.037 & 0.111 & 0.122 & 0.937 \\
  &               & MRS    
      & 0.150 & 0.176 & 0.167 & 0.956 
      & 0.155 & 0.176 & 0.166 & 0.953 \\
  & \multirow{2}{*}{W6}
    & Coef   
      & 0.050 & 0.101 & 0.139 & 0.868 
      & 0.049 & 0.101 & 0.139 & 0.868 \\
  &               & MRS    
      & 0.148 & 0.177 & 0.168 & 0.958 
      & 0.153 & 0.176 & 0.166 & 0.952 \\
\toprule
& & 
  & \multicolumn{4}{c}{v-iATE}
  & \multicolumn{4}{c}{v-cATE} \\
  \cmidrule(lr){4-7}\cmidrule(lr){8-11}
 & & 
 & RBias & AESE & MCSD & CP 
 & RBias & AESE & MCSD & CP \\
  & & 
 & \multicolumn{4}{c}{\(\tau_{\mathrm{I}}^{\mathrm{v}}=2.472\)} 
 & \multicolumn{4}{c}{\(\tau_{\mathrm{C}}^{\mathrm{v}}=2.472\)} \\
 \midrule
\multirow{18}{*}{Vertical}
  & \textbackslash & UNADJ 
    & 0.314 & 0.266 & 0.245 & 0.942 
    & 0.220 & 0.269 & 0.253 & 0.944 \\
  & \textbackslash & ANCOVA  
    & 0.138 & 0.123 & 0.190 & 0.821 
    & 0.171 & 0.125 & 0.189 & 0.828 \\
  & \multirow{2}{*}{W1}
    & Coef   
      & 0.145 & 0.139 & 0.154 & 0.931 
      & 0.145 & 0.139 & 0.154 & 0.931 \\
  &               & MRS    
      & 0.150 & 0.176 & 0.167 & 0.953 
      & 0.177 & 0.175 & 0.167 & 0.953 \\
  & \multirow{2}{*}{W2}
    & Coef   
      & 0.151 & 0.112 & 0.168 & 0.830 
      & 0.151 & 0.112 & 0.168 & 0.830 \\
  &               & MRS    
      & 0.151 & 0.177 & 0.168 & 0.953 
      & 0.175 & 0.176 & 0.167 & 0.950 \\
  & \multirow{2}{*}{W3}
    & Coef   
      & 0.064 & 0.114 & 0.123 & 0.940 
      & 0.064 & 0.114 & 0.123 & 0.940 \\
  &               & MRS    
      & 0.154 & 0.176 & 0.168 & 0.952 
      & 0.180 & 0.175 & 0.167 & 0.947 \\
  & \multirow{2}{*}{W4}
    & Coef   
      & 0.081 & 0.104 & 0.140 & 0.879 
      & 0.081 & 0.104 & 0.140 & 0.879 \\
  &               & MRS    
      & 0.153 & 0.177 & 0.168 & 0.951 
      & 0.177 & 0.175 & 0.167 & 0.950 \\
  & \multirow{2}{*}{W5}
    & Coef   
      & 0.038 & 0.111 & 0.122 & 0.937 
      & 0.038 & 0.111 & 0.122 & 0.937 \\
  &               & MRS    
      & 0.154 & 0.176 & 0.168 & 0.952 
      & 0.180 & 0.175 & 0.167 & 0.948 \\
  & \multirow{2}{*}{W6}
    & Coef   
      & 0.064 & 0.101 & 0.139 & 0.866 
      & 0.064 & 0.101 & 0.139 & 0.866 \\
  &               & MRS    
      & 0.152 & 0.177 & 0.168 & 0.952 
      & 0.101 & 0.180 & 0.163 & 0.941 \\
\bottomrule
\end{tabular}
\end{table}

\begin{table}[ht!]
\centering
\caption{
Simulation results in Scenario C2 for estimating four estimands under a continuous outcome with $I=10$ clusters and $J=11$ periods (switching one cluster per sequence) using covariate-adjusted working models (W1)-(W6) with the Coef, MRS and ANCOVA estimators. MRS: proposed augmented estimator; Coef: treatment-effect coefficients from covariate-adjusted working model; ANCOVA: model-assisted ANCOVA estimator; UNADJ: nonparametric estimator. RBias (\%): absolute relative bias in percent; AESE: average estimated standard error; MCSD: Monte Carlo standard deviation; CP: 95\% confidence interval coverage.
}
\label{tab:sim_C2_small_seq}
\begin{tabular}{cclrrrrrrrr}
\toprule
\multirow{2}{*}{Direction} & \multirow{2}{*}{Working model}
  & \multirow{2}{*}{Method}
  & \multicolumn{4}{c}{h-iATE}
  & \multicolumn{4}{c}{h-cATE} \\
\cmidrule(lr){4-7}\cmidrule(lr){8-11}
 & & 
 & RBias & AESE & MCSD & CP 
 & RBias & AESE & MCSD & CP \\
  & & 
 & \multicolumn{4}{c}{\(\tau_{\mathrm{I}}^{\mathrm{h}}=8.135\)} 
 & \multicolumn{4}{c}{\(\tau_{\mathrm{C}}^{\mathrm{h}}=7.617\)} \\
\midrule
\multirow{18}{*}{Horizontal}
  & \textbackslash & UNADJ 
    & 5.398 & 1.264 & 1.133 & 0.918 
    & 4.670 & 1.261 & 1.136 & 0.925 \\
  & \textbackslash & ANCOVA  
    & 5.444 & 0.823 & 1.133 & 0.825 
    & 4.711 & 0.821 & 1.135 & 0.834 \\
  & \multirow{2}{*}{W1}
    & Coef   
      & 2.682 & 0.753 & 0.762 & 0.934 
      & 0.059 & 0.753 & 0.762 & 0.936 \\
  &               & MRS    
      & 5.525 & 0.979 & 0.812 & 0.944 
      & 4.760 & 0.978 & 0.812 & 0.948 \\
  & \multirow{2}{*}{W2}
    & Coef   
      & 5.561 & 0.647 & 0.815 & 0.852 
      & 2.914 & 0.643 & 0.815 & 0.877 \\
  &               & MRS    
      & 5.561 & 0.990 & 0.815 & 0.947 
      & 4.794 & 0.989 & 0.814 & 0.948 \\
  & \multirow{2}{*}{W3}
    & Coef   
      & 4.896 & 0.975 & 0.979 & 0.932 
      & 2.216 & 0.975 & 0.979 & 0.937 \\
  &               & MRS    
      & 5.533 & 0.979 & 0.813 & 0.943 
      & 4.788 & 0.979 & 0.813 & 0.948 \\
  & \multirow{2}{*}{W4}
    & Coef   
      & 7.401 & 0.910 & 1.028 & 0.881 
      & 4.797 & 0.908 & 1.030 & 0.911 \\
  &               & MRS    
      & 5.559 & 0.992 & 0.816 & 0.947 
      & 4.813 & 0.991 & 0.816 & 0.950 \\
  & \multirow{2}{*}{W5}
    & Coef   
      & 25.468 & 0.707 & 0.721 & 0.286 
      & 23.369 & 0.707 & 0.721 & 0.399 \\
  &               & MRS    
      & 5.581 & 0.973 & 0.815 & 0.941 
      & 4.824 & 0.972 & 0.816 & 0.946 \\
  & \multirow{2}{*}{W6}
    & Coef   
      & 23.156 & 0.651 & 0.787 & 0.325 
      & 20.979 & 0.646 & 0.788 & 0.408 \\
  &               & MRS    
      & 5.580 & 0.982 & 0.817 & 0.943 
      & 4.820 & 0.981 & 0.818 & 0.950 \\
\toprule
& & 
  & \multicolumn{4}{c}{v-iATE}
  & \multicolumn{4}{c}{v-cATE} \\
  \cmidrule(lr){4-7}\cmidrule(lr){8-11}
 & & 
 & RBias & AESE & MCSD & CP 
 & RBias & AESE & MCSD & CP \\
  & & 
 & \multicolumn{4}{c}{\(\tau_{\mathrm{I}}^{\mathrm{v}}=8.134\)} 
 & \multicolumn{4}{c}{\(\tau_{\mathrm{C}}^{\mathrm{v}}=6.011\)} \\
 \midrule
\multirow{18}{*}{Vertical}
  & \textbackslash & UNADJ 
    & 7.604 & 1.272 & 1.127 & 0.901 
    & 0.424 & 1.183 & 1.077 & 0.936 \\
  & \textbackslash & ANCOVA  
    & 7.652 & 0.822 & 1.128 & 0.791 
    & 0.413 & 0.787 & 1.074 & 0.858 \\
  & \multirow{2}{*}{W1}
    & Coef   
      & 2.672 & 0.753 & 0.762 & 0.934 
      & 31.681 & 0.753 & 0.762 & 0.403 \\
  &               & MRS    
      & 7.770 & 0.982 & 0.814 & 0.915 
      & 0.590 & 0.930 & 0.771 & 0.958 \\
  & \multirow{2}{*}{W2}
    & Coef   
      & 7.828 & 0.645 & 0.816 & 0.793 
      & 24.705 & 0.645 & 0.816 & 0.495 \\
  &               & MRS    
      & 7.828 & 0.994 & 0.816 & 0.919 
      & 0.557 & 0.941 & 0.773 & 0.958 \\
  & \multirow{2}{*}{W3}
    & Coef   
      & 4.886 & 0.975 & 0.979 & 0.932 
      & 28.686 & 0.975 & 0.979 & 0.679 \\
  &               & MRS    
      & 7.778 & 0.983 & 0.815 & 0.913 
      & 0.577 & 0.930 & 0.772 & 0.958 \\
  & \multirow{2}{*}{W4}
    & Coef   
      & 9.672 & 0.909 & 1.029 & 0.852 
      & 22.210 & 0.909 & 1.029 & 0.748 \\
  &               & MRS    
      & 7.826 & 0.995 & 0.817 & 0.921 
      & 0.559 & 0.942 & 0.774 & 0.958 \\
  & \multirow{2}{*}{W5}
    & Coef   
      & 25.461 & 0.707 & 0.721 & 0.286 
      & 0.849 & 0.707 & 0.721 & 0.940 \\
  &               & MRS    
      & 7.824 & 0.976 & 0.816 & 0.910 
      & 0.521 & 0.921 & 0.771 & 0.954 \\
  & \multirow{2}{*}{W6}
    & Coef   
      & 25.496 & 0.641 & 0.777 & 0.241 
      & 0.801 & 0.641 & 0.777 & 0.897 \\
  &               & MRS    
      & 7.843 & 0.986 & 0.818 & 0.913 
      & 0.543 & 0.931 & 0.773 & 0.955 \\
\bottomrule
\end{tabular}
\end{table}

\begin{table}[ht!]
\centering
\caption{
Simulation results in Scenario C3 for estimating four estimands under a continuous outcome with $I=10$ clusters and $J=11$ periods (switching one cluster per sequence) using covariate-adjusted working models (W1)-(W6) with the Coef, MRS and ANCOVA estimators. MRS: proposed augmented estimator; Coef: treatment-effect coefficients from covariate-adjusted working model; ANCOVA: model-assisted ANCOVA estimator; UNADJ: nonparametric estimator. RBias (\%): absolute relative bias in percent; AESE: average estimated standard error; MCSD: Monte Carlo standard deviation; CP: 95\% confidence interval coverage.
}
\label{tab:sim_C3_small_seq}
\begin{tabular}{cclrrrrrrrr}
\toprule
\multirow{2}{*}{Direction} & \multirow{2}{*}{Working model}
  & \multirow{2}{*}{Method}
  & \multicolumn{4}{c}{h-iATE}
  & \multicolumn{4}{c}{h-cATE} \\
\cmidrule(lr){4-7}\cmidrule(lr){8-11}
 & & 
 & RBias & AESE & MCSD & CP 
 & RBias & AESE & MCSD & CP \\
  & & 
 & \multicolumn{4}{c}{\(\tau_{\mathrm{I}}^{\mathrm{h}}=19.552\)} 
 & \multicolumn{4}{c}{\(\tau_{\mathrm{C}}^{\mathrm{h}}=19.106\)} \\
\midrule
\multirow{18}{*}{Horizontal}
  & \textbackslash & UNADJ 
    & 4.519 & 2.486 & 2.235 & 0.919 
    & 3.895 & 2.480 & 2.240 & 0.928 \\
  & \textbackslash & ANCOVA  
    & 4.541 & 1.630 & 2.239 & 0.820 
    & 3.914 & 1.625 & 2.243 & 0.833 \\
  & \multirow{2}{*}{W1}
    & Coef   
      & 2.246 & 1.483 & 1.494 & 0.931 
      & 0.031 & 1.483 & 1.494 & 0.940 \\
  &               & MRS    
      & 4.609 & 1.929 & 1.591 & 0.943 
      & 3.956 & 1.928 & 1.590 & 0.951 \\
  & \multirow{2}{*}{W2}
    & Coef   
      & 4.640 & 1.276 & 1.595 & 0.849 
      & 2.429 & 1.269 & 1.595 & 0.874 \\
  &               & MRS    
      & 4.640 & 1.952 & 1.595 & 0.945 
      & 3.984 & 1.950 & 1.594 & 0.952 \\
  & \multirow{2}{*}{W3}
    & Coef   
      & 4.084 & 1.949 & 1.953 & 0.933 
      & 1.849 & 1.949 & 1.953 & 0.939 \\
  &               & MRS    
      & 4.617 & 1.931 & 1.593 & 0.945 
      & 3.979 & 1.930 & 1.592 & 0.949 \\
  & \multirow{2}{*}{W4}
    & Coef   
      & 6.162 & 1.820 & 2.049 & 0.883 
      & 3.981 & 1.815 & 2.052 & 0.906 \\
  &               & MRS    
      & 4.638 & 1.955 & 1.597 & 0.946 
      & 4.000 & 1.954 & 1.597 & 0.953 \\
  & \multirow{2}{*}{W5}
    & Coef   
      & 21.380 & 1.393 & 1.413 & 0.270 
      & 19.548 & 1.393 & 1.413 & 0.373 \\
  &               & MRS    
      & 4.657 & 1.917 & 1.597 & 0.941 
      & 4.009 & 1.917 & 1.597 & 0.948 \\
  & \multirow{2}{*}{W6}
    & Coef   
      & 19.414 & 1.283 & 1.545 & 0.304 
      & 17.526 & 1.275 & 1.547 & 0.389 \\
  &               & MRS    
      & 4.656 & 1.936 & 1.601 & 0.945 
      & 4.006 & 1.935 & 1.601 & 0.948 \\
\toprule
& & 
  & \multicolumn{4}{c}{v-iATE}
  & \multicolumn{4}{c}{v-cATE} \\
  \cmidrule(lr){4-7}\cmidrule(lr){8-11}
 & & 
 & RBias & AESE & MCSD & CP 
 & RBias & AESE & MCSD & CP \\
  & & 
 & \multicolumn{4}{c}{\(\tau_{\mathrm{I}}^{\mathrm{v}}=19.550\)} 
 & \multicolumn{4}{c}{\(\tau_{\mathrm{C}}^{\mathrm{v}}=15.312\)} \\
 \midrule
\multirow{18}{*}{Vertical}
  & \textbackslash & UNADJ 
    & 6.356 & 2.502 & 2.225 & 0.895 
    & 0.300 & 2.318 & 2.117 & 0.933 \\
  & \textbackslash & ANCOVA  
    & 6.378 & 1.627 & 2.230 & 0.788 
    & 0.298 & 1.557 & 2.118 & 0.854 \\
  & \multirow{2}{*}{W1}
    & Coef   
      & 2.238 & 1.483 & 1.494 & 0.931 
      & 24.818 & 1.483 & 1.494 & 0.394 \\
  &               & MRS    
      & 6.477 & 1.937 & 1.596 & 0.913 
      & 0.435 & 1.831 & 1.509 & 0.956 \\
  & \multirow{2}{*}{W2}
    & Coef   
      & 6.525 & 1.273 & 1.601 & 0.795 
      & 19.345 & 1.273 & 1.601 & 0.488 \\
  &               & MRS    
      & 6.525 & 1.961 & 1.601 & 0.915 
      & 0.409 & 1.852 & 1.514 & 0.956 \\
  & \multirow{2}{*}{W3}
    & Coef   
      & 4.076 & 1.949 & 1.953 & 0.933 
      & 22.471 & 1.949 & 1.953 & 0.675 \\
  &               & MRS    
      & 6.484 & 1.938 & 1.598 & 0.913 
      & 0.424 & 1.831 & 1.511 & 0.956 \\
  & \multirow{2}{*}{W4}
    & Coef   
      & 8.053 & 1.817 & 2.052 & 0.853 
      & 17.394 & 1.817 & 2.052 & 0.744 \\
  &               & MRS    
      & 6.524 & 1.964 & 1.602 & 0.917 
      & 0.410 & 1.854 & 1.515 & 0.957 \\
  & \multirow{2}{*}{W5}
    & Coef   
      & 21.373 & 1.393 & 1.413 & 0.270 
      & 0.387 & 1.393 & 1.413 & 0.938 \\
  &               & MRS    
      & 6.523 & 1.925 & 1.601 & 0.909 
      & 0.380 & 1.812 & 1.510 & 0.952 \\
  & \multirow{2}{*}{W6}
    & Coef   
      & 21.361 & 1.264 & 1.527 & 0.223 
      & 0.402 & 1.264 & 1.527 & 0.899 \\
  &               & MRS    
      & 6.538 & 1.945 & 1.605 & 0.912 
      & 0.398 & 1.832 & 1.513 & 0.956 \\
\bottomrule
\end{tabular}
\end{table}

\section{Simulation results for binary outcome in Scenario B1 and B2} \label{sim:bin_adj}

Table \ref{tab:sim_B1_results} presents the simulation results under Scenario  B1 and non-informative cluster size. Regardless of the choice of working models (W7-W12), the proposed MRS estimator consistently gives nearly unbiased estimates with empirical coverage close to the nominal level. 
In contrast, the Coef estimators can exhibit noticeable bias, particularly under GLMMs, which results in under-coverage. In GLMM models with only a cluster-level random effect (W9-W10), the model-based variance estimator also dramatically underestimates the true variability. 
Interestingly, the GEE Coef estimators (W7-W8) lead to small bias, with generally closer to nominal coverage compared to their GLMM counterparts. 
Finally, the unadjusted estimator is also unbiased; in this simulation without informative cluster size, its Monte Carlo standard deviation is generally comparable to (only slightly larger than) that of MRS.

Table \ref{tab:sim_B2_results} summarizes the results under Scenario B2, where the cluster-period sizes are informative. Overall, the MRS estimator continues to exhibit low bias and nominal coverage across all estimands and working outcome models, demonstrating it model-robustness property for inferring potential outcomes estimands. Although its efficiency varies moderately by model specification (with GEE being less efficient than GLMM), MRS also consistently outperforms the unadjusted estimator due to baseline adjustment. In contrast, the Coef estimators can correspond to substantial bias, especially for the cluster-average estimands across all working models. This bias appears to be the most substantial under GLMM (W9-W12). On the other hand, the Coef estimators under both independence GEE and GLMM appear to have much smaller (but still some) bias for estimating the individual-average estimands. Except under GLMM with a cluster-level random effect (W9-W10), the coverage probability of the Coef interval estimators is at least 90\% for estimating individual-average estimands. For binary outcomes, similarly divergent performance of Coef estimators for cluster-average versus individual-average estimands were discussed in \citet{li2025model} in simpler parallel-arm CRTs.

\begin{table}[ht!]
\centering
\caption{
Simulation results in Scenario B1 for estimating four estimands under a binary outcome with $I=30$ clusters and $J=4$ periods using covariate-adjusted working models (W7)--(W12) with the Coef, MRS and nonparametric (UNADJ) estimators. MRS: proposed augmented estimator; Coef: treatment‐effect coefficients from covariate‐adjusted working model; UNADJ: nonparametric estimator. RBias: absolute bias; AESE: average estimated standard error; MCSD: Monte Carlo standard deviation; CP: 95\% confidence interval coverage.
}
\label{tab:sim_B1_results}
\begin{tabular}{cclrrrrrrrr}
\toprule
\multirow{2}{*}{Direction} & \multirow{2}{*}{Working model}
  & \multirow{2}{*}{Method}
  & \multicolumn{4}{c}{h-iATE}
  & \multicolumn{4}{c}{h-cATE} \\
\cmidrule(lr){4-7}\cmidrule(lr){8-11}
 & & 
 & RBias & AESE & MCSD & CP 
 & RBias & AESE & MCSD & CP \\
 & & 
 & \multicolumn{4}{c}{\(\Delta_{\mathrm{I}}^{\mathrm{h}}=0.884\)} 
 & \multicolumn{4}{c}{\(\Delta_{\mathrm{C}}^{\mathrm{h}}=0.884\)} \\
\midrule
\multirow{13}{*}{Horizontal}
  & \textbackslash & UNADJ 
    & 1.029 & 0.255 & 0.253 & 0.952 
    & 0.385 & 0.251 & 0.251 & 0.955 \\
  & \multirow{2}{*}{W7}
    & Coef   
      & 4.474 & 0.240 & 0.260 & 0.937 
      & 4.479 & 0.240 & 0.260 & 0.937 \\
  &               & MRS    
      & 1.162 & 0.257 & 0.252 & 0.952 
      & 0.556 & 0.252 & 0.250 & 0.955 \\
  & \multirow{2}{*}{W8}
    & Coef   
      & 4.767 & 0.238 & 0.261 & 0.933 
      & 4.781 & 0.238 & 0.261 & 0.932 \\
  &               & MRS    
      & 1.146 & 0.258 & 0.252 & 0.954 
      & 0.545 & 0.252 & 0.250 & 0.957 \\
  & \multirow{2}{*}{W9}
    & Coef   
      & 11.412 & 0.153 & 0.243 & 0.771 
      & 11.417 & 0.153 & 0.243 & 0.772 \\
  &               & MRS    
      & 1.125 & 0.257 & 0.252 & 0.954 
      & 0.511 & 0.252 & 0.250 & 0.958 \\
  & \multirow{2}{*}{W10}
    & Coef   
      & 11.685 & 0.154 & 0.244 & 0.776 
      & 11.696 & 0.154 & 0.244 & 0.773 \\
  &               & MRS    
      & 1.111 & 0.257 & 0.252 & 0.954 
      & 0.501 & 0.252 & 0.250 & 0.957 \\
  & \multirow{2}{*}{W11}
    & Coef   
      & 15.170 & 0.230 & 0.237 & 0.923 
      & 15.176 & 0.230 & 0.237 & 0.923 \\
  &               & MRS    
      & 1.140 & 0.257 & 0.252 & 0.955 
      & 0.530 & 0.252 & 0.250 & 0.957 \\
  & \multirow{2}{*}{W12}
    & Coef   
      & 15.385 & 0.229 & 0.239 & 0.922 
      & 15.390 & 0.229 & 0.239 & 0.921 \\
  &               & MRS    
      & 1.128 & 0.257 & 0.252 & 0.955 
      & 0.522 & 0.252 & 0.250 & 0.958 \\
\toprule
& & 
  & \multicolumn{4}{c}{v-iATE}
  & \multicolumn{4}{c}{v-cATE} \\
  \cmidrule(lr){4-7}\cmidrule(lr){8-11}
 & & 
 & RBias & AESE & MCSD & CP 
 & RBias & AESE & MCSD & CP \\
& & 
 & \multicolumn{4}{c}{\(\Delta_{\mathrm{I}}^{\mathrm{v}}=0.884\)} 
 & \multicolumn{4}{c}{\(\Delta_{\mathrm{C}}^{\mathrm{v}}=0.884\)} \\
 \midrule
\multirow{13}{*}{Vertical}
  & \textbackslash & UNADJ 
    & 0.968 & 0.256 & 0.253 & 0.947 
    & 0.262 & 0.249 & 0.252 & 0.948 \\
  & \multirow{2}{*}{W7}
    & Coef   
      & 4.474 & 0.240 & 0.260 & 0.937 
      & 4.445 & 0.240 & 0.260 & 0.937 \\
  &               & MRS    
      & 1.086 & 0.258 & 0.251 & 0.950 
      & 0.433 & 0.251 & 0.251 & 0.948 \\
  & \multirow{2}{*}{W8}
    & Coef   
      & 4.689 & 0.238 & 0.261 & 0.930 
      & 4.660 & 0.238 & 0.261 & 0.930 \\
  &               & MRS    
      & 1.069 & 0.258 & 0.252 & 0.952 
      & 0.420 & 0.251 & 0.251 & 0.947 \\
  & \multirow{2}{*}{W9}
    & Coef   
      & 11.412 & 0.153 & 0.243 & 0.771 
      & 11.381 & 0.153 & 0.243 & 0.772 \\
  &               & MRS    
      & 1.051 & 0.257 & 0.252 & 0.952 
      & 0.368 & 0.250 & 0.251 & 0.950 \\
  & \multirow{2}{*}{W10}
    & Coef   
      & 11.570 & 0.154 & 0.244 & 0.774 
      & 11.539 & 0.154 & 0.244 & 0.774 \\
  &               & MRS    
      & 1.036 & 0.258 & 0.252 & 0.953 
      & 0.358 & 0.250 & 0.251 & 0.951 \\
  & \multirow{2}{*}{W11}
    & Coef   
      & 15.170 & 0.230 & 0.237 & 0.923 
      & 15.138 & 0.230 & 0.237 & 0.923 \\
  &               & MRS    
      & 1.066 & 0.257 & 0.252 & 0.953 
      & 0.402 & 0.250 & 0.251 & 0.949 \\
  & \multirow{2}{*}{W12}
    & Coef   
      & 15.267 & 0.229 & 0.238 & 0.920 
      & 15.235 & 0.229 & 0.238 & 0.920 \\
  &               & MRS    
      & 1.052 & 0.257 & 0.252 & 0.954 
      & 0.392 & 0.250 & 0.251 & 0.949 \\
\bottomrule
\end{tabular}
\end{table}

\begin{table}[ht!]
\centering
\caption{
Simulation results in Scenario B2 for estimating four estimands under a binary outcome with $I=30$ clusters and $J=4$ periods using covariate-adjusted working models (W7)--(W12) with the Coef, MRS and nonparametric (UNADJ) estimators. MRS: proposed augmented estimator; Coef: treatment‐effect coefficients from covariate‐adjusted working model; UNADJ: nonparametric estimator. RBias (\%): absolute relative bias in percent; AESE: average estimated standard error; MCSD: Monte Carlo standard deviation; CP: 95\% confidence interval coverage.
}
\label{tab:sim_B2_results}
\begin{tabular}{cclrrrrrrrr}
\toprule
\multirow{2}{*}{Direction} & \multirow{2}{*}{Working model}
  & \multirow{2}{*}{Method}
  & \multicolumn{4}{c}{h-iATE}
  & \multicolumn{4}{c}{h-cATE} \\
\cmidrule(lr){4-7}\cmidrule(lr){8-11}
 & & 
 & RBias & AESE & MCSD & CP 
 & RBias & AESE & MCSD & CP \\
 & & 
 & \multicolumn{4}{c}{\(\tau_{\mathrm{I}}^{\mathrm{h}}=1.834\)} 
 & \multicolumn{4}{c}{\(\tau_{\mathrm{C}}^{\mathrm{h}}=1.596\)} \\
\midrule
\multirow{13}{*}{Horizontal}
  & \textbackslash & UNADJ 
    & 0.728 & 0.329 & 0.321 & 0.952 
    & 0.463 & 0.331 & 0.321 & 0.949 \\
  & \multirow{2}{*}{W7}
    & Coef   
      & 2.935 & 0.283 & 0.300 & 0.927 
      & 18.283 & 0.283 & 0.300 & 0.826 \\
  &               & MRS    
      & 0.761 & 0.307 & 0.296 & 0.949 
      & 0.412 & 0.310 & 0.298 & 0.959 \\
  & \multirow{2}{*}{W8}
    & Coef   
      & 3.766 & 0.283 & 0.305 & 0.922 
      & 19.192 & 0.283 & 0.305 & 0.811 \\
  &               & MRS    
      & 0.665 & 0.307 & 0.297 & 0.950 
      & 0.513 & 0.311 & 0.299 & 0.958 \\
  & \multirow{2}{*}{W9}
    & Coef   
      & 6.920 & 0.181 & 0.308 & 0.727 
      & 22.862 & 0.181 & 0.308 & 0.492 \\
  &               & MRS    
      & 0.783 & 0.304 & 0.293 & 0.948 
      & 0.354 & 0.308 & 0.296 & 0.958 \\
  & \multirow{2}{*}{W10}
    & Coef   
      & 7.716 & 0.184 & 0.312 & 0.726 
      & 23.733 & 0.184 & 0.312 & 0.479 \\
  &               & MRS    
      & 0.695 & 0.305 & 0.294 & 0.951 
      & 0.444 & 0.309 & 0.296 & 0.956 \\
  & \multirow{2}{*}{W11}
    & Coef   
      & 7.616 & 0.295 & 0.307 & 0.933 
      & 23.662 & 0.295 & 0.307 & 0.772 \\
  &               & MRS    
      & 0.748 & 0.302 & 0.290 & 0.950 
      & 0.371 & 0.306 & 0.294 & 0.954 \\
  & \multirow{2}{*}{W12}
    & Coef   
      & 8.351 & 0.296 & 0.310 & 0.920 
      & 24.475 & 0.296 & 0.310 & 0.760 \\
  &               & MRS    
      & 0.674 & 0.302 & 0.291 & 0.952 
      & 0.446 & 0.306 & 0.294 & 0.953 \\
\toprule
& & 
  & \multicolumn{4}{c}{v-iATE}
  & \multicolumn{4}{c}{v-cATE} \\
  \cmidrule(lr){4-7}\cmidrule(lr){8-11}
 & & 
 & RBias & AESE & MCSD & CP 
 & RBias & AESE & MCSD & CP \\
& & 
  & \multicolumn{4}{c}{\(\tau_{\mathrm{I}}^{\mathrm{v}}=1.834\)} 
  & \multicolumn{4}{c}{\(\tau_{\mathrm{C}}^{\mathrm{v}}=1.372\)} \\
  \midrule
\multirow{13}{*}{Vertical}
  & \textbackslash & UNADJ 
    & 1.093 & 0.330 & 0.322 & 0.949 
    & 0.596 & 0.317 & 0.307 & 0.953 \\
  & \multirow{2}{*}{W7}
    & Coef   
      & 2.947 & 0.283 & 0.300 & 0.927 
      & 37.572 & 0.283 & 0.300 & 0.597 \\
  &               & MRS    
      & 1.117 & 0.308 & 0.297 & 0.952 
      & 0.734 & 0.301 & 0.286 & 0.957 \\
  & \multirow{2}{*}{W8}
    & Coef   
      & 3.398 & 0.282 & 0.305 & 0.920 
      & 38.174 & 0.282 & 0.305 & 0.586 \\
  &               & MRS    
      & 1.023 & 0.308 & 0.298 & 0.952 
      & 0.834 & 0.301 & 0.287 & 0.958 \\
  & \multirow{2}{*}{W9}
    & Coef   
      & 6.932 & 0.181 & 0.308 & 0.727 
      & 42.897 & 0.181 & 0.308 & 0.250 \\
  &               & MRS    
      & 1.138 & 0.305 & 0.294 & 0.950 
      & 0.711 & 0.299 & 0.284 & 0.957 \\
  & \multirow{2}{*}{W10}
    & Coef   
      & 7.299 & 0.183 & 0.311 & 0.731 
      & 43.388 & 0.183 & 0.311 & 0.247 \\
  &               & MRS    
      & 1.053 & 0.306 & 0.295 & 0.951 
      & 0.809 & 0.300 & 0.285 & 0.956 \\
  & \multirow{2}{*}{W11}
    & Coef   
      & 7.629 & 0.295 & 0.307 & 0.933 
      & 43.828 & 0.295 & 0.307 & 0.500 \\
  &               & MRS    
      & 1.102 & 0.303 & 0.291 & 0.953 
      & 0.773 & 0.297 & 0.283 & 0.961 \\
  & \multirow{2}{*}{W12}
    & Coef   
      & 7.886 & 0.295 & 0.310 & 0.921 
      & 44.172 & 0.295 & 0.310 & 0.496 \\
  &               & MRS    
      & 1.030 & 0.303 & 0.292 & 0.954 
      & 0.851 & 0.298 & 0.283 & 0.959 \\
\bottomrule
\end{tabular}
\end{table}


\section{Simulation results for binary outcome with non-covariate-adjusted model}
\label{sim:bin_unadj}

Tables \ref{tab:sim_B1_unadj}-\ref{tab:sim_B3_unadj} present the results for the Coef estimators under Scenarios B1 to B3 using six working models without covariate adjustment. Similar to the adjusted models, when informative cluster size exists, all models show a large bias and undercoverage for all four estimands when informative cluster size exists. Overall, the efficiency of the unadjusted models is generally lower than that of the adjusted models.

\begin{table}[ht!]
\centering
\caption{
Simulation results in Scenario B1 for estimating four estimands under a binary outcome with $I=30$ clusters and $J=4$ periods using non‐covariate‐adjusted working models (W7)--(W12) with the Coef estimators. RBias: absolute bias; AESE: average estimated standard error; MCSD: Monte Carlo standard deviation; CP: 95\% confidence interval coverage.
}
\label{tab:sim_B1_unadj}
\begin{tabular}{ccrrrrrrrr}
\toprule
\multirow{2}{*}{Direction} & \multirow{2}{*}{Working model}
  & \multicolumn{4}{c}{h-iATE}
  & \multicolumn{4}{c}{h-cATE} \\
\cmidrule(lr){3-6}\cmidrule(lr){7-10}
 & 
 & RBias & AESE & MCSD & CP 
 & RBias & AESE & MCSD & CP \\
 & 
 & \multicolumn{4}{c}{\(\Delta_{\mathrm{I}}^{\mathrm{h}}=0.884\)} 
 & \multicolumn{4}{c}{\(\Delta_{\mathrm{C}}^{\mathrm{h}}=0.884\)} \\
\midrule
\multirow{6}{*}{Horizontal}
  & UNADJ-W7  
    & 1.199 & 0.233 & 0.253 & 0.940 
    & 1.205 & 0.233 & 0.253 & 0.940 \\
  & UNADJ-W8  
    & 1.490 & 0.231 & 0.255 & 0.940 
    & 1.500 & 0.231 & 0.255 & 0.938 \\
  & UNADJ-W9  
    & 7.380 & 0.149 & 0.236 & 0.795 
    & 7.385 & 0.149 & 0.236 & 0.794 \\
  & UNADJ-W10 
    & 7.638 & 0.150 & 0.236 & 0.792 
    & 7.645 & 0.150 & 0.237 & 0.795 \\
  & UNADJ-W11 
    & 10.834 & 0.222 & 0.230 & 0.939 
    & 10.839 & 0.222 & 0.230 & 0.939 \\
  & UNADJ-W12 
    & 11.022 & 0.221 & 0.231 & 0.938 
    & 11.024 & 0.221 & 0.231 & 0.937 \\
\toprule
& 
  & \multicolumn{4}{c}{v-iATE}
  & \multicolumn{4}{c}{v-cATE} \\
  \cmidrule(lr){3-6}\cmidrule(lr){7-10}
 & 
 & RBias & AESE & MCSD & CP 
 & RBias & AESE & MCSD & CP \\
& 
 & \multicolumn{4}{c}{\(\Delta_{\mathrm{I}}^{\mathrm{v}}=0.884\)} 
 & \multicolumn{4}{c}{\(\Delta_{\mathrm{C}}^{\mathrm{v}}=0.884\)} \\
 \midrule
\multirow{6}{*}{Vertical}
  & UNADJ-W7  
    & 1.199 & 0.233 & 0.253 & 0.940 
    & 1.171 & 0.233 & 0.253 & 0.940 \\
  & UNADJ-W8  
    & 1.430 & 0.231 & 0.254 & 0.934 
    & 1.401 & 0.231 & 0.254 & 0.934 \\
  & UNADJ-W9  
    & 7.380 & 0.149 & 0.236 & 0.795 
    & 7.350 & 0.149 & 0.236 & 0.795 \\
  & UNADJ-W10 
    & 7.545 & 0.150 & 0.236 & 0.797 
    & 7.516 & 0.150 & 0.236 & 0.797 \\
  & UNADJ-W11 
    & 10.834 & 0.222 & 0.230 & 0.939 
    & 10.803 & 0.222 & 0.230 & 0.939 \\
  & UNADJ-W12 
    & 10.929 & 0.221 & 0.230 & 0.937 
    & 10.898 & 0.221 & 0.230 & 0.939 \\
\bottomrule
\end{tabular}
\end{table}

\begin{table}[ht!]
\centering
\caption{
Simulation results in Scenario B2 for estimating h-iATE and h-cATE under a binary outcome with $I=30$ clusters and $J=4$ periods using non–covariate–adjusted working models (W7)--(W12) with the Coef estimators. RBias: absolute bias; AESE: average estimated standard error; MCSD: Monte Carlo standard deviation; CP: 95\% confidence interval coverage.
}
\label{tab:sim_B2_unadj}
\begin{tabular}{ccrrrrrrrr}
\toprule
\multirow{2}{*}{Direction} & \multirow{2}{*}{Working model}
  & \multicolumn{4}{c}{h-iATE}
  & \multicolumn{4}{c}{h-cATE} \\
\cmidrule(lr){3-6}\cmidrule(lr){7-10}
 & 
 & RBias & AESE & MCSD & CP 
 & RBias & AESE & MCSD & CP \\
 & 
 & \multicolumn{4}{c}{\(\tau_{\mathrm{I}}^{{\mathrm{h}}}=1.834\)} 
 & \multicolumn{4}{c}{\(\tau_{\mathrm{C}}^{\mathrm{h}}=1.596\)} \\
\midrule
\multirow{6}{*}{Horizontal}
  & UNADJ-W7  
    & 0.315 & 0.297 & 0.317 & 0.936 
    & 14.548 & 0.297 & 0.317 & 0.869 \\
  & UNADJ-W8  
    & 0.569 & 0.300 & 0.323 & 0.933 
    & 15.524 & 0.300 & 0.323 & 0.867 \\
  & UNADJ-W9  
    & 2.579 & 0.176 & 0.324 & 0.734 
    & 17.874 & 0.176 & 0.324 & 0.571 \\
  & UNADJ-W10 
    & 3.528 & 0.179 & 0.331 & 0.730 
    & 18.926 & 0.179 & 0.331 & 0.549 \\
  & UNADJ-W11 
    & 3.403 & 0.313 & 0.339 & 0.940 
    & 18.820 & 0.313 & 0.339 & 0.855 \\
  & UNADJ-W12 
    & 4.206 & 0.313 & 0.343 & 0.931 
    & 19.721 & 0.313 & 0.343 & 0.841 \\
\toprule
& 
  & \multicolumn{4}{c}{v-iATE}
  & \multicolumn{4}{c}{v-cATE} \\
  \cmidrule(lr){3-6}\cmidrule(lr){7-10}
 & 
 & RBias & AESE & MCSD & CP 
 & RBias & AESE & MCSD & CP \\
 & 
 & \multicolumn{4}{c}{\(\tau_{\mathrm{I}}^{\mathrm{v}}=1.834\)} 
 & \multicolumn{4}{c}{\(\tau_{\mathrm{C}}^{\mathrm{v}}=1.372\)} \\
 \midrule
\multirow{6}{*}{Vertical}
  & UNADJ-W7  
    & 0.303 & 0.297 & 0.317 & 0.936 
    & 33.228 & 0.297 & 0.317 & 0.705 \\
  & UNADJ-W8  
    & 0.194 & 0.299 & 0.324 & 0.931 
    & 33.892 & 0.299 & 0.324 & 0.693 \\
  & UNADJ-W9  
    & 2.591 & 0.176 & 0.324 & 0.734 
    & 37.096 & 0.176 & 0.324 & 0.332 \\
  & UNADJ-W10 
    & 3.102 & 0.179 & 0.330 & 0.730 
    & 37.779 & 0.179 & 0.330 & 0.331 \\
  & UNADJ-W11 
    & 3.415 & 0.313 & 0.339 & 0.940 
    & 38.196 & 0.313 & 0.339 & 0.640 \\
  & UNADJ-W12 
    & 3.737 & 0.313 & 0.343 & 0.934 
    & 38.628 & 0.313 & 0.343 & 0.631 \\
\bottomrule
\end{tabular}
\end{table}

\begin{table}[ht!]
\centering
\caption{
Simulation results in Scenario B3 for estimating h-iATE and h-cATE under a binary outcome with $I=30$ clusters and $J=4$ periods using non-covariate-adjusted working models (W7)--(W12) with the Coef estimators. RBias: absolute bias; AESE: average estimated standard error; MCSD: Monte Carlo standard deviation; CP: 95\% confidence interval coverage.
}
\label{tab:sim_B3_unadj}
\begin{tabular}{ccrrrrrrrr}
\toprule
\multirow{2}{*}{Direction} & \multirow{2}{*}{Working model}
  & \multicolumn{4}{c}{h-iATE}
  & \multicolumn{4}{c}{h-cATE} \\
\cmidrule(lr){3-6}\cmidrule(lr){7-10}
 & 
 & RBias & AESE & MCSD & CP 
 & RBias & AESE & MCSD & CP \\
 & 
 & \multicolumn{4}{c}{\(\Delta_{\mathrm{I}}^{\mathrm{h}}=3.195\)} 
 & \multicolumn{4}{c}{\(\Delta_{\mathrm{C}}^{\mathrm{h}}=2.879\)} \\
\midrule
\multirow{6}{*}{Horizontal}
  & UNADJ-W7  
    & 4.534 & 0.435 & 0.496 & 0.925 
    & 16.002 & 0.435 & 0.496 & 0.836 \\
  & UNADJ-W8  
    & 6.158 & 0.431 & 0.527 & 0.902 
    & 17.842 & 0.431 & 0.528 & 0.796 \\
  & UNADJ-W9  
    & 8.401 & 0.215 & 0.511 & 0.592 
    & 20.294 & 0.215 & 0.511 & 0.379 \\
  & UNADJ-W10 
    & 9.795 & 0.229 & 0.545 & 0.574 
    & 21.883 & 0.229 & 0.546 & 0.371 \\
  & UNADJ-W11 
    & 18.700 & 0.406 & 0.508 & 0.700 
    & 31.722 & 0.406 & 0.508 & 0.431 \\
  & UNADJ-W12 
    & 20.246 & 0.406 & 0.526 & 0.666 
    & 33.481 & 0.406 & 0.527 & 0.405 \\
\toprule
& 
  & \multicolumn{4}{c}{v-iATE}
  & \multicolumn{4}{c}{v-cATE} \\
  \cmidrule(lr){3-6}\cmidrule(lr){7-10}
 & 
 & RBias & AESE & MCSD & CP 
 & RBias & AESE & MCSD & CP \\
 & 
 & \multicolumn{4}{c}{\(\Delta_{\mathrm{I}}^{\mathrm{v}}=3.172\)} 
 & \multicolumn{4}{c}{\(\Delta_{\mathrm{C}}^{\mathrm{v}}=2.524\)} \\
 \midrule
\multirow{6}{*}{Vertical}
  & UNADJ-W7  
    & 5.282 & 0.435 & 0.496 & 0.924 
    & 32.297 & 0.435 & 0.496 & 0.595 \\
  & UNADJ-W8  
    & 5.882 & 0.431 & 0.526 & 0.897 
    & 33.051 & 0.431 & 0.526 & 0.579 \\
  & UNADJ-W9  
    & 9.177 & 0.215 & 0.511 & 0.579 
    & 37.192 & 0.215 & 0.511 & 0.131 \\
  & UNADJ-W10 
    & 9.490 & 0.229 & 0.544 & 0.574 
    & 37.586 & 0.229 & 0.544 & 0.151 \\
  & UNADJ-W11 
    & 19.549 & 0.406 & 0.508 & 0.689 
    & 50.225 & 0.406 & 0.508 & 0.186 \\
  & UNADJ-W12 
    & 19.875 & 0.405 & 0.529 & 0.673 
    & 50.636 & 0.405 & 0.529 & 0.182 \\
\bottomrule
\end{tabular}
\end{table}

\section{Simulation results for binary outcome in Scenario B1-B3 with smaller number of clusters \(I=9\) }
\label{bin_small}

Tables \ref{tab:sim_B1_small}-\ref{tab:sim_B3_small} report the small-sample binary outcome results for Scenarios B1-B3 with $I = 9$ clusters and $J = 4$ periods. Across all three scenarios, the proposed MRS estimator shows small relative bias and empirical coverage close to 95\% for all four estimands and working models. In the simpler settings (Scenarios B1 and B2), its Monte Carlo variability is very similar, but in some cases slightly larger, than that of the nonparametric UNADJ estimator. The modest loss of efficiency is largely attributable to the greater instability of the working models and the increased likelihood of convergence issues in such small samples. In the more complex setting with informative cluster-period sizes and time-varying treatment effects (Scenario B3), MRS achieves modest efficiency gains, with Monte Carlo standard deviations roughly 10-15\% smaller than those of UNADJ. By contrast, the Coef estimators show substantial and persistent bias with undercoverage, especially for cluster-average estimands and GLMM working models (W9-W12), even in this small-$I$ design. For numerical stability in this small-sample setting, the outcome model coefficients in Scenario B3 were slightly modified from the main-text specification to avoid near-degenerate patterns with all 0 or all 1 outcomes within some clusters or periods. Tables \ref{tab:sim_B1_small_seq}--\ref{tab:sim_B3_small_seq} show a very similar pattern for the case where each sequence includes only one cluster switch.

\section{Simulation for informative cluster size} \label{sim:ics}


To investigate the sensitivity of the proposed test for informative cluster size introduced in main paper Section 4, we conducted additional simulation studies under both continuous and binary outcomes. The objective was to evaluate how the rejection rate of the test responds to increasing degrees of informativeness in cluster-period sizes. The design for continuous outcomes builds upon Scenario \ref{sc:C1}, while the binary outcome simulations extend Scenario  B1, both of which assume non-informative cluster sizes under the null. To induce informativeness, we modified the data generating mechanism so that the individual-level treatment effect explicitly depends on the size of the corresponding cluster-period.

For continuous outcomes, the treatment effect for subject \( k \) in cluster \( i \), period \( j \), was defined as
\[
\theta_{ijk} = 1 + \sin(X_{ijk,1}) + \exp(-X_{ijk,2}) + \delta \left( \frac{\log(N_{ij})N_{ij}^2}{\mathbb{E}(N_{ij})^2} \right), \quad N_{ij} \sim \mathcal{U}(20,100),
\]
The final term introduces dependence on cluster-period size, modulated by the parameter \( \delta \): when \( \delta = 0 \), the design is non-informative; as \( \delta \) increases, larger cluster sizes give systematically stronger treatment effects, thereby creating an informative cluster size structure. Consequently, rejection rates at \( \delta = 0 \) represent the Type I error rate, while higher \( \delta >0 \) values give the power of the test for informative cluster size.

For binary outcomes, we adopted a similar structure with a distinct functional form:
\[
\theta_{ijk} = 0.5 + 0.5\sin(\pi X_{ijk,1}) + \log\left(1 + \frac{1}{3}X_{ijk,1} + X_{ijk,2}^2\right) + \delta\left(\frac{\log(N_{ij})N_{ij}^2}{\mathbb{E}(N_{ij})^2}\right), \quad N_{ij} \sim \mathcal{U}(5,50).
\]
In both outcome settings, we evaluated the performance of three tests: (1) a horizontal test comparing h-iATE vs h-cATE, (2) a vertical test comparing v-iATE vs v-cATE, and (3) a global omnibus test assessing all four estimands simultaneously. All the estimands are estimated using the MRS estimators under working models summarized in Table \ref{tab:working-models}. 1000 Monte Carlo iterations were conducted. The empirical rejection rates for these tests are reported.

\begin{figure}[ht!]
    \centering
    \includegraphics[width=0.8\linewidth]{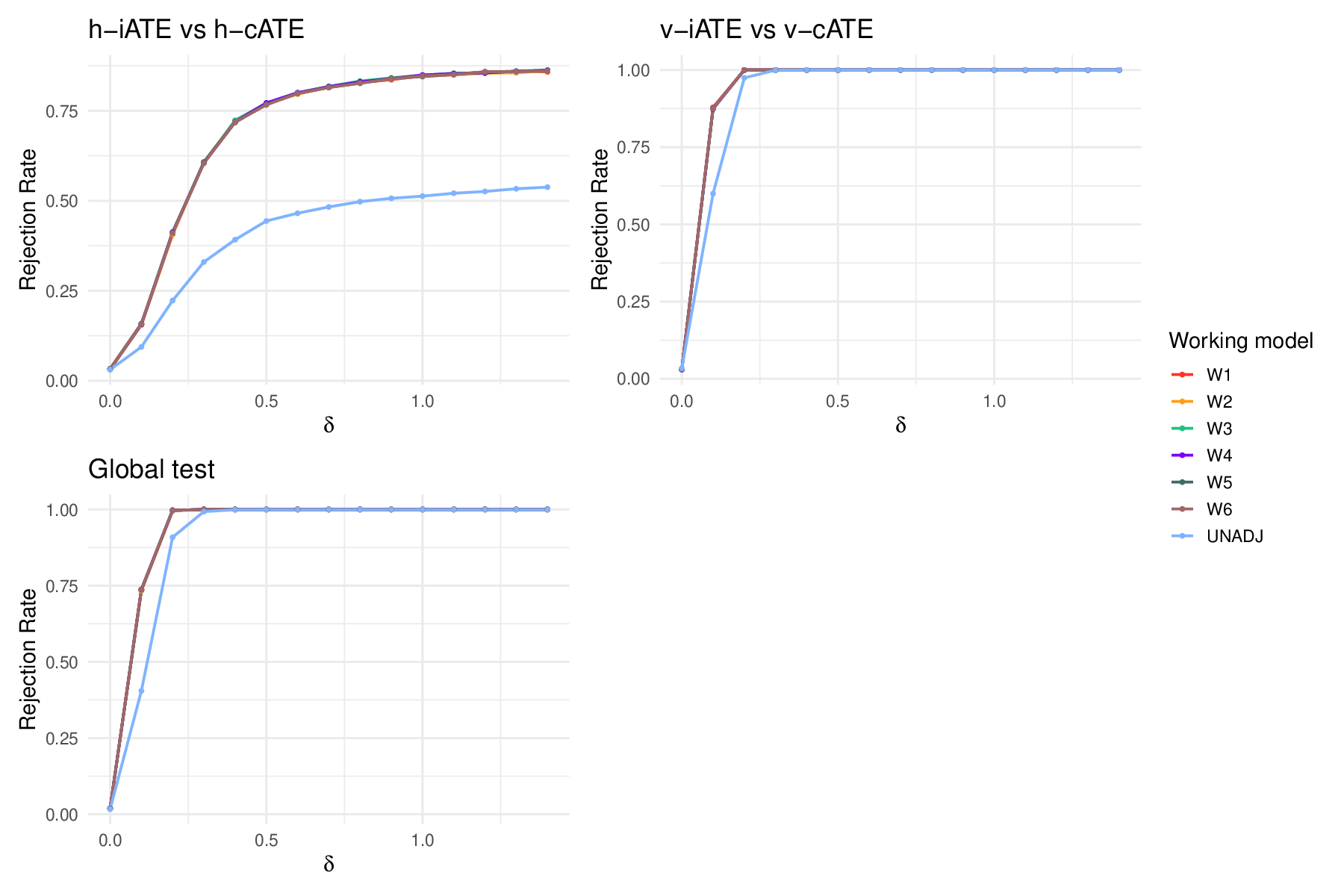}
    \caption{Empirical rejection rates for the pairwise comparison tests for h-iATE versus h-cATE, v-iATE versus v-cATE and omnibus test of informative cluster size under different values of \(\delta\), based on continuous outcomes and working models (W1) through (W6), as well as the unadjusted estimator (UNADJ).}
    \label{fig:info_power_cont}
\end{figure}

Figure \ref{fig:info_power_cont} summarizes the empirical rejection rates of the three tests across increasing values of $\delta$ for continuous outcomes, evaluated under a range of working models. The horizontal test (top-left panel) shows increasing power for all covariate-adjusted estimators (W1-W6) as $\delta$ rises, with rejection rates exceeding 80\% once $\delta > 0.75$. In contrast, the unadjusted method (UNADJ) remains underpowered throughout, plateauing below 55\%. The vertical test (top-right panel) shows high sensitivity to violations of non-informative cluster size vertically, achieving near-perfect rejection (power $\approx$ 95\%) at moderate informativeness levels across all models. The global test (bottom panel), which aggregates evidence across both horizontal and vertical contrasts, shows uniformly strong performance under all working models, with rejection rates approaching 1 as $\delta$ increases.

\begin{figure}[ht!]
    \centering
    \includegraphics[width=0.8\linewidth]{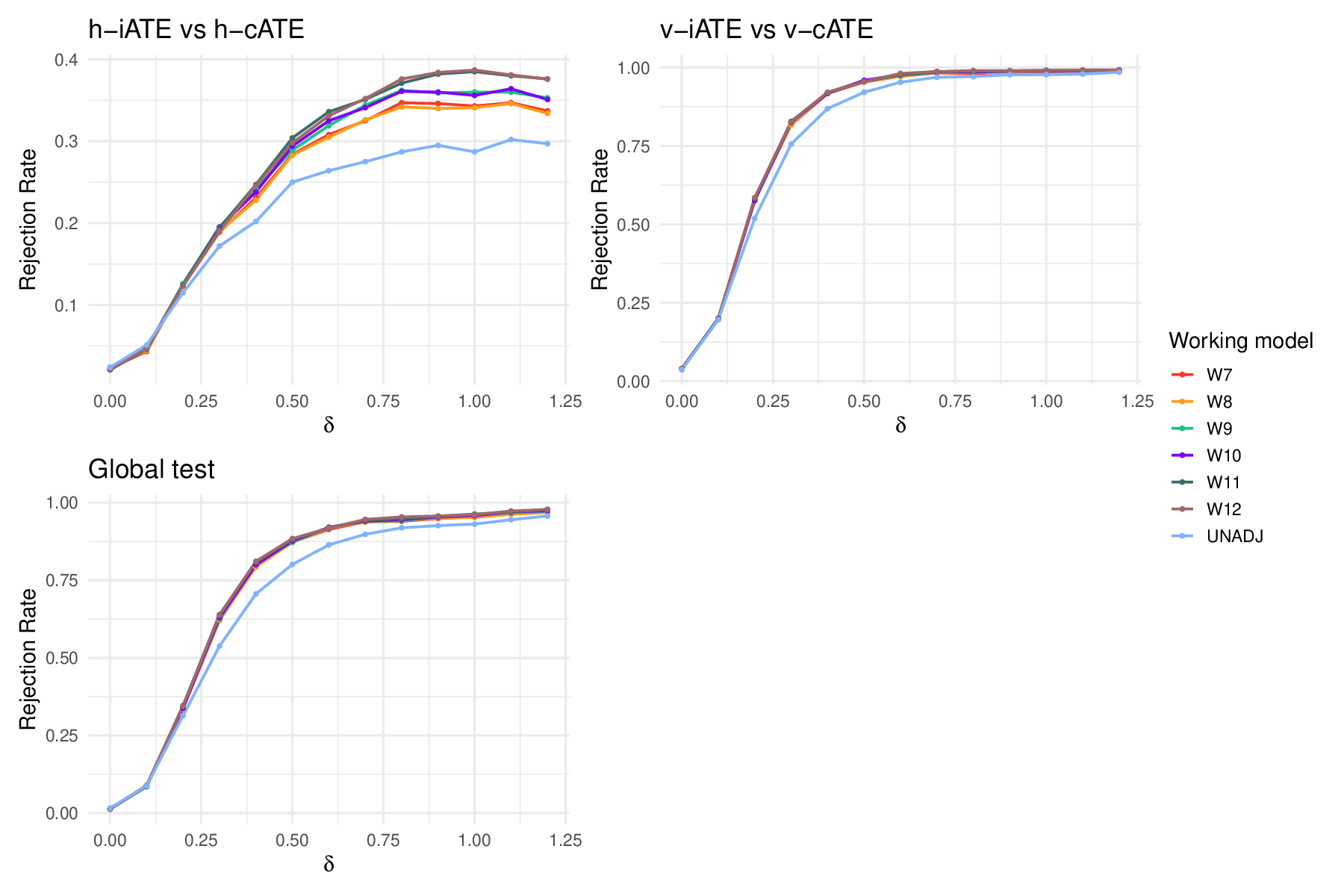}
    \caption{Empirical rejection rates for the pairwise comparison tests for h-iATE versus h-cATE, v-iATE versus v-cATE and omnibus test of informative cluster size under different values of \(\delta\), based on binary outcomes and working models (W7) through (W12), as well as the unadjusted estimator (UNADJ).}
    \label{fig:info_power_bin}
\end{figure}

Figure \ref{fig:info_power_bin} presents the corresponding rejection rates for binary outcomes across all working models (W7-W12). The overall pattern is qualitatively similar, though the horizontal test (top-left panel) exhibits more modest power gains. The vertical test (top-right panel) remains higher sensitive across all settings, showing strong power with small values of $\delta$, and the global test (bottom panel) continues to offer robust performance across all working models and the rejection rate increases until \(1\) as the increase of \(\delta\).

\section{Informative cluster size test of two completed SW-CRTs}
\label{ics:dat_analysis}

\subsection{TSOS data analysis} \label{ics:tsos}
Table \ref{tab:tsos_ics_tests} summarizes the informative cluster size test $p$-values for the TSOS trial under the unadjusted estimator and working models (W1) through (W6). Each model includes three tests: a pairwise comparison between individual- and cluster-average estimands on the horizontal direction (h-iATE vs. h-cATE), a similar comparison on the vertical direction (v-iATE vs. v-cATE), and an omnibus test assessing whether cluster size is informative across all four estimands. All tests were performed using $t$-statistics with 24 degrees of freedom. Across all models, the $p$-values are well above 0.05, indicating no statistically significant evidence of informative cluster size under any model specification. In addition, we also used the ANCOVA III estimators proposed by \citet{chen2024model} The results show no statistically significant treatment effects across any of the four estimands. The point estimates ranged from 0.85 to 1.93, with standard errors between 2.09 and 2.72. All 95\% confidence intervals included zero. For h-iATE, the estimate is 0.95 (SE = 2.12; 95\% CI: $[-3.41,\ 5.32]$); for v-iATE, the estimate is 0.85 (SE = 2.09; 95\% CI: $[-3.47,\ 5.16]$), and for v-cATE, the estimate is 1.93 (SE = 2.72; 95\% CI: $[-3.69,\ 7.54]$). These results are consistent with the results from MRS estimators.

\begin{table}[ht]
\centering
\caption{Informative cluster size test $p$-values for the TSOS trial under working models (W1) through (W6) and unadjusted estimator (UNAJD). Each model includes three tests: a pairwise comparison between horizontal direction estimands (h-iATE vs. h-cATE), a comparison between vertical direction estimands (v-iATE vs.\ v-cATE), and a global omnibus test across all four estimands. All tests are based on $t$-statistics with 24 degrees of freedom. }
\label{tab:tsos_ics_tests}
\begin{tabular}{lccc}
\toprule
\textbf{Model} & \textbf{h-iATE vs h-cATE} & \textbf{v-iATE vs v-cATE} & \textbf{Global test} \\
\midrule
UNAJD & 0.830 & 0.395 & 626\\
(W1)  & 0.950 & 0.408 & 0.743 \\
(W2)  & 0.950 & 0.423 & 0.756 \\
(W3)  & 0.925 & 0.414 & 0.705 \\
(W4)  & 0.926 & 0.440 & 0.738 \\
(W5)  & 0.913 & 0.400 & 0.681 \\
(W6)  & 0.912 & 0.409 & 0.690 \\
\bottomrule
\end{tabular}
\end{table}

\subsection{ACS-QUIK data analysis}
Table \ref{tab:acs_ics_tests} presents the $p$-values from the informative cluster size tests conducted in the ACS-QUIK trial. For each model, three comparisons were performed: a horizontal comparison (h-iATE vs. h-cATE), a vertical comparison (v-iATE vs. v-cATE), and an omnibus test across all four estimands. All tests used $t$-statistics with 62 degrees of freedom. Across all models, the $p$-values were greater than 0.05, indicating no statistically significant evidence of informative cluster size. 

\begin{table}[ht]
\centering
\caption{Informative cluster size test $p$-values for the ACS-QUIK trial under working models (W7) through (W12) and unadjusted estimator (UNAJD), For each model, three tests were conducted: (1) a pairwise comparison between the individual-average and cluster-average estimands on the horizontal direction (h-iATE vs h-cATE), (2) a pairwise comparison on the vertical direction (v-iATE vs. v-cATE), and (3) an omnibus test across all four estimands. All tests were conducted using $t$-statistics with 62 degrees of freedom.}
\label{tab:acs_ics_tests}
\begin{tabular}{lccc}
\toprule
\textbf{Working model} & \textbf{h-iATE vs h-cATE} & \textbf{v-iATE vs v-cATE} & \textbf{Global test} \\
\midrule
UNADJ & 0.474 & 0.501 & 651 \\
(W7)  & 0.384 & 0.442 & 0.517 \\
(W8)  & 0.396 & 0.455 & 0.529 \\
(W9)  & 0.362 & 0.431 & 0.493 \\
(W10) & 0.376 & 0.448 & 0.507 \\
(W11) & 0.370 & 0.439 & 0.498 \\
(W12) & 0.381 & 0.447 & 0.510 \\
\bottomrule
\end{tabular}
\end{table}

\section{Generalization to crossover CRT and parallel-arm CRT} \label{xo_parallel}

\begin{figure}[htbp!]
  \centering
  \begin{adjustbox}{angle=90}
    \begin{minipage}{0.95\textheight}
      \centering
      \includegraphics[width=0.85\textwidth]{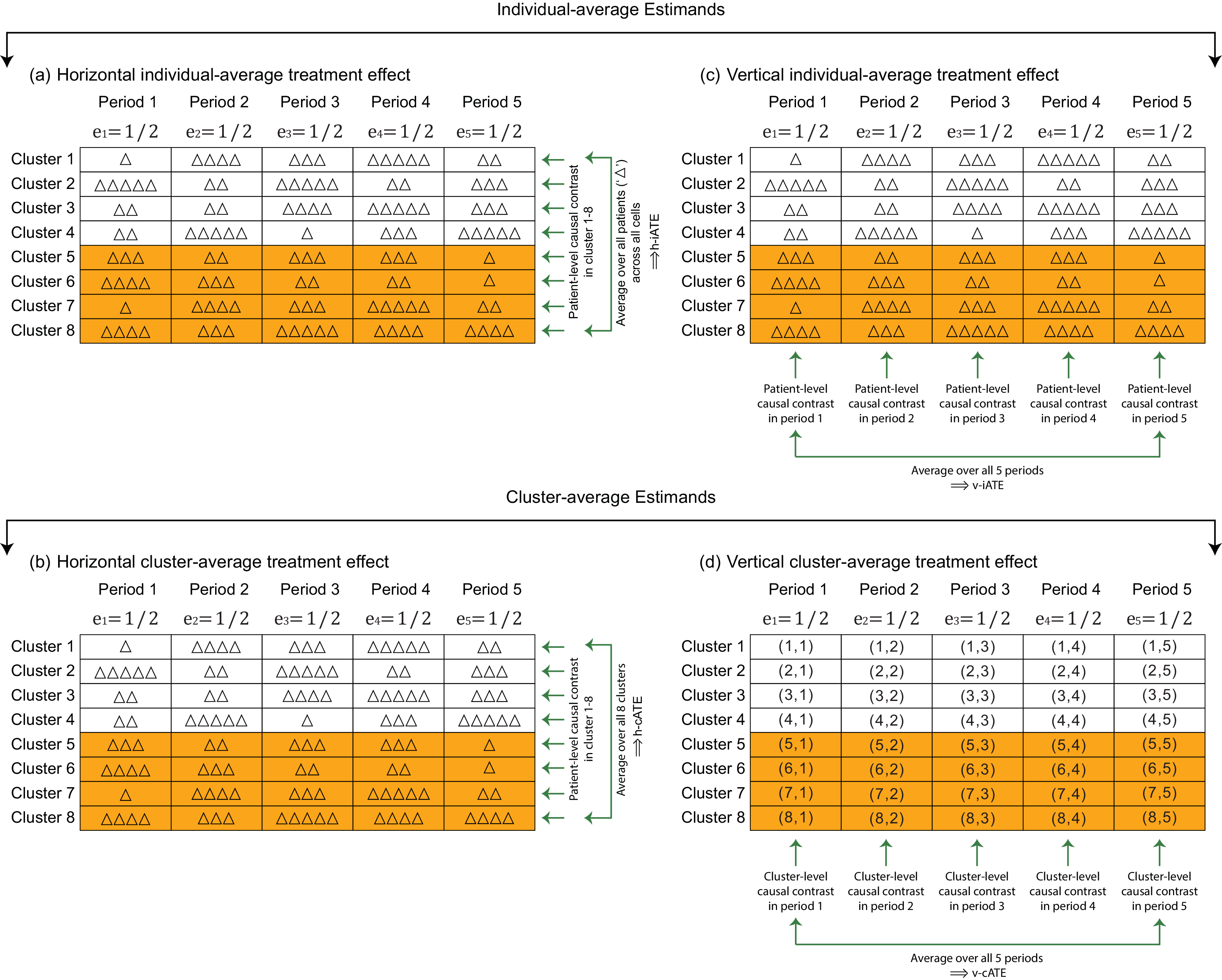}

     \begingroup
    \refstepcounter{figure}
    \tiny         
    \parbox{0.87\textwidth}{%
      \textbf{Figure \thefigure.  \label{fig:paralle_ATE}} A schematic illustration of different treatment effect definitions in a hypothetical two-arm parallel CRT with 8 clusters and 5 periods. White cells in periods 1-5 indicate control cluster-periods and colored cells in periods 1-5 indicate intervention cluster-periods. (a) Horizontal individual-average treatment effect (h-iATE): the average is computed across all individuals in eligible cluster-periods, treating each individual equally. (b) Horizontal cluster-average treatment effect (h-cATE): outcomes are averaged within each cluster across periods, and then across clusters, giving each cluster equal weight. (c) Vertical individual-average treatment effect (v-iATE): individual-average treatment effects are first computed within each period and then averaged across periods, giving equal weight to each period. (d) Vertical cluster-average treatment effect (v-cATE): cluster-period means are averaged across all cluster-periods, giving equal weight to each cluster-period.
    }
  \endgroup

    \end{minipage}
  \end{adjustbox}
\end{figure}

Although Sections \ref{estimand}--\ref{method} are motivated by cross-sectional SW-CRT, the weighted estimand class in \eqref{weight_causal_effect}--\eqref{cp_mean} and the model-robust standardization estimator in \eqref{aug_mu} extend directly to a broader class of longitudinal cluster-randomized trials (LCRTs) in which treatment may vary over calendar periods and is randomized at the cluster level. Two common examples are parallel-arm LCRTs, in which each cluster is randomized once and remains in its assigned treatment throughout follow-up, and cluster-randomized cross-over designs, in which clusters are randomized to a prespecified treatment sequence over periods so that both treatment conditions are observed within each period and, over time, within each cluster. In these designs, the period-specific treatment probabilities \(e_j=\Pr(Z_{ij}=1)\) are known by design and typically satisfy \(e_j\wedge(1-e_j)>0\) for all \(j\), implying that all periods \(j=1,\ldots,J\) are eligible in the definitions of \(\mu_\omega(z)\) and \(\tau_\omega\), in contrast to SW-CRTs where periods with \(e_j\in\{0,1\}\) at \(j=1\) and \(j=J\) are excluded. Figures \ref{fig:paralle_ATE} and \ref{fig:xo_ATE} provide schematic illustrations of the four estimands under parallel-arm and cross-over LCRTs, respectively, and we provide more explicit explantion in the following two subsections.

\subsection{Parallel-arm longitudinal CRTs}\label{app:lcrt_parallel}
In a parallel-arm longitudinal CRT, clusters are randomized once at baseline and remain in their assigned condition throughout follow-up, so that \(Z_{ij}\) is constant in \(j\) for each \(i\) (equivalently, \(Z_{ij}=A_i\) for all \(j=1,\ldots,J\)). The weighted mean potential outcome in \eqref{weight_potential} differs from the SW-CRT definition only in the period range. We denote the individual- and cluster-period weights by \(\omega_{ijk}^{\mathrm{par}}\) and \(\omega_{ij}^{\mathrm{par}}=\sum_{k=1}^{N_{ij}}\omega_{ijk}^{\mathrm{par}}\), respectively. Then, for \(z\in\{0,1\}\),
\[
\mu_{\omega}^{\mathrm{par}}(z)=\frac{\mathbb{E}\!\left[ \sum_{j=1}^{J} \omega_{ij}^{\mathrm{par}} \,\overline{Y}_{ij}(z) \right]}{\mathbb{E}\!\left[\sum_{j=1}^{J} \omega_{ij}^{\mathrm{par}} \right]},
\qquad 
\tau_{\omega}^{\mathrm{par}}=f\{\mu_{\omega}^{\mathrm{par}}(1),\mu_{\omega}^{\mathrm{par}}(0)\},
\]
with \(\overline{Y}_{ij}(z)\) defined in \eqref{cp_mean}. The four estimands in Section \ref{sec:estimands_def} are obtained by the same four weight specifications over \(j=1,\ldots,J\). Specifically, for h-iATE let \(\omega_{ijk}^{\mathrm{par}}=1\), so \(\omega_{ij}^{\mathrm{par}}=N_{ij}\), which each individual contributes equally. For h-cATE, we sets \(\omega_{ijk}^{\mathrm{par}}=1/\sum_{s=1}^{J}N_{is}\), so \(\omega_{ij}^{\mathrm{par}}=N_{ij}/\sum_{s=1}^{J}N_{is}\), which leads to \(\sum_{j=1}^{J}\omega_{ij}^{\mathrm{par}}=1\) and provides a cluster-average estimand in which each cluster contributes equally regardless of its size. For v-iATE, let \(\omega_{ijk}^{\mathrm{par}}=1/\{I\,\mathbb{E}(N_{ij})\}\), so \(\omega_{ij}^{\mathrm{par}}=N_{ij}/\{I\,\mathbb{E}(N_{ij})\}\). This produces a period average of the individual means. For v-cATE we set \(\omega_{ijk}^{\mathrm{par}}=1/N_{ij}\), so \(\omega_{ij}^{\mathrm{par}}=1\), which assigns equal weight to each cluster-period cell.

Under additional conditions, the four estimands may simplify. In particular, if \(\mathbb{E}[N_{ij}]=\overline{N}\) for all \(j\), then the h-iATE and v-iATE coincide, \(\tau_{\mathrm I}^{\mathrm h}=\tau_{\mathrm I}^{\mathrm v}\). If cluster-period sizes are constant within cluster across follow-up (i.e., \(N_{ij}=N_i\) for all \(j\)), then the h-cATE and v-cATE coincide, \(\tau_{\mathrm C}^{\mathrm h}=\tau_{\mathrm C}^{\mathrm v}\), and the four estimands collapse to two, consistent with the estimand provided in \citet{li2025model}.

The model-robust standardization estimator in \eqref{aug_mu} is extended by the same replacement. Writing \(\omega_{ij} = \omega_{ij}^{\mathrm{par}}\) and \(\omega_j=\sum_{i=1}^{I}\omega_{ij}\), the period-specific augmented estimator and its across-period aggregation are
\begin{equation}\label{eq:aug_parallel}
\widehat{\mu}_{\omega,j}^{\text{aug}}(z)=\frac{\sum_{i=1}^{I }\omega_{ij} \, m_{zj}(\underline{\bm{X}}_{ij},N_{ij})}{\omega_j}
+\frac{\sum_{i=1}^{I}\omega_{ij}\mathbb{I}(Z_{ij}=z)\{\overline{Y}_{ij}-m_{zj}(\underline{\bm{X}}_{ij},N_{ij})\}}{\sum_{i=1}^{I}\omega_{ij}\mathbb{I}(Z_{ij}=z)},
\end{equation}
and \(\widehat{\mu}_\omega^{\text{aug}}(z)=\frac{\sum_{j=1}^{J}\omega_j\widehat{\mu}_{\omega,j}^{\text{aug}}(z)}{\sum_{j=1}^{J}\omega_j}\), with \(\widehat{\tau}_\omega^{\text{aug}}=f\{\widehat{\mu}_\omega^{\text{aug}}(1),\widehat{\mu}_\omega^{\text{aug}}(0)\}\). As in Section \ref{method}, \(\widehat{\mu}_\omega^{\text{aug}}(z)\) is consistent for any prespecified working regression \(m_{zj}(\underline{\bm X}_{ij},N_{ij})\).

\subsection{Cluster-randomized cross-over designs}\label{app:lcrt_xo}
\begin{figure}[htbp!]
  \centering
  \begin{adjustbox}{angle=90}
    \begin{minipage}{0.95\textheight}
      \centering
      \includegraphics[width=0.85\textwidth]{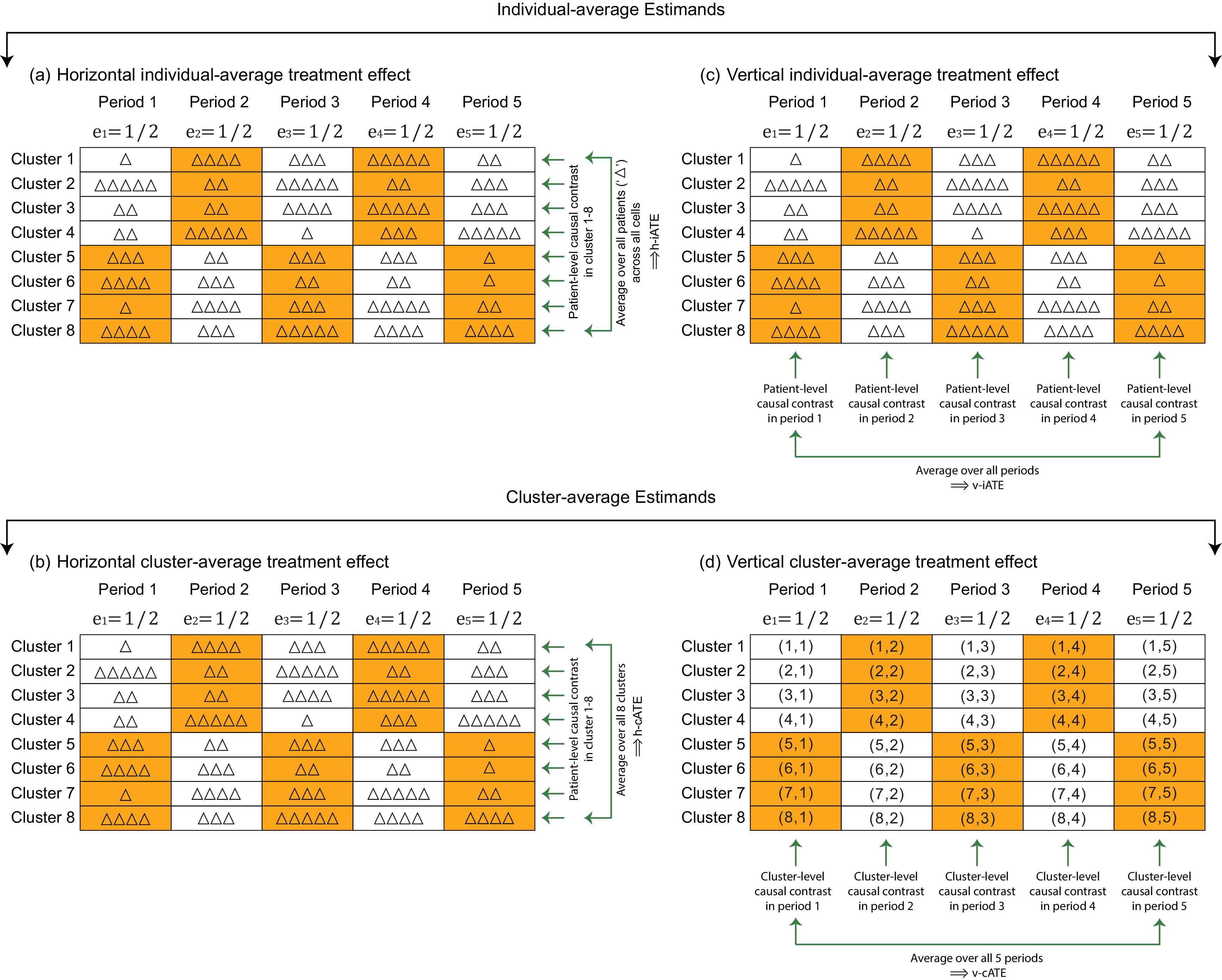}

     \begingroup
    \refstepcounter{figure}
    \tiny         
    \parbox{0.87\textwidth}{%
      \textbf{Figure \thefigure.  \label{fig:xo_ATE}} A schematic illustration of different treatment effect definitions in a hypothetical multiple-period cross-over CRT with 8 clusters and 5 periods. White cells in periods 1-5 indicate control cluster-periods and colored cells in periods 1-5 indicate intervention cluster-periods. (a) Horizontal individual-average treatment effect (h-iATE): the average is computed across all individuals in eligible cluster-periods, treating each individual equally. (b) Horizontal cluster-average treatment effect (h-cATE): outcomes are averaged within each cluster across periods, and then across clusters, giving each cluster equal weight. (c) Vertical individual-average treatment effect (v-iATE): individual-average treatment effects are first computed within each period and then averaged across periods, giving equal weight to each period. (d) Vertical cluster-average treatment effect (v-cATE): cluster-period means are averaged across all cluster-periods, giving equal weight to each cluster-period.
    }
  \endgroup

    \end{minipage}
  \end{adjustbox}
\end{figure}

In a cluster-randomized cross-over LCRT, clusters are randomized to a prespecified treatment sequence over periods such that each period contains both treated and untreated clusters, and clusters contribute observations under both treatment conditions over time. In this design, \(A_i\) indexes the randomized treatment sequence, and such a design ensures \(e_j=\Pr(Z_{ij}=1)=1/2\) for all \(j=1,\ldots,J\)). As a result, all periods contributes in the definition of the estimand, analogous to \eqref{weight_potential}. Denoting the weights by \(\omega_{ijk}^{\mathrm{xo}}\) and \(\omega_{ij}^{\mathrm{xo}}=\sum_{k=1}^{N_{ij}}\omega_{ijk}^{\mathrm{xo}}\), the estimand is
\[
\mu_{\omega}^{\mathrm{xo}}(z)=\frac{\mathbb{E}\!\left[ \sum_{j=1}^{J} \omega_{ij}^{\mathrm{xo}} \,\overline{Y}_{ij}(z) \right]}{\mathbb{E}\!\left[\sum_{j=1}^{J} \omega_{ij}^{\mathrm{xo}} \right]},
\qquad 
\tau_{\omega}^{\mathrm{xo}}=f\{\mu_{\omega}^{\mathrm{xo}}(1),\mu_{\omega}^{\mathrm{xo}}(0)\}.
\]
The four estimands are generated by the same four weight specifications used above (h-iATE, h-cATE, v-iATE, and v-cATE), now interpreted in the presence of within-cluster treatment alternation. Figure \ref{fig:xo_ATE} provides a schematic illustration. The cross-over feature does not change the form of the estimands. The vertical estimands (v-iATE and v-cATE) may be viewed as uniformly averaging period-specific causal contrasts. Even though \(e_j\) is constant across \(j\), the distinction between horizontal and vertical aggregation remains meaningful whenever cluster-period sizes are informative or treatment effects vary over calendar time.

Estimation proceeds exactly as in the parallel-arm case with \(j=1,\ldots,J\). In particular, the model-robust standardization estimator is given with \(\omega_{ij} = \omega_{ij}^{\mathrm{xo}}\) and \(\omega_j=\sum_{i=1}^{I}\omega_{ij}\), yielding \(\widehat{\tau}_{\omega}^{\text{aug}}=f\{\widehat{\mu}_{\omega}^{\text{aug}}(1),\widehat{\mu}_{\omega}^{\text{aug}}(0)\}\). The summary of the estimands in these three type of design are summarized in Table \ref{tab:three_designs_four_estimands}.

\begin{sidewaystable}[htbp]
\centering
\caption{Summary of the four estimands under three longitudinal cluster-randomized trial designs.}
\label{tab:three_designs_four_estimands}
\renewcommand{\arraystretch}{2}
\setlength{\tabcolsep}{10pt}
\begin{tabular}{>{\centering\arraybackslash}m{1.2cm}
                >{\centering\arraybackslash}m{5.3cm}
                >{\centering\arraybackslash}m{5.3cm}
                >{\centering\arraybackslash}m{5.3cm}}
\toprule
\textbf{Estimands} & \textbf{SW-CRT} & \textbf{Parallel-arm LCRT} & \textbf{Cross-over LCRT} \\
\midrule

\textbf{\(e_j\)} &
\(\begin{aligned}
&e_1=0,\ e_J=1,\\
&e_j\in(0,1),\ j=2,\ldots,J-1
\end{aligned}\) &
\(\begin{aligned}
&e_j=\frac{1}{2},\ j=1,\ldots,J
\end{aligned}\) &
\(\begin{aligned}
&e_j=\frac{1}{2},\ j=1,\ldots,J
\end{aligned}\) \\
\cdashline{1-4}

\textbf{h-iATE} &
\(\begin{aligned}
&\omega_{ijk}=1,\ \omega_{ij}=N_{ij},\ \omega_j=\sum_{i=1}^{I}N_{ij};\\
&\mu_{\mathrm I}^{\mathrm h}(z)=
\frac{\sum_{j=2}^{J-1}\mathbb E\!\left[\sum_{k=1}^{N_{ij}}Y_{ijk}(z)\right]}
{\sum_{j=2}^{J-1}\mathbb E[N_{ij}] }.
\end{aligned}\) &
\(\begin{aligned}
&\omega_{ijk}=1,\ \omega_{ij}=N_{ij},\ \omega_j=\sum_{i=1}^{I}N_{ij};\\
&\mu_{\mathrm I}^{\mathrm h}(z)=
\frac{\sum_{j=1}^{J}\mathbb E\!\left[\sum_{k=1}^{N_{ij}}Y_{ijk}(z)\right]}
{\sum_{j=1}^{J}\mathbb E[N_{ij}] }.
\end{aligned}\) &
\(\begin{aligned}
&\omega_{ijk}=1,\ \omega_{ij}=N_{ij},\ \omega_j=\sum_{i=1}^{I}N_{ij};\\
&\mu_{\mathrm I}^{\mathrm h}(z)=
\frac{\sum_{j=1}^{J}\mathbb E\!\left[\sum_{k=1}^{N_{ij}}Y_{ijk}(z)\right]}
{\sum_{j=1}^{J}\mathbb E[N_{ij}] }.
\end{aligned}\) \\
\cdashline{1-4}

\textbf{h-cATE} &
\(\begin{aligned}
&\omega_{ijk}=\frac{1}{\sum_{s=2}^{J-1}N_{is}},\ 
\omega_{ij}=\frac{N_{ij}}{\sum_{s=2}^{J-1}N_{is}},\\
&\omega_j=\sum_{i=1}^{I}\frac{N_{ij}}{\sum_{s=2}^{J-1}N_{is}};\\
&\mu_{\mathrm C}^{\mathrm h}(z)=
\mathbb E\!\left[\frac{\sum_{j=2}^{J-1}N_{ij}\overline Y_{ij}(z)}
{\sum_{j=2}^{J-1}N_{ij}}\right].
\end{aligned}\) &
\(\begin{aligned}
&\omega_{ijk}=\frac{1}{\sum_{s=1}^{J}N_{is}},\ 
\omega_{ij}=\frac{N_{ij}}{\sum_{s=1}^{J}N_{is}},\\
&\omega_j=\sum_{i=1}^{I}\frac{N_{ij}}{\sum_{s=1}^{J}N_{is}};\\
&\mu_{\mathrm C}^{\mathrm h}(z)=
\mathbb E\!\left[\frac{\sum_{j=1}^{J}N_{ij}\overline Y_{ij}(z)}
{\sum_{j=1}^{J}N_{ij}}\right].
\end{aligned}\) &
\(\begin{aligned}
&\omega_{ijk}=\frac{1}{\sum_{s=1}^{J}N_{is}},\ 
\omega_{ij}=\frac{N_{ij}}{\sum_{s=1}^{J}N_{is}},\\
&\omega_j=\sum_{i=1}^{I}\frac{N_{ij}}{\sum_{s=1}^{J}N_{is}};\\
&\mu_{\mathrm C}^{\mathrm h}(z)=
\mathbb E\!\left[\frac{\sum_{j=1}^{J}N_{ij}\overline Y_{ij}(z)}
{\sum_{j=1}^{J}N_{ij}}\right].
\end{aligned}\) \\
\cdashline{1-4}

\textbf{v-iATE} &
\(\begin{aligned}
&\omega_{ijk}=\frac{1}{I\,\mathbb E[N_{ij}]},\ 
\omega_{ij}=\frac{N_{ij}}{I\,\mathbb E[N_{ij}]},\ 
\omega_j=1;\\
&\mu_{\mathrm I}^{\mathrm v}(z)=\frac{1}{J-2}\sum_{j=2}^{J-1}
\frac{\mathbb E\!\left[\sum_{k=1}^{N_{ij}}Y_{ijk}(z)\right]}{\mathbb E[N_{ij}]}.
\end{aligned}\) &
\(\begin{aligned}
&\omega_{ijk}=\frac{1}{I\,\mathbb E[N_{ij}]},\ 
\omega_{ij}=\frac{N_{ij}}{I\,\mathbb E[N_{ij}]},\ 
\omega_j=1;\\
&\mu_{\mathrm I}^{\mathrm v}(z)=\frac{1}{J}\sum_{j=1}^{J}
\frac{\mathbb E\!\left[\sum_{k=1}^{N_{ij}}Y_{ijk}(z)\right]}{\mathbb E[N_{ij}]}.
\end{aligned}\) &
\(\begin{aligned}
&\omega_{ijk}=\frac{1}{I\,\mathbb E[N_{ij}]},\ 
\omega_{ij}=\frac{N_{ij}}{I\,\mathbb E[N_{ij}]},\ 
\omega_j=1;\\
&\mu_{\mathrm I}^{\mathrm v}(z)=\frac{1}{J}\sum_{j=1}^{J}
\frac{\mathbb E\!\left[\sum_{k=1}^{N_{ij}}Y_{ijk}(z)\right]}{\mathbb E[N_{ij}]}.
\end{aligned}\) \\
\cdashline{1-4}

\textbf{v-cATE} &
\(\begin{aligned}
&\omega_{ijk}=\frac{1}{N_{ij}},\ 
\omega_{ij}=1,\ 
\omega_j=1;\\
&\mu_{\mathrm C}^{\mathrm v}(z)=\frac{1}{J-2}\sum_{j=2}^{J-1}\mathbb E\!\left[\overline Y_{ij}(z)\right].
\end{aligned}\) &
\(\begin{aligned}
&\omega_{ijk}=\frac{1}{N_{ij}},\ 
\omega_{ij}=1,\ 
\omega_j=1;\\
&\mu_{\mathrm C}^{\mathrm v}(z)=\frac{1}{J}\sum_{j=1}^{J}\mathbb E\!\left[\overline Y_{ij}(z)\right].
\end{aligned}\) &
\(\begin{aligned}
&\omega_{ijk}=\frac{1}{N_{ij}},\ 
\omega_{ij}=1,\ 
\omega_j=1;\\
&\mu_{\mathrm C}^{\mathrm v}(z)=\frac{1}{J}\sum_{j=1}^{J}\mathbb E\!\left[\overline Y_{ij}(z)\right].
\end{aligned}\) \\
\bottomrule
\end{tabular}
\end{sidewaystable}

\begin{table}[ht!]
\centering
\caption{
Simulation results in Scenario B1 for estimating four estimands under a binary outcome with $I=9$ clusters and $J=4$ periods using covariate-adjusted working models (W7)--(W12) with the Coef, MRS and nonparametric (UNADJ) estimators. MRS: proposed augmented estimator; Coef: treatment‐effect coefficients from covariate‐adjusted working model; UNADJ: nonparametric estimator. RBias: absolute bias; AESE: average estimated standard error; MCSD: Monte Carlo standard deviation; CP: 95\% confidence interval coverage.
}
\label{tab:sim_B1_small}
\begin{tabular}{cclrrrrrrrr}
\toprule
\multirow{2}{*}{Direction} & \multirow{2}{*}{Working model}
  & \multirow{2}{*}{Method}
  & \multicolumn{4}{c}{h-iATE}
  & \multicolumn{4}{c}{h-cATE} \\
\cmidrule(lr){4-7}\cmidrule(lr){8-11}
 & & 
 & RBias & AESE & MCSD & CP 
 & RBias & AESE & MCSD & CP \\
 & & 
 & \multicolumn{4}{c}{\(\Delta_{\mathrm{I}}^{\mathrm{h}}=0.884\)} 
 & \multicolumn{4}{c}{\(\Delta_{\mathrm{C}}^{\mathrm{h}}=0.884\)} \\
\midrule
\multirow{13}{*}{Horizontal}
  & \textbackslash & UNADJ 
    & 3.015 & 0.497 & 0.471 & 0.944 
    & 2.962 & 0.488 & 0.466 & 0.947 \\
  & \multirow{2}{*}{W7}
    & Coef   
      & 8.154 & 0.460 & 0.502 & 0.932 
      & 8.160 & 0.460 & 0.502 & 0.932 \\
  &               & MRS    
      & 3.180 & 0.507 & 0.478 & 0.949 
      & 3.143 & 0.501 & 0.475 & 0.953 \\
  & \multirow{2}{*}{W8}
    & Coef   
      & 9.486 & 0.454 & 0.516 & 0.911 
      & 9.491 & 0.454 & 0.515 & 0.913 \\
  &               & MRS    
      & 3.064 & 0.507 & 0.477 & 0.950 
      & 3.044 & 0.501 & 0.475 & 0.952 \\
  & \multirow{2}{*}{W9}
    & Coef   
      & 15.175 & 0.291 & 0.479 & 0.840 
      & 15.181 & 0.291 & 0.479 & 0.840 \\
  &               & MRS    
      & 2.948 & 0.506 & 0.478 & 0.949 
      & 2.881 & 0.498 & 0.474 & 0.950 \\
  & \multirow{2}{*}{W10}
    & Coef   
      & 15.948 & 0.296 & 0.487 & 0.839 
      & 15.969 & 0.296 & 0.485 & 0.836 \\
  &               & MRS    
      & 2.884 & 0.506 & 0.478 & 0.952 
      & 2.824 & 0.499 & 0.474 & 0.953 \\
  & \multirow{2}{*}{W11}
    & Coef   
      & 17.239 & 0.397 & 0.472 & 0.934 
      & 17.245 & 0.397 & 0.472 & 0.934 \\
  &               & MRS    
      & 3.073 & 0.505 & 0.477 & 0.950 
      & 3.031 & 0.497 & 0.473 & 0.950 \\
  & \multirow{2}{*}{W12}
    & Coef   
      & 17.813 & 0.392 & 0.483 & 0.922 
      & 17.823 & 0.392 & 0.482 & 0.922 \\
  &               & MRS    
      & 3.016 & 0.505 & 0.477 & 0.953 
      & 2.989 & 0.498 & 0.473 & 0.953 \\
\toprule
& & 
  & \multicolumn{4}{c}{v-iATE}
  & \multicolumn{4}{c}{v-cATE} \\
  \cmidrule(lr){4-7}\cmidrule(lr){8-11}
 & & 
 & RBias & AESE & MCSD & CP 
 & RBias & AESE & MCSD & CP \\
& & 
 & \multicolumn{4}{c}{\(\Delta_{\mathrm{I}}^{\mathrm{v}}=0.884\)} 
 & \multicolumn{4}{c}{\(\Delta_{\mathrm{C}}^{\mathrm{v}}=0.884\)} \\
 \midrule
\multirow{13}{*}{Vertical}
  & \textbackslash & UNADJ 
    & 3.023 & 0.499 & 0.469 & 0.946 
    & 2.706 & 0.485 & 0.465 & 0.953 \\
  & \multirow{2}{*}{W7}
    & Coef   
      & 8.154 & 0.460 & 0.502 & 0.932 
      & 8.124 & 0.460 & 0.502 & 0.932 \\
  &               & MRS    
      & 3.170 & 0.509 & 0.475 & 0.952 
      & 2.843 & 0.497 & 0.475 & 0.957 \\
  & \multirow{2}{*}{W8}
    & Coef   
      & 9.469 & 0.454 & 0.513 & 0.919 
      & 9.438 & 0.454 & 0.513 & 0.919 \\
  &               & MRS    
      & 3.040 & 0.509 & 0.475 & 0.951 
      & 2.725 & 0.497 & 0.475 & 0.959 \\
  & \multirow{2}{*}{W9}
    & Coef   
      & 15.175 & 0.291 & 0.479 & 0.840 
      & 15.143 & 0.291 & 0.479 & 0.840 \\
  &               & MRS    
      & 2.978 & 0.508 & 0.475 & 0.952 
      & 2.701 & 0.493 & 0.474 & 0.959 \\
  & \multirow{2}{*}{W10}
    & Coef   
      & 15.885 & 0.297 & 0.484 & 0.839 
      & 15.853 & 0.297 & 0.484 & 0.840 \\
  &               & MRS    
      & 2.905 & 0.508 & 0.475 & 0.951 
      & 2.635 & 0.494 & 0.474 & 0.957 \\
  & \multirow{2}{*}{W11}
    & Coef   
      & 17.239 & 0.397 & 0.472 & 0.934 
      & 17.206 & 0.397 & 0.472 & 0.934 \\
  &               & MRS    
      & 3.090 & 0.507 & 0.474 & 0.952 
      & 2.863 & 0.492 & 0.473 & 0.956 \\
  & \multirow{2}{*}{W12}
    & Coef   
      & 17.775 & 0.393 & 0.480 & 0.923 
      & 17.743 & 0.393 & 0.480 & 0.924 \\
  &               & MRS    
      & 3.027 & 0.508 & 0.474 & 0.952 
      & 2.820 & 0.493 & 0.472 & 0.955 \\
\bottomrule
\end{tabular}
\end{table}

\begin{table}[ht!]
\centering
\caption{
Simulation results in Scenario B2 for estimating four estimands under a binary outcome with $I=9$ clusters and $J=4$ periods using covariate-adjusted working models (W7)--(W12) with the Coef, MRS and nonparametric (UNADJ) estimators. MRS: proposed augmented estimator; Coef: treatment‐effect coefficients from covariate‐adjusted working model; UNADJ: nonparametric estimator. RBias: absolute bias; AESE: average estimated standard error; MCSD: Monte Carlo standard deviation; CP: 95\% confidence interval coverage.
}
\label{tab:sim_B2_small}
\begin{tabular}{cclrrrrrrrr}
\toprule
\multirow{2}{*}{Direction} & \multirow{2}{*}{Working model}
  & \multirow{2}{*}{Method}
  & \multicolumn{4}{c}{h-iATE}
  & \multicolumn{4}{c}{h-cATE} \\
\cmidrule(lr){4-7}\cmidrule(lr){8-11}
 & & 
 & RBias & AESE & MCSD & CP 
 & RBias & AESE & MCSD & CP \\
 & & 
 & \multicolumn{4}{c}{\(\Delta_{\mathrm{I}}^{\mathrm{h}}=1.834\)} 
 & \multicolumn{4}{c}{\(\Delta_{\mathrm{C}}^{\mathrm{h}}=1.596\)} \\
\midrule
\multirow{13}{*}{Horizontal}
  & \textbackslash & UNADJ 
    & 2.390 & 0.660 & 0.554 & 0.954 
    & 1.404 & 0.654 & 0.559 & 0.959 \\
  & \multirow{2}{*}{W7}
    & Coef   
      & 6.133 & 0.556 & 0.569 & 0.957 
      & 21.957 & 0.556 & 0.569 & 0.918 \\
  &               & MRS    
      & 2.365 & 0.631 & 0.523 & 0.964 
      & 1.532 & 0.628 & 0.531 & 0.962 \\
  & \multirow{2}{*}{W8}
    & Coef   
      & 9.070 & 0.589 & 0.604 & 0.946 
      & 25.234 & 0.588 & 0.604 & 0.920 \\
  &               & MRS    
      & 2.307 & 0.628 & 0.523 & 0.964 
      & 1.609 & 0.629 & 0.532 & 0.962 \\
  & \multirow{2}{*}{W9}
    & Coef   
      & 9.456 & 0.347 & 0.587 & 0.830 
      & 25.776 & 0.347 & 0.587 & 0.746 \\
  &               & MRS    
      & 2.259 & 0.632 & 0.518 & 0.964 
      & 1.560 & 0.629 & 0.528 & 0.965 \\
  & \multirow{2}{*}{W10}
    & Coef   
      & 12.339 & 0.365 & 0.619 & 0.830 
      & 29.005 & 0.364 & 0.620 & 0.737 \\
  &               & MRS    
      & 2.194 & 0.629 & 0.518 & 0.965 
      & 1.678 & 0.630 & 0.529 & 0.962 \\
  & \multirow{2}{*}{W11}
    & Coef   
      & 9.365 & 0.502 & 0.594 & 0.919 
      & 25.671 & 0.502 & 0.594 & 0.886 \\
  &               & MRS    
      & 2.132 & 0.624 & 0.516 & 0.968 
      & 1.713 & 0.622 & 0.526 & 0.963 \\
  & \multirow{2}{*}{W12}
    & Coef   
      & 12.298 & 0.498 & 0.625 & 0.899 
      & 28.998 & 0.498 & 0.626 & 0.868 \\
  &               & MRS    
      & 2.079 & 0.621 & 0.516 & 0.966 
      & 1.800 & 0.623 & 0.526 & 0.962 \\
\toprule
& & 
  & \multicolumn{4}{c}{v-iATE}
  & \multicolumn{4}{c}{v-cATE} \\
  \cmidrule(lr){4-7}\cmidrule(lr){8-11}
 & & 
 & RBias & AESE & MCSD & CP 
 & RBias & AESE & MCSD & CP \\
& & 
 & \multicolumn{4}{c}{\(\Delta_{\mathrm{I}}^{\mathrm{v}}=1.834\)} 
 & \multicolumn{4}{c}{\(\Delta_{\mathrm{C}}^{\mathrm{v}}=1.372\)} \\
 \midrule
\multirow{13}{*}{Vertical}
  & \textbackslash & UNADJ 
    & 3.706 & 0.669 & 0.558 & 0.953 
    & 3.192 & 0.630 & 0.553 & 0.962 \\
  & \multirow{2}{*}{W7}
    & Coef   
      & 6.145 & 0.556 & 0.569 & 0.957 
      & 41.845 & 0.556 & 0.569 & 0.857 \\
  &               & MRS    
      & 3.732 & 0.639 & 0.527 & 0.962 
      & 3.295 & 0.610 & 0.531 & 0.961 \\
  & \multirow{2}{*}{W8}
    & Coef   
      & 7.517 & 0.585 & 0.602 & 0.946 
      & 43.679 & 0.585 & 0.602 & 0.868 \\
  &               & MRS    
      & 3.699 & 0.635 & 0.527 & 0.961 
      & 3.327 & 0.615 & 0.531 & 0.963 \\
  & \multirow{2}{*}{W9}
    & Coef   
      & 9.469 & 0.347 & 0.587 & 0.830 
      & 46.287 & 0.347 & 0.587 & 0.614 \\
  &               & MRS    
      & 3.616 & 0.640 & 0.522 & 0.962 
      & 3.364 & 0.609 & 0.529 & 0.962 \\
  & \multirow{2}{*}{W10}
    & Coef   
      & 10.735 & 0.363 & 0.614 & 0.833 
      & 47.979 & 0.363 & 0.614 & 0.624 \\
  &               & MRS    
      & 3.573 & 0.637 & 0.522 & 0.961 
      & 3.450 & 0.615 & 0.529 & 0.962 \\
  & \multirow{2}{*}{W11}
    & Coef   
      & 9.378 & 0.502 & 0.594 & 0.919 
      & 46.165 & 0.502 & 0.594 & 0.812 \\
  &               & MRS    
      & 3.505 & 0.631 & 0.521 & 0.965 
      & 3.523 & 0.604 & 0.527 & 0.962 \\
  & \multirow{2}{*}{W12}
    & Coef   
      & 10.534 & 0.496 & 0.619 & 0.906 
      & 47.710 & 0.496 & 0.619 & 0.796 \\
  &               & MRS    
      & 3.471 & 0.628 & 0.521 & 0.966 
      & 3.571 & 0.609 & 0.527 & 0.962 \\
\bottomrule
\end{tabular}
\end{table}

\begin{table}[ht!]
\centering
\caption{
Simulation results in Scenario B3 for estimating four estimands under a binary outcome with $I=9$ clusters and $J=4$ periods using covariate-adjusted working models (W7)--(W12) with the Coef, MRS and nonparametric (UNADJ) estimators. MRS: proposed augmented estimator; Coef: treatment‐effect coefficients from covariate‐adjusted working model; UNADJ: nonparametric estimator. RBias: absolute bias; AESE: average estimated standard error; MCSD: Monte Carlo standard deviation; CP: 95\% confidence interval coverage.
}
\label{tab:sim_B3_small}
\begin{tabular}{cclrrrrrrrr}
\toprule
\multirow{2}{*}{Direction} & \multirow{2}{*}{Working model}
  & \multirow{2}{*}{Method}
  & \multicolumn{4}{c}{h-iATE}
  & \multicolumn{4}{c}{h-cATE} \\
\cmidrule(lr){4-7}\cmidrule(lr){8-11}
 & & 
 & RBias & AESE & MCSD & CP 
 & RBias & AESE & MCSD & CP \\
 & & 
 & \multicolumn{4}{c}{\(\Delta_{\mathrm{I}}^{\mathrm{h}}=2.123\)} 
 & \multicolumn{4}{c}{\(\Delta_{\mathrm{C}}^{\mathrm{h}}=1.897\)} \\
\midrule
\multirow{13}{*}{Horizontal}
  & \textbackslash & UNADJ 
    & 2.539 & 0.610 & 0.529 & 0.943 
    & 0.038 & 0.601 & 0.527 & 0.955 \\
  & \multirow{2}{*}{W7}
    & Coef   
      & 7.501 & 0.498 & 0.495 & 0.951 
      & 20.322 & 0.498 & 0.495 & 0.904 \\
  &               & MRS    
      & 2.262 & 0.562 & 0.471 & 0.965 
      & 0.556 & 0.556 & 0.471 & 0.966 \\
  & \multirow{2}{*}{W8}
    & Coef   
      & 10.381 & 0.541 & 0.535 & 0.954 
      & 23.562 & 0.541 & 0.535 & 0.917 \\
  &               & MRS    
      & 1.981 & 0.560 & 0.468 & 0.960 
      & 0.857 & 0.556 & 0.469 & 0.965 \\
  & \multirow{2}{*}{W9}
    & Coef   
      & 9.561 & 0.281 & 0.558 & 0.754 
      & 22.628 & 0.281 & 0.558 & 0.649 \\
  &               & MRS    
      & 2.191 & 0.564 & 0.470 & 0.963 
      & 0.479 & 0.556 & 0.470 & 0.967 \\
  & \multirow{2}{*}{W10}
    & Coef   
      & 12.234 & 0.296 & 0.597 & 0.749 
      & 25.648 & 0.296 & 0.598 & 0.632 \\
  &               & MRS    
      & 1.879 & 0.562 & 0.468 & 0.962 
      & 0.802 & 0.556 & 0.468 & 0.962 \\
  & \multirow{2}{*}{W11}
    & Coef   
      & 15.108 & 0.465 & 0.551 & 0.878 
      & 28.836 & 0.465 & 0.551 & 0.808 \\
  &               & MRS    
      & 1.687 & 0.553 & 0.462 & 0.964 
      & 1.057 & 0.547 & 0.464 & 0.963 \\
  & \multirow{2}{*}{W12}
    & Coef   
      & 18.505 & 0.465 & 0.583 & 0.854 
      & 32.670 & 0.465 & 0.583 & 0.791 \\
  &               & MRS    
      & 1.575 & 0.547 & 0.459 & 0.967 
      & 1.132 & 0.543 & 0.461 & 0.959 \\
\toprule
& & 
  & \multicolumn{4}{c}{v-iATE}
  & \multicolumn{4}{c}{v-cATE} \\
  \cmidrule(lr){4-7}\cmidrule(lr){8-11}
 & & 
 & RBias & AESE & MCSD & CP 
 & RBias & AESE & MCSD & CP \\
& & 
 & \multicolumn{4}{c}{\(\Delta_{\mathrm{I}}^{\mathrm{v}}=2.107\)} 
 & \multicolumn{4}{c}{\(\Delta_{\mathrm{C}}^{\mathrm{v}}=1.660\)} \\
 \midrule
\multirow{13}{*}{Vertical}
  & \textbackslash & UNADJ 
    & 3.861 & 0.625 & 0.543 & 0.930 
    & 1.645 & 0.587 & 0.524 & 0.947 \\
  & \multirow{2}{*}{W7}
    & Coef   
      & 8.335 & 0.498 & 0.495 & 0.952 
      & 37.524 & 0.498 & 0.495 & 0.831 \\
  &               & MRS    
      & 3.514 & 0.572 & 0.479 & 0.948 
      & 2.326 & 0.543 & 0.466 & 0.954 \\
  & \multirow{2}{*}{W8}
    & Coef   
      & 8.912 & 0.534 & 0.524 & 0.955 
      & 38.256 & 0.534 & 0.524 & 0.857 \\
  &               & MRS    
      & 3.258 & 0.570 & 0.475 & 0.956 
      & 2.605 & 0.546 & 0.464 & 0.956 \\
  & \multirow{2}{*}{W9}
    & Coef   
      & 10.412 & 0.281 & 0.558 & 0.749 
      & 40.160 & 0.281 & 0.558 & 0.492 \\
  &               & MRS    
      & 3.432 & 0.572 & 0.477 & 0.956 
      & 2.374 & 0.544 & 0.462 & 0.963 \\
  & \multirow{2}{*}{W10}
    & Coef   
      & 10.650 & 0.295 & 0.584 & 0.758 
      & 40.462 & 0.295 & 0.584 & 0.515 \\
  &               & MRS    
      & 3.146 & 0.570 & 0.474 & 0.958 
      & 2.654 & 0.546 & 0.461 & 0.963 \\
  & \multirow{2}{*}{W11}
    & Coef   
      & 16.001 & 0.465 & 0.551 & 0.875 
      & 47.256 & 0.465 & 0.551 & 0.707 \\
  &               & MRS    
      & 2.936 & 0.560 & 0.467 & 0.958 
      & 2.825 & 0.535 & 0.455 & 0.965 \\
  & \multirow{2}{*}{W12}
    & Coef   
      & 16.602 & 0.461 & 0.568 & 0.868 
      & 48.018 & 0.461 & 0.568 & 0.707 \\
  &               & MRS    
      & 2.836 & 0.553 & 0.463 & 0.966 
      & 2.793 & 0.533 & 0.453 & 0.963 \\
\bottomrule
\end{tabular}
\end{table}

\begin{table}[ht!]
\centering
\caption{
Simulation results in Scenario B1 for estimating four estimands under a binary outcome with $I=9$ clusters and $J=10$ periods (switching one cluster per sequence) using covariate-adjusted working models (W7)--(W12) with the Coef, MRS and nonparametric (UNADJ) estimators. MRS: proposed augmented estimator; Coef: treatment‐effect coefficients from covariate‐adjusted working model; UNADJ: nonparametric estimator. RBias: absolute bias; AESE: average estimated standard error; MCSD: Monte Carlo standard deviation; CP: 95\% confidence interval coverage.
}
\label{tab:sim_B1_small_seq}
\begin{tabular}{cclrrrrrrrr}
\toprule
\multirow{2}{*}{Direction} & \multirow{2}{*}{Working model}
  & \multirow{2}{*}{Method}
  & \multicolumn{4}{c}{h-iATE}
  & \multicolumn{4}{c}{h-cATE} \\
\cmidrule(lr){4-7}\cmidrule(lr){8-11}
 & & 
 & RBias & AESE & MCSD & CP 
 & RBias & AESE & MCSD & CP \\
 & & 
 & \multicolumn{4}{c}{\(\Delta_{\mathrm{I}}^{\mathrm{h}}=0.849\)} 
 & \multicolumn{4}{c}{\(\Delta_{\mathrm{C}}^{\mathrm{h}}=0.849\)} \\
\midrule
\multirow{13}{*}{Horizontal}
  & \textbackslash & UNADJ 
    & 0.348 & 0.375 & 0.340 & 0.961 
    & 0.375 & 0.374 & 0.338 & 0.958 \\
  & \multirow{2}{*}{W7}
    & Coef   
      & 6.985 & 0.340 & 0.359 & 0.938 
      & 6.985 & 0.340 & 0.359 & 0.938 \\
  &               & MRS    
      & 0.078 & 0.378 & 0.340 & 0.956 
      & 0.137 & 0.376 & 0.338 & 0.958 \\
  & \multirow{2}{*}{W8}
    & Coef   
      & 10.030 & 0.282 & 0.388 & 0.964 
      & 10.001 & 0.281 & 0.388 & 0.964 \\
  &               & MRS    
      & 0.159 & 0.378 & 0.339 & 0.955 
      & 0.212 & 0.376 & 0.337 & 0.958 \\
  & \multirow{2}{*}{W9}
    & Coef   
      & 11.430 & 0.165 & 0.258 & 0.840 
      & 11.430 & 0.165 & 0.258 & 0.840 \\
  &               & MRS    
      & 0.120 & 0.377 & 0.339 & 0.959 
      & 0.181 & 0.375 & 0.337 & 0.958 \\
  & \multirow{2}{*}{W10}
    & Coef   
      & 14.127 & 0.184 & 0.279 & 0.836 
      & 14.096 & 0.184 & 0.280 & 0.837 \\
  &               & MRS    
      & 0.184 & 0.377 & 0.339 & 0.960 
      & 0.236 & 0.376 & 0.337 & 0.958 \\
  & \multirow{2}{*}{W11}
    & Coef   
      & 16.326 & 0.241 & 0.256 & 0.933 
      & 16.326 & 0.241 & 0.256 & 0.933 \\
  &               & MRS    
      & 0.142 & 0.376 & 0.339 & 0.959 
      & 0.199 & 0.375 & 0.336 & 0.958 \\
  & \multirow{2}{*}{W12}
    & Coef   
      & 17.758 & 0.246 & 0.280 & 0.923 
      & 17.722 & 0.246 & 0.280 & 0.923 \\
  &               & MRS    
      & 0.205 & 0.377 & 0.339 & 0.958 
      & 0.254 & 0.375 & 0.337 & 0.959 \\
\toprule
& & 
  & \multicolumn{4}{c}{v-iATE}
  & \multicolumn{4}{c}{v-cATE} \\
  \cmidrule(lr){4-7}\cmidrule(lr){8-11}
 & & 
 & RBias & AESE & MCSD & CP 
 & RBias & AESE & MCSD & CP \\
& & 
 & \multicolumn{4}{c}{\(\Delta_{\mathrm{I}}^{\mathrm{v}}=0.849\)} 
 & \multicolumn{4}{c}{\(\Delta_{\mathrm{C}}^{\mathrm{v}}=0.849\)} \\
 \midrule
\multirow{13}{*}{Vertical}
  & \textbackslash & UNADJ 
    & 0.253 & 0.375 & 0.339 & 0.956 
    & 0.218 & 0.366 & 0.334 & 0.956 \\
  & \multirow{2}{*}{W7}
    & Coef   
      & 6.985 & 0.340 & 0.359 & 0.938 
      & 6.976 & 0.340 & 0.359 & 0.939 \\
  &               & MRS    
      & 0.037 & 0.377 & 0.338 & 0.955 
      & 0.004 & 0.368 & 0.334 & 0.960 \\
  & \multirow{2}{*}{W8}
    & Coef   
      & 10.289 & 0.284 & 0.386 & 0.967 
      & 10.280 & 0.284 & 0.386 & 0.967 \\
  &               & MRS    
      & 0.048 & 0.377 & 0.338 & 0.955 
      & 0.102 & 0.368 & 0.334 & 0.959 \\
  & \multirow{2}{*}{W9}
    & Coef   
      & 11.431 & 0.165 & 0.258 & 0.840 
      & 11.422 & 0.165 & 0.258 & 0.840 \\
  &               & MRS    
      & 0.012 & 0.376 & 0.338 & 0.959 
      & 0.040 & 0.367 & 0.333 & 0.961 \\
  & \multirow{2}{*}{W10}
    & Coef   
      & 14.220 & 0.186 & 0.278 & 0.849 
      & 14.211 & 0.186 & 0.278 & 0.849 \\
  &               & MRS    
      & 0.073 & 0.377 & 0.338 & 0.956 
      & 0.116 & 0.367 & 0.333 & 0.956 \\
  & \multirow{2}{*}{W11}
    & Coef   
      & 16.326 & 0.241 & 0.256 & 0.933 
      & 16.317 & 0.241 & 0.256 & 0.933 \\
  &               & MRS    
      & 0.030 & 0.376 & 0.337 & 0.956 
      & 0.070 & 0.366 & 0.333 & 0.961 \\
  & \multirow{2}{*}{W12}
    & Coef   
      & 17.873 & 0.247 & 0.279 & 0.923 
      & 17.863 & 0.247 & 0.279 & 0.923 \\
  &               & MRS    
      & 0.091 & 0.376 & 0.338 & 0.956 
      & 0.137 & 0.367 & 0.333 & 0.955 \\
\bottomrule
\end{tabular}
\end{table}

\begin{table}[ht!]
\centering
\caption{
Simulation results in Scenario B2 for estimating four estimands under a binary outcome with $I=9$ clusters and $J=10$ periods (switching one cluster per sequence) using covariate-adjusted working models (W7)--(W12) with the Coef, MRS and nonparametric (UNADJ) estimators. MRS: proposed augmented estimator; Coef: treatment‐effect coefficients from covariate-adjusted working model; UNADJ: nonparametric estimator. RBias: absolute bias; AESE: average estimated standard error; MCSD: Monte Carlo standard deviation; CP: 95\% confidence interval coverage.
}
\label{tab:sim_B2_small_seq}
\begin{tabular}{cclrrrrrrrr}
\toprule
\multirow{2}{*}{Direction} & \multirow{2}{*}{Working model}
  & \multirow{2}{*}{Method}
  & \multicolumn{4}{c}{h-iATE}
  & \multicolumn{4}{c}{h-cATE} \\
\cmidrule(lr){4-7}\cmidrule(lr){8-11}
 & & 
 & RBias & AESE & MCSD & CP 
 & RBias & AESE & MCSD & CP \\
 & & 
 & \multicolumn{4}{c}{\(\Delta_{\mathrm{I}}^{\mathrm{h}}=1.364\)} 
 & \multicolumn{4}{c}{\(\Delta_{\mathrm{C}}^{\mathrm{h}}=1.329\)} \\
\midrule
\multirow{13}{*}{Horizontal}
  & \textbackslash & UNADJ 
    & 7.665 & 0.402 & 0.358 & 0.953 
    & 6.789 & 0.400 & 0.357 & 0.953 \\
  & \multirow{2}{*}{W7}
    & Coef   
      & 4.438 & 0.343 & 0.355 & 0.954 
      & 7.205 & 0.343 & 0.355 & 0.947 \\
  &               & MRS    
      & 7.262 & 0.388 & 0.341 & 0.959 
      & 6.376 & 0.386 & 0.340 & 0.959 \\
  & \multirow{2}{*}{W8}
    & Coef   
      & 6.007 & 0.341 & 0.384 & 0.969 
      & 8.782 & 0.342 & 0.384 & 0.972 \\
  &               & MRS    
      & 6.691 & 0.386 & 0.341 & 0.957 
      & 5.809 & 0.384 & 0.340 & 0.961 \\
  & \multirow{2}{*}{W9}
    & Coef   
      & 8.817 & 0.178 & 0.302 & 0.796 
      & 11.700 & 0.178 & 0.302 & 0.760 \\
  &               & MRS    
      & 7.199 & 0.390 & 0.344 & 0.958 
      & 6.318 & 0.388 & 0.343 & 0.962 \\
  & \multirow{2}{*}{W10}
    & Coef   
      & 9.788 & 0.200 & 0.320 & 0.823 
      & 12.686 & 0.200 & 0.320 & 0.793 \\
  &               & MRS    
      & 6.626 & 0.388 & 0.344 & 0.954 
      & 5.749 & 0.386 & 0.343 & 0.961 \\
  & \multirow{2}{*}{W11}
    & Coef   
      & 7.029 & 0.273 & 0.300 & 0.960 
      & 9.865 & 0.273 & 0.300 & 0.951 \\
  &               & MRS    
      & 7.098 & 0.390 & 0.343 & 0.959 
      & 6.220 & 0.387 & 0.342 & 0.962 \\
  & \multirow{2}{*}{W12}
    & Coef   
      & 9.331 & 0.279 & 0.320 & 0.942 
      & 12.225 & 0.279 & 0.320 & 0.935 \\
  &               & MRS    
      & 6.546 & 0.387 & 0.342 & 0.963 
      & 5.673 & 0.385 & 0.341 & 0.966 \\
\toprule
& & 
  & \multicolumn{4}{c}{v-iATE}
  & \multicolumn{4}{c}{v-cATE} \\
  \cmidrule(lr){4-7}\cmidrule(lr){8-11}
 & & 
 & RBias & AESE & MCSD & CP 
 & RBias & AESE & MCSD & CP \\
& & 
 & \multicolumn{4}{c}{\(\Delta_{\mathrm{I}}^{\mathrm{v}}=1.364\)} 
 & \multicolumn{4}{c}{\(\Delta_{\mathrm{C}}^{\mathrm{v}}=1.064\)} \\
 \midrule
\multirow{13}{*}{Vertical}
  & \textbackslash & UNADJ 
    & 9.923 & 0.402 & 0.358 & 0.952 
    & 2.431 & 0.383 & 0.345 & 0.954 \\
  & \multirow{2}{*}{W7}
    & Coef   
      & 4.456 & 0.343 & 0.355 & 0.954 
      & 34.063 & 0.343 & 0.355 & 0.827 \\
  &               & MRS    
      & 9.551 & 0.391 & 0.341 & 0.949 
      & 2.258 & 0.373 & 0.331 & 0.963 \\
  & \multirow{2}{*}{W8}
    & Coef   
      & 3.597 & 0.341 & 0.384 & 0.973 
      & 32.960 & 0.341 & 0.384 & 0.589 \\
  &               & MRS    
      & 8.970 & 0.388 & 0.341 & 0.946 
      & 1.764 & 0.373 & 0.332 & 0.962 \\
  & \multirow{2}{*}{W9}
    & Coef   
      & 8.836 & 0.178 & 0.302 & 0.795 
      & 39.684 & 0.178 & 0.302 & 0.485 \\
  &               & MRS    
      & 9.492 & 0.393 & 0.345 & 0.950 
      & 2.154 & 0.373 & 0.332 & 0.961 \\
  & \multirow{2}{*}{W10}
    & Coef   
      & 7.391 & 0.201 & 0.320 & 0.836 
      & 37.829 & 0.201 & 0.320 & 0.581 \\
  &               & MRS    
      & 8.906 & 0.391 & 0.344 & 0.952 
      & 1.672 & 0.373 & 0.332 & 0.961 \\
  & \multirow{2}{*}{W11}
    & Coef   
      & 7.048 & 0.273 & 0.300 & 0.960 
      & 37.390 & 0.273 & 0.300 & 0.772 \\
  &               & MRS    
      & 9.402 & 0.392 & 0.344 & 0.952 
      & 2.129 & 0.372 & 0.331 & 0.963 \\
  & \multirow{2}{*}{W12}
    & Coef   
      & 6.765 & 0.279 & 0.319 & 0.948 
      & 37.026 & 0.279 & 0.319 & 0.784 \\
  &               & MRS    
      & 8.833 & 0.390 & 0.343 & 0.953 
      & 1.647 & 0.372 & 0.331 & 0.961 \\
\bottomrule
\end{tabular}
\end{table}

\begin{table}[ht!]
\centering
\caption{
Simulation results in Scenario B3 for estimating four estimands under a binary outcome with $I=9$ clusters and $J=10$ periods (switching one cluster per sequence) using covariate-adjusted working models (W7)--(W12) with the Coef, MRS and nonparametric (UNADJ) estimators. MRS: proposed augmented estimator; Coef: treatment‐effect coefficients from covariate‐adjusted working model; UNADJ: nonparametric estimator. RBias: absolute bias; AESE: average estimated standard error; MCSD: Monte Carlo standard deviation; CP: 95\% confidence interval coverage.
}
\label{tab:sim_B3_small_seq}
\begin{tabular}{cclrrrrrrrr}
\toprule
\multirow{2}{*}{Direction} & \multirow{2}{*}{Working model}
  & \multirow{2}{*}{Method}
  & \multicolumn{4}{c}{h-iATE}
  & \multicolumn{4}{c}{h-cATE} \\
\cmidrule(lr){4-7}\cmidrule(lr){8-11}
 & & 
 & RBias & AESE & MCSD & CP 
 & RBias & AESE & MCSD & CP \\
 & & 
 & \multicolumn{4}{c}{\(\Delta_{\mathrm{I}}^{\mathrm{h}}=1.321\)} 
 & \multicolumn{4}{c}{\(\Delta_{\mathrm{C}}^{\mathrm{h}}=1.297\)} \\
\midrule
\multirow{13}{*}{Horizontal}
  & \textbackslash & UNADJ 
    & 1.112 & 0.291 & 0.252 & 0.961 
    & 0.674 & 0.292 & 0.253 & 0.963 \\
  & \multirow{2}{*}{W7}
    & Coef   
      & 11.296 & 0.290 & 0.267 & 0.947 
      & 13.410 & 0.290 & 0.267 & 0.929 \\
  &               & MRS    
      & 1.605 & 0.294 & 0.250 & 0.964 
      & 0.983 & 0.294 & 0.250 & 0.961 \\
  & \multirow{2}{*}{W8}
    & Coef   
      & 15.531 & 0.294 & 0.278 & 0.943 
      & 17.700 & 0.294 & 0.278 & 0.934 \\
  &               & MRS    
      & 1.457 & 0.295 & 0.250 & 0.965 
      & 0.848 & 0.296 & 0.250 & 0.963 \\
  & \multirow{2}{*}{W9}
    & Coef   
      & 12.174 & 0.114 & 0.238 & 0.626 
      & 14.304 & 0.114 & 0.238 & 0.591 \\
  &               & MRS    
      & 1.563 & 0.297 & 0.252 & 0.961 
      & 0.964 & 0.298 & 0.252 & 0.962 \\
  & \multirow{2}{*}{W10}
    & Coef   
      & 16.937 & 0.119 & 0.255 & 0.540 
      & 19.137 & 0.119 & 0.255 & 0.510 \\
  &               & MRS    
      & 1.456 & 0.300 & 0.254 & 0.961 
      & 0.866 & 0.300 & 0.254 & 0.965 \\
  & \multirow{2}{*}{W11}
    & Coef   
      & 5.835 & 0.218 & 0.231 & 0.962 
      & 7.845 & 0.218 & 0.231 & 0.953 \\
  &               & MRS    
      & 1.230 & 0.300 & 0.254 & 0.965 
      & 0.618 & 0.300 & 0.254 & 0.961 \\
  & \multirow{2}{*}{W12}
    & Coef   
      & 18.653 & 0.213 & 0.264 & 0.816 
      & 20.883 & 0.213 & 0.264 & 0.790 \\
  &               & MRS    
      & 1.474 & 0.301 & 0.256 & 0.957 
      & 0.861 & 0.302 & 0.256 & 0.961 \\
\toprule
& & 
  & \multicolumn{4}{c}{v-iATE}
  & \multicolumn{4}{c}{v-cATE} \\
  \cmidrule(lr){4-7}\cmidrule(lr){8-11}
 & & 
 & RBias & AESE & MCSD & CP 
 & RBias & AESE & MCSD & CP \\
& & 
 & \multicolumn{4}{c}{\(\Delta_{\mathrm{I}}^{\mathrm{v}}=1.163\)} 
 & \multicolumn{4}{c}{\(\Delta_{\mathrm{C}}^{\mathrm{v}}=1.025\)} \\
 \midrule
\multirow{13}{*}{Vertical}
  & \textbackslash & UNADJ 
    & 3.389 & 0.282 & 0.259 & 0.953 
    & 1.418 & 0.274 & 0.253 & 0.957 \\
  & \multirow{2}{*}{W7}
    & Coef   
      & 22.615 & 0.290 & 0.267 & 0.891 
      & 43.446 & 0.290 & 0.267 & 0.770 \\
  &               & MRS    
      & 3.000 & 0.279 & 0.248 & 0.966 
      & 1.854 & 0.273 & 0.244 & 0.960 \\
  & \multirow{2}{*}{W8}
    & Coef   
      & 14.680 & 0.290 & 0.264 & 0.905 
      & 34.162 & 0.290 & 0.264 & 0.806 \\
  &               & MRS    
      & 2.678 & 0.281 & 0.248 & 0.966 
      & 1.626 & 0.274 & 0.242 & 0.959 \\
  & \multirow{2}{*}{W9}
    & Coef   
      & 23.582 & 0.114 & 0.238 & 0.447 
      & 44.577 & 0.114 & 0.238 & 0.216 \\
  &               & MRS    
      & 2.948 & 0.282 & 0.251 & 0.965 
      & 1.904 & 0.275 & 0.245 & 0.961 \\
  & \multirow{2}{*}{W10}
    & Coef   
      & 15.920 & 0.122 & 0.238 & 0.623 
      & 35.613 & 0.122 & 0.238 & 0.349 \\
  &               & MRS    
      & 2.639 & 0.285 & 0.252 & 0.967 
      & 1.643 & 0.277 & 0.244 & 0.958 \\
  & \multirow{2}{*}{W11}
    & Coef   
      & 16.599 & 0.218 & 0.231 & 0.901 
      & 36.407 & 0.218 & 0.231 & 0.707 \\
  &               & MRS    
      & 2.469 & 0.284 & 0.251 & 0.963 
      & 2.170 & 0.277 & 0.246 & 0.956 \\
  & \multirow{2}{*}{W12}
    & Coef   
      & 16.408 & 0.209 & 0.244 & 0.878 
      & 36.183 & 0.209 & 0.244 & 0.679 \\
  &               & MRS    
      & 2.536 & 0.286 & 0.253 & 0.963 
      & 1.572 & 0.278 & 0.246 & 0.953 \\
\bottomrule
\end{tabular}
\end{table}

\clearpage
\section{Example code for the MRStdLCRT R package} \label{pack_ex}

We demonstrate the use of the \texttt{mrstdlcrt\_fit} function in the \texttt{MRStdLCRT} package to compute model-robust standardized estimators of four treatment effects in stepped-wedge cluster-randomized trials, namely \texttt{h-iATE}, \texttt{h-cATE}, \texttt{v-iATE}, and \texttt{v-cATE}. The function reports both unadjusted and augmented (covariate-adjusted) estimators on a user-selected scale (risk difference, log relative risk, or log odds ratio), and quantifies uncertainty via a delete-one-cluster jackknife variance estimator. The accompanying \texttt{summary} method additionally reports key design diagnostics (e.g., the mixed periods used for aggregation and stepped-wedge edge patterns) and, by default, a global linear-contrast F-test assessing discrepancies among the four estimands.

The function has the following declaration:
\begin{verbatim}
mrstdlcrt_fit(
  data, formula,
  cluster_id = "cluster",
  period     = "period",
  trt        = "trt",
  method     = c("gee","lmer","glmer"),
  family     = c("gaussian","binomial"),
  corstr     = "independence",
  scale      = c("RD","RR","OR")
)
\end{verbatim}

with arguments:
\begin{itemize}
  \item \texttt{data}: A data frame with one row per subject.
  \item \texttt{formula}: A model formula for the outcome; fixed effects are supported for all methods, and random effects may be included when using \texttt{lmer}/\texttt{glmer}.
  \item \texttt{cluster\_id}: A string giving the cluster ID column name.
  \item \texttt{period}: A string giving the period column name (values $1,2,\dots,J$).
  \item \texttt{trt}: A string giving the binary treatment indicator column name (0/1).
  \item \texttt{method}: A string specifying the modeling approach: \texttt{"gee"}, \texttt{"lmer"} (continuous outcomes), or \texttt{"glmer"} (binary outcomes).
  \item \texttt{family}: Outcome family, \texttt{"gaussian"} or \texttt{"binomial"}.
  \item \texttt{corstr}: (GEE only) working correlation structure, e.g., \texttt{"independence"} or \texttt{"exchangeable"}.
  \item \texttt{scale}: For binary outcomes, the estimand scale: \texttt{"RD"}, \texttt{"RR"} (log risk ratio), or \texttt{"OR"} (log odds ratio).
\end{itemize}

\subsubsection*{Example}

We illustrate the function using the example dataset \texttt{sw\_c} shipped with the package:
\begin{verbatim}
> suppressPackageStartupMessages(library(MRStdLCRT))
> data("sw_c", package = "MRStdLCRT")

> fml_c <- y ~ 0 + factor(period) + trt + x1 + x2 + (1 | cluster)

> fit_c <- mrstdlcrt_fit(
+   data       = sw_c,
+   formula    = fml_c,
+   cluster_id = "cluster",
+   period     = "period",
+   trt        = "trt",
+   method     = "gee",
+   family     = "gaussian",
+   scale      = "RD"
+ )

> summary(fit_c)

MRS-LCRT Summary
========================================================================
Method: gee   Family: gaussian   Scale: RD
GEE corstr: independence
Clusters: 30   Periods kept: 4 (of 6 total)
Kept periods: 2, 3, 4, 5
CIs: 95.0% (t, df = 29)
Stepped-wedge edge pattern: YES (first period 1 all control; last period 6 all treatment)

ICS hypothesis test (linear-contrast F test)
------------------------------------------------------------------------
  method_type df1 df2        F p_value
1  unadjusted   3  29 13.14064 1.3e-05
2    adjusted   3  29 13.42656 1.1e-05

h-iATE
           Estimate       SE      LCL      UCL
unadjusted 8.202067 1.248330 5.648944 10.75519
adjusted   8.269544 1.249144 5.714757 10.82433

h-cATE
           Estimate       SE      LCL      UCL
unadjusted 7.361880 1.345551 4.609919 10.11384
adjusted   7.418796 1.332380 4.693772 10.14382

v-iATE
           Estimate       SE      LCL      UCL
unadjusted 8.131903 1.238139 5.599625 10.66418
adjusted   8.200374 1.239530 5.665251 10.73550

v-cATE
           Estimate       SE      LCL      UCL
unadjusted 5.499351 1.126920 3.194540 7.804162
adjusted   5.584761 1.108609 3.317401 7.852122
\end{verbatim}

\section{Estimands with implementation periods}\label{imp_est}
\begin{figure}[htbp!]
    \centering
    \includegraphics[width=0.8\linewidth]{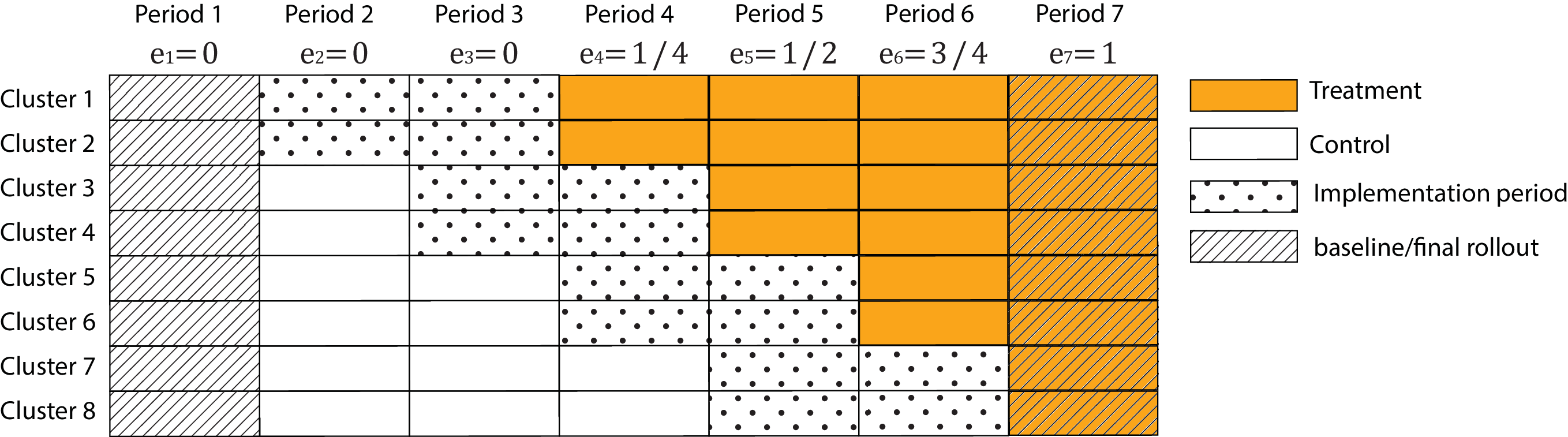}
    \caption{Illustration of a stepped-wedge cluster-randomized trial with an implementation period. The dashed columns indicate the baseline and final rollout periods, which are excluded from the estimand definition. Colored cells indicate intervention exposure after a cluster adopts the intervention; the dotted cells represent the implementation period immediately following adoption and are excluded from the estimand definition. Only non-dotted, non-dashed cluster-periods during the rollout contribute to the treatment effect estimands.}
    \label{fig:sw_imp}
\end{figure}

In many SW-CRTs, intervention adoption at time $A_i$ does not coincide with immediate steady-state delivery. To define treatment effects for the implemented intervention (see Figure \ref{fig:sw_imp}), we allow an implementation phase whose duration may vary across clusters. Specifically, in addition to the adoption time $A_i$, we define for each cluster $i$ a full-implementation start time $B_i\in\{A_i,\ldots,J+1\}$, with $B_i\ge A_i$, where $B_i$ is the first period at which the cluster is regarded as fully implemented. Under this, the implementation periods for cluster $i$ are $\{A_i,\ldots,B_i-1\}$, which are excluded from the estimand definition. We retain the general estimand class in \eqref{weight_causal_effect}, and \eqref{weight_potential}, but implement the exclusion by assigning zero weight to cluster-periods in the implementation phase. Define the indicator
\(G_{ij}=\mathbb{I}\{2\le j\le J-1\}\,\mathbb{I}\{j<A_i\ \text{or}\ j\ge B_i\}\). All estimands redefined by specifying $\omega_{ijk}^{\mathrm{imp}}$ and the induced $\omega_{ij}^{\mathrm{imp}}=\sum_{k=1}^{N_{ij}}\omega_{ijk}^{\mathrm{imp}}$, with $G_{ij}$ ensuring that baseline ($j=1$), final rollout ($j=J$), and implementation-period cluster-periods contribute zero weight. The implementation-adjusted weighted mean potential outcome is
\[
\mu_{\omega}^{\mathrm{imp}}(z)=\frac{\mathbb{E}\left[\sum_{j=2}^{J-1}\omega_{ij}^{\mathrm{imp}}\,\overline{Y}_{ij}(z)\right]}{\mathbb{E}\left[\sum_{j=2}^{J-1}\omega_{ij}^{\mathrm{imp}}\right]}, \qquad z\in\{0,1\},
\]
and the corresponding estimand is $\tau_{\omega}^{\mathrm{imp}}=f\{\mu_{\omega}^{\mathrm{imp}}(1),\mu_{\omega}^{\mathrm{imp}}(0)\}$. The h-iATE excluding implementation periods pools all eligible individuals across all eligible cluster-periods, assigning each eligible individual equal weight. This is obtained by setting \(\omega_{ijk}^{\text{imp}}=G_{ij}\), and \(\omega_{ij}^{\text{imp}}=G_{ij}N_{ij}\).
Under these weights, the estimand targets the mean contrast over the super-population of individuals represented in non-excluded cluster-periods during roll-out.

The h-cATE excluding implementation periods aggregates within each cluster over eligible periods and then averages equally over clusters, so that each cluster contributes the same total weight regardless of its eligible size. Let the eligible total size for cluster $i$ be \(N_i^{\star}=\sum_{j=2}^{J-1} G_{ij}N_{ij}\). This estimand is defined with \(\omega_{ijk}^{\text{imp}}=\frac{G_{ij}}{N_i^{\star}}\), and \(\omega_{ij}^{\text{imp}}=\frac{G_{ij}N_{ij}}{N_i^{\star}}\), which ensures that $\sum_{j=2}^{J-1}\omega_{ij}^{\text{imp}}=1$ for each cluster.

The v-iATE excluding implementation periods constructs an individual-average contrast within each calendar period among eligible cluster-periods and then averages uniformly across periods. To express this under the super-population expectation used in \eqref{weight_potential}, set
\(\omega_{ijk}^{\text{imp}}=\frac{G_{ij}}{I\mathbb{E}\left[G_{ij}N_{ij}\right]} \), and \(\omega_{ij}^{\text{imp}}=\frac{G_{ij}N_{ij}}{I \mathbb{E}\left[G_{ij}N_{ij}\right]}\).
With these weights, $\mu_{\omega}^{\mathrm{imp}}(z)$ corresponds to a uniform average over $j=2,\ldots,J-1$ of the period-specific mean potential outcome among eligible observations.

The v-cATE excluding implementation periods targets the mean of cluster-period mean potential outcomes across eligible cluster-periods, giving each eligible cluster-period equal weight. This is obtained by
\( \omega_{ijk}^{\text{imp}}=\frac{G_{ij}}{N_{ij}} \), \(\omega_{ij}^{\text{imp}}=G_{ij}\).
This assigns total weight one to each eligible cluster-period and thus defines a cluster-period average estimand over the roll-out after excluding the implementation phase $\{A_i,\ldots,B_i-1\}$ for each cluster. The definitions of the four estimands are summarized in Table \ref{tab:estimands_imp}.

\begin{table}[htbp]
    \centering
    \caption{Summary of four treatment effect estimands $\tau_{\omega}^{\mathrm{imp}}$ in SW-CRTs excluding implementation-phase cluster-periods.}
    \label{tab:estimands_imp}
\resizebox{\textwidth}{!}{%
\begin{tabular}{ccccp{3.6cm}}
\toprule
\textbf{Aggregation} & \textbf{Estimand} & \textbf{Weights} & \textbf{Definition} & \textbf{Remarks} \\
\midrule
\multirow{8}{*}{Horizontal}
& \multirow{4}{*}{h-iATE}
& $\omega_{ijk}^{\mathrm{imp}} = G_{ij}$
& \multirow{2}{*}{$\mu_{\mathrm{I}}^{\mathrm{h,imp}}(z)=
\frac{\mathbb{E}\left[\sum_{j=2}^{J-1} G_{ij} N_{ij}\,\overline{Y}_{ij}(z)\right]}
{\mathbb{E}\left[\sum_{j=2}^{J-1} G_{ij} N_{ij}\right]}$}
& \multirow{4}{3.6cm}{Equal weight for each eligible individual} \\
& & $\omega_{ij}^{\mathrm{imp}} = G_{ij}N_{ij}$ & & \\
& & $\omega_{j}^{\mathrm{imp}} = \sum_{i=1}^{I} G_{ij}N_{ij}$
& \multirow{2}{*}{$\tau_{\mathrm{I}}^{\mathrm{h,imp}}=f\{\mu_{\mathrm{I}}^{\mathrm{h,imp}}(1),\mu_{\mathrm{I}}^{\mathrm{h,imp}}(0)\}$} & \\
& & $\omega_{i}^{\mathrm{imp}} = \sum_{j=2}^{J-1} G_{ij}N_{ij}$ & & \\
\cdashline{2-5}
& \multirow{4}{*}{h-cATE}
& $\omega_{ijk}^{\mathrm{imp}} = G_{ij}/N_i^\star$
& \multirow{2}{*}{$\mu_{\mathrm{C}}^{\mathrm{h,imp}}(z)=
\mathbb{E}\left[\frac{\sum_{j=2}^{J-1} G_{ij} N_{ij}\,\overline{Y}_{ij}(z)}{N_i^\star}\right]$}
& \multirow{4}{3.6cm}{Equal weight for each cluster over eligible periods} \\
& & $\omega_{ij}^{\mathrm{imp}} = G_{ij}N_{ij}/N_i^\star$ & & \\
& & $\omega_{j}^{\mathrm{imp}} = \sum_{i=1}^{I}\left(G_{ij}N_{ij}/N_i^\star\right)$
& \multirow{2}{*}{$\tau_{\mathrm{C}}^{\mathrm{h,imp}}=f\{\mu_{\mathrm{C}}^{\mathrm{h,imp}}(1),\mu_{\mathrm{C}}^{\mathrm{h,imp}}(0)\}$} & \\
& & $\omega_{i}^{\mathrm{imp}} = 1$ & & \\
\midrule
\multirow{8}{*}{Vertical}
& \multirow{4}{*}{v-iATE}
& $\omega_{ijk}^{\mathrm{imp}} = \dfrac{G_{ij}}{I\,\mathbb{E}[G_{ij}N_{ij}]}$
& \multirow{2}{*}{$\mu_{\mathrm{I}}^{\mathrm{v,imp}}(z)=
\frac{1}{J-2}\sum_{j=2}^{J-1}\frac{\mathbb{E}\left[G_{ij}\sum_{k=1}^{N_{ij}}Y_{ijk}(z)\right]}{\mathbb{E}[G_{ij}N_{ij}]}$}
& \multirow{4}{3.6cm}{Equal weight for each period among eligible cluster-periods} \\
& & $\omega_{ij}^{\mathrm{imp}} = \dfrac{G_{ij}N_{ij}}{I\,\mathbb{E}[G_{ij}N_{ij}]}$ & & \\
& & $\omega_{j}^{\mathrm{imp}} = 1$
& \multirow{2}{*}{$\tau_{\mathrm{I}}^{\mathrm{v,imp}}=f\{\mu_{\mathrm{I}}^{\mathrm{v,imp}}(1),\mu_{\mathrm{I}}^{\mathrm{v,imp}}(0)\}$} & \\
& & $\omega_{i}^{\mathrm{imp}} = \sum_{j=2}^{J-1}\dfrac{G_{ij}N_{ij}}{I\,\mathbb{E}[G_{ij}N_{ij}]}$ & & \\
\cdashline{2-5}
& \multirow{4}{*}{v-cATE}
& $\omega_{ijk}^{\mathrm{imp}} = G_{ij}/N_{ij}$
& \multirow{2}{*}{$\mu_{\mathrm{C}}^{\mathrm{v,imp}}(z)=
\frac{1}{J-2}\sum_{j=2}^{J-1}\mathbb{E}\left[G_{ij}\overline{Y}_{ij}(z)\right]\Big/\mathbb{E}[G_{ij}]$}
& \multirow{4}{3.6cm}{Equal weight for each eligible cluster-period .} \\
& & $\omega_{ij}^{\mathrm{imp}} = G_{ij}$ & & \\
& & $\omega_{j}^{\mathrm{imp}} = \sum_{i=1}^{I} G_{ij}$
& \multirow{2}{*}{$\tau_{\mathrm{C}}^{\mathrm{v,imp}}=f\{\mu_{\mathrm{C}}^{\mathrm{v,imp}}(1),\mu_{\mathrm{C}}^{\mathrm{v,imp}}(0)\}$} & \\
& & $\omega_{i}^{\mathrm{imp}} = \sum_{j=2}^{J-1} G_{ij}$ & & \\
\bottomrule
\end{tabular}
}
\end{table}

To estimate the estimands defined above, our proposed model-robust standardization estimators can be implemented in exactly the same manner as described in Section \ref{method} of the main paper. The unadjusted estimator in \eqref{non_adj} is obtained by the substitution \(\omega_{ij}\) with \(\omega_{ij}^{\mathrm{imp}}\). Likewise, the model-robust standardization estimators in \eqref{aug_mu} are implemented by the same substitution, with the same model-robustness properties as described in Section \ref{method}. The working model $m_{zj}(\underline{\bm{X}}_{ij},N_{ij})$ enters through the augmentation term and can therefore be specified using standard regression in the main paper. Because the target estimand excludes implementation-phase cluster-periods, the direct way is to fit $m_{zj}(\underline{\bm{X}}_{ij},N_{ij})$ among eligible cluster-period, i.e., $\mathbb{E}\{\overline{Y}_{ij}(z)\mid \underline{\bm{X}}_{ij},N_{ij},G_{ij}=1\}$. However, this restriction is not necessary for consistency, since the estimator remains consistent for $\mu_{\omega}^{\mathrm{imp}}(z)$ for any working model $m_{zj}(\cdot)$, regardless of whether the regression is fit using only eligible cluster-periods or using all observed cluster-periods.

\end{document}